\def\isarxiv{1}

\ifdefined\isarxiv
\documentclass[11pt]{article}
\else
\documentclass{article}
\fi

\usepackage{tabu}
\usepackage{array}
\usepackage{amsmath}
\usepackage{amsthm}
\usepackage{amssymb}
\usepackage{algorithm}

\usepackage{subfig}
\usepackage{color}
\usepackage[english]{babel}
\usepackage{graphicx}
\usepackage{grffile}
\usepackage{wrapfig,epsfig}
\usepackage{epstopdf}
\usepackage{url}
\usepackage{color}
\usepackage{epstopdf}
\usepackage{algpseudocode}
\usepackage[T1]{fontenc}
\usepackage{bbm}
\usepackage{comment}
\usepackage{dsfont}
\usepackage{mathtools}
\usepackage{enumitem}

\usepackage{tikz}
\usepackage{hyperref}  
\hypersetup{
    colorlinks=true,
    citecolor=red,
    linkcolor=blue,
    filecolor=cyan,
    urlcolor=magenta,
}
\usetikzlibrary{arrows}
\ifdefined\isarxiv
\usepackage[margin=1in]{geometry}
\else
\fi
\graphicspath{{./figs/}}

\usepackage[nameinlink,capitalize]{cleveref}
\usepackage{textcomp}
\usepackage{multirow}
\newtheorem{theorem}{Theorem}[section]
\newtheorem{lemma}[theorem]{Lemma}
\newtheorem{definition}[theorem]{Definition}

\newtheorem{corollary}[theorem]{Corollary}

\newtheorem{fact}[theorem]{Fact}

\newcommand{\wh}{\widehat}

\newcommand{\ov}{\overline}

\newcommand{\A}{\mathcal{A}}

\newcommand{\F}{\mathcal{F}}
\newcommand{\R}{\mathbb{R}}
\renewcommand{\P}{\mathbb{P}}

\renewcommand{\d}{\mathrm{d}}

\renewcommand{\hat}{\wh}

\newcommand{\dist}{\mathrm{dist}}
\renewcommand{\d}{\mathrm{d}}

\DeclareMathOperator*{\E}{{\mathbb{E}}}

\DeclareMathOperator{\rank}{rank}

\newcommand{\vect}{\mathrm{vec}}

\definecolor{mygreen}{RGB}{80,180,0}
\definecolor{b2}{RGB}{51,153,255}

\newcommand{\mat}[1]{\ensuremath{\mathbf{#1}}}

\newcommand{\I}{\mat{I}}

\DeclareMathOperator*{\argmax}{arg\,max}

\newcommand{\maxip}{\mathsf{Max}\text{-}\mathsf{IP}}
\newcommand{\maxmatnorm}{\mathsf{Max}\text{-}\mathsf{Mat}\mathsf{Norm}} 
\newcommand{\nn}{\mathsf{NN}} 
\newcommand{\ann}{\mathsf{ANN}} 
\newcommand{\lsh}{$\mathsf{LSH}$}
\newcommand{\mdp}{$\rm{MDP}$}
\newcommand{\core}{\mathrm{core}}
\newcommand{\fail}{$\mathsf{fail}$}

\title{Sublinear Least-Squares Value Iteration via\\ Locality Sensitive Hashing}

\ifdefined\isarxiv
\author{Anshumali Shrivastava\thanks{\texttt{anshumali@rice.edu}. Rice University.}
\and 
Zhao Song\thanks{\texttt{zhaos@ias.edu}. Institute for Advanced Study, Princeton University.}
\and
Zhaozhuo Xu\thanks{\texttt{zx22@rice.edu}. Rice University.}
}
\else

\fi
\date{}

\allowdisplaybreaks

\begin{document}

\ifdefined\isarxiv
\begin{titlepage}
  \maketitle
  \begin{abstract}

We present the first provable Least-Squares Value Iteration (LSVI) algorithms that achieves runtime complexity sublinear in the number of actions.  We formulate the value function estimation procedure in value iteration as an approximate maximum inner product search problem and propose a  locality sensitive hashing ({\lsh}) [Indyk and Motwani STOC'98, Andoni and Razenshteyn STOC'15, Andoni,  Laarhoven, Razenshteyn and Waingarten SODA'17] type data structure to solve this problem with sublinear time complexity. Moreover, we build the connections between the theory of approximate maximum inner product search and the regret analysis of reinforcement learning. We prove that, with our choice of approximation factor, our Sublinear LSVI algorithms maintain the same regret as the original LSVI algorithms while reducing the runtime complexity to sublinear in the number of actions. To the best of our knowledge, this is the first work that combines {\lsh} with reinforcement learning resulting in provable improvements.  We hope that our novel way of combining data structures and iterative algorithm will open the door for further study into cost reduction in optimization.
  \end{abstract}
  \thispagestyle{empty}
\end{titlepage}

\else

\maketitle
  \begin{abstract}

We present the first provable Least-Squares Value Iteration (LSVI) algorithms that achieves runtime complexity sublinear in the number of actions.  We formulate the value function estimation procedure in value iteration as an approximate maximum inner product search problem and propose a  locality sensitive hashing ({\lsh}) [Indyk and Motwani STOC'98, Andoni and Razenshteyn STOC'15, Andoni,  Laarhoven, Razenshteyn and Waingarten SODA'17] type data structure to solve this problem with sublinear time complexity. Moreover, we build the connections between the theory of approximate maximum inner product search and the regret analysis of reinforcement learning. We prove that, with our choice of approximation factor, our Sublinear LSVI algorithms maintain the same regret as the original LSVI algorithms while reducing the runtime complexity to sublinear in the number of actions. To the best of our knowledge, this is the first work that combines {\lsh} with reinforcement learning resulting in provable improvements.  We hope that our novel way of combining data structures and iterative algorithm will open the door for further study into cost reduction in optimization.
  \end{abstract}

\fi

\section{Introduction}

Reinforcement learning (RL) is an essential problem in machine learning that targets maximizing the cumulative reward when an agent is taking actions within an unknown environment~\cite{s18}. RL is a trending topic over the last few years. We have seen
a remarkable growth of RL applications in Go~\cite{shm+16}, robotics~\cite{kbp13}, dialogue systems~\cite{lmr+16} and recommendation~\cite{zzz+18}.  In practical RL, most approaches~\cite{wd92,sld+14,jazbj18} perform iterative-type algorithms that modify the choice of actions at each step based on the agent iteration with the environment. This iterative natural causes the training of RL algorithms to be expensive. For instance, it takes around three weeks to train the agent in AlphaGo~\cite{shm+16}. Moreover, the training is conducted on 50 GPUs, which means the training of RL on limited computational resources is almost infeasible.

Given the efficiency bottleneck of RL algorithms, it is natural to ask the following question.
\begin{center}
{\it
Are there some TCS techniques that could apply to iterative-type RL algorithms and improve their running time efficiency? 
}
\end{center}

The practical success of a typical TCS technique, Locality sensitive hashing ({\lsh}), shed lights on answering the question. {\lsh} is a randomized data structure with provable efficiency in approximate nearest neighbor search (${\ann}$)~\cite{im98,c02,diim04,sdi05,ai08,a09,ainr14,ar15,ailrs15,arm17,c17,r17,air18,w19,dirw20}. Meanwhile, {\lsh} could also be extended to maximum inner product search ($\maxip$)~\cite{sl14}. Moreover, in practical machine learning (ML), {\lsh} has been widely used in many fundamental learning problems to improve the practical running time of iterative-type algorithms such as gradient descents~\cite{cxs19}, back-propagation~\cite{cmfgts20,dmzs21,clp+21} and  MCMC sampling~\cite{ls19}.  However, the current empirical combination of {\lsh} with iterative-type algorithms does not have theoretical support. It is unknown to give a provable guarantee for the impact of {\lsh} over the total number of iterations and per cost iteration of iterative-type algorithms.

Inspired by a large number of successes about using {\lsh} to tackle efficiency bottlenecks in practice, it is natural to ask the following question.

\begin{center}
{\it
Is there an interesting regime (e.g., some iterative-type algorithms) where we can apply {\lsh} to give provable improvement?
}
\end{center}

In this work, we answer both questions by proposing a theoretical framework that combines {\lsh} with RL. We focus on Q-learning~\cite{wd92}, a simple and flexible type of RL framework that directly optimizes the maximum expected reward based on the outcome of actions that the agent taken at each step. Theoretical analysis also suggests that Q-learning is proved to be sample efficient~\cite{jazbj18}. However, the running time efficiency of Q-learning requires improvement in practical scenarios. We identify that the runtime complexity of Q-learning is dominated by the value function estimation procedure. Value function estimation requires a linear scan over all the actions at each step, which is unscalable in real RL tasks. For instance, in news recommendation systems, the action of an RL agent is recommending an article to the users. The iterative-type Q-learning algorithm scan over all articles at each iteration to find the action that maximizes the expected reward. In practice, this search space is too large so that linear scan is prohibitive. Therefore, reducing the enormous overhead in value function estimation over the large action space becomes a significant research problem in Q-learning.

We focus on applying {\lsh} techniques to reduce this value function estimation overhead in the iterative-type Q-learning algorithm. However, combing {\lsh} with any iterative-type algorithm in Q-learning is challenging due to four major reasons: (1) It remains unknown whether the linear scan over all possible actions in Q-learning could be formulated as an $\ann$ or $\maxip$ problem  (2) {\lsh} accelerate this linear scan by introducing an error in estimating value function.  This approximation error would accumulate in the value iteration and break the current upper bound for regret. (3) Although {\lsh} has demonstrated success in practical ML, its theoretical efficiency guarantee in RL remains unknown. (4) The Q-learning algorithm would query {\lsh} at each step. As the query in each step depends on the previous step, the total failure probability of {\lsh} over this adaptive query sequence could not be union bounded due to correlations.

In this work, we solve these challenges affirmatively by presenting a Q-learning algorithm that uses {\lsh} type approximate $\maxip$ data structure. We focus on the Least-Squares Value Iteration (LSVI)~\cite{bb96} and its extensions with UCB exploration (LSVI-UCB~\cite{jywj20}). We also discuss LSVI-UCB under policy switch limitation~\cite{gxdy21} or model-free setting~\cite{wdys20}.  We connect the theory of $\maxip$ with reinforcement learning by formulating the value function estimation in LSVI and LSVI-UCB as an approximate $\maxip$ problem. Then, we propose Sublinear LSVI and Sublinear LSVI-UCB, two algorithms with {\lsh} that have value iteration running time sublinear in the number of actions. For LSVI-UCB, we extend the {\lsh} type $\maxip$ data structure to approximate maximum matrix norm search so that Sublinear LSVI-UCB could also enjoy the sublinear value iteration complexity over actions. Moreover, we theoretically prove that, with our choice of approximation factor, both Sublinear LSVI and Sublinear LSVI-UCB achieve the same regret with their original versions. Furthermore, we identify the potential risks of {\lsh} type approximate $\maxip$ data structure in iterative-type algorithm and proposes a series of techniques to reduce them.

\section{Related Work}\label{sec:related}
\paragraph{Approximate Maximum Inner Product Search} \emph{Maximum Inner Product Search} ($\maxip$) is a fundamental yet challenging problem in theoretical computer science~\cite{w05,arw17,c18,cw19,w18}.  Given a query $x\in \R^d$ and a dataset $Y \subset \R^d$ with $n$ vectors, the goal of $\maxip$ is to retrieve a $z\in Y$ so that $x^\top z =\arg\max_{z\in Y} x^\top y$. The brute-force algorithm solves $\maxip$ in $O(dn)$ time for $x$ by linear scanning over all elements in $Y$.  To improve the $\maxip$ efficiency in practice, approximation methods are proposed to achieve sublinear query time complexity by returning point with a multiplicative approximation ratio to the  $\maxip$ solution.

Chen~\cite{c18} show that for bichromatic $\maxip$\footnote{Given two $n$-point set $A\in \R^d$ and $B\in \R^d$, the goal of bichromatic $\maxip$ is to find $b\in B$ that maximize inner product for every $a\in A$.} with two set of $n$ vectors from $\{0,1\}^d$, there is a $n^{2-\Omega(1)}$ time algorithm with  $(d/\log n)^{\Omega(1)}$ approximation ratio. Moreover, Chen~\cite{c18} show that this algorithm is conditional optimal as such a  $(d/\log n)^{o(1)}$ approximation algorithm would refute Strong Exponential Time Hypothesis ($\mathsf{SETH}$) \cite{ip01}\footnote{SETH (Strong Exponential Time Hypothesis) states that for every $\epsilon >0$  there is a $k$ such that $k$-SAT cannot be solved in $O((2-\epsilon)^n)$ time.}.

Most previous approximate  $\maxip$ approaches reduce the $\maxip$ to nearest neighbor ($\nn$) search problem and apply approximate nearest neighbor ($\ann$) data structures such as locality sensitive hashing ({\lsh})~\cite{sl14,sl15_uai,ns15,sl15_www,yldcc18}. Given a query $x\in \R^d$ and a dataset $Y \subset \R^d$ with $n$ vectors, the goal of $(\ov{c},r)$-$\ann$ with $\ov{c}>1$ is to retrieve a $z\in Y$ so that $\|x-z\|_2\leq \ov{c}\cdot r$ if there $\min_{y\in Y} \|x-y\|_2\leq r$.  The {\lsh} solves this problem with query time in $O(d\cdot n^{\rho+o(1)})$. Here, $\rho<1$ and it depends on $\ov{c}$. For randomized {\lsh} that is independent of data, Antoni, Indyk and Razenshteyn~\cite{air18} show that $\rho\geq 1/\ov{c}^2$. To further reduce $\rho$, Antoni and Razenshteyn~\cite{ar15} proposes a data-dependent {\lsh} that achieves $\rho=1/(2\ov{c}^2-1)$ with preprocessing time and space in $O(n^{1+\rho}+dn)$. Andoni, Laarhoven, Razenshteyn and Waingarten~\cite{alrw17} propose a improved proposes a data-dependent {\lsh} that solves $(\ov{c},r)$-$\ann$ with query time $O(d\cdot n^{\rho_q+o(1)})$, space $O(n^{1+\rho_u+o(1)}+dn)$ and preprocessing time $O(dn^{1+\rho_u+o(1)})$. Andoni, Laarhoven, Razenshteyn and Waingarten~\cite{alrw17} also states that for $\ov{c}>1$, $r>0$, $\rho_u\geq 0$ and $\rho_q\geq 0$, we have
$
    \ov{c}^2\sqrt{\rho_q}+(\ov{c}^2-1)\sqrt{\rho_u}\geq \sqrt{2\ov{c}^2-1}
$. 
Moreover, if we achieve $\rho_u=0$, we could reduce the preprocessing overhead to $O(n^{1+o(1)}+dn)$ while achieving $\rho_q=\frac{2}{\ov{c}^2}-\frac{1}{\ov{c}^4}$. These {\lsh} approaches have concise theoretical guarantees on the trade-off between search quality and query time. Thus, they could solve approximate $\maxip$ efficiently.

Meanwhile, other non-reduction approximate $\maxip$ approaches build efficient data structures such as quantization codebooks~\cite{gkcs16,gslgsck20}, alias tables~\cite{yhld17,dyh19}, trees~\cite{rg12} and graphs~\cite{mb18,ztxl19,tzxl19}. However, there exists no theoretical guarantee on these non-reduction approaches so that their evaluation is totally empirical.

\paragraph{Locality Sensitive Hashing Applications}

In practice, well-implemented {\lsh} algorithms are developed~\cite{ljw+07,ailrs15} and have demonstrated their superiority in tackling efficiency bottlenecks in practical applications. In optimization, \cite{cxs19} proposes a {\lsh} based approach to estimate gradients in large scale linear models. Moreover, this idea has been extended to neural network training~\cite{cmfgts20,clp+21}. Further more,
Besides deep learning, \cite{ls19} also proposes a {\lsh} method for efficient MCMC sampling. \cite{cs17,bcis18,srb+19,biw19,ckns20} use {\lsh} for efficient kernel density estimation. \cite{znvkr20} proposes a {\lsh} based approach for kernel ridge regression. \cite{yrkpds21} proposes an {\lsh} algorithm for efficient linear bandits.

\paragraph{Provable Efficient Reinforcement Learning}
The theoretical analysis on the efficiency of modern reinforcement learning (RL) approaches has drawn a lot of attention recently~\cite{jazbj18,bxjw19,ss19,jywj20,yw20,cyjw20,wzd+20,zzj20,wdys20,dkwy20,fwyd+20,dkl+21,xsd21}. \cite{jazbj18} presents the first Q-learning with UCB exploration algorithm with provable sublinear regret. \cite{jywj20} proposes a provable RL algorithm with linear function approximation that achieves both polynomial runtime and polynomial sample complexity. There also exist other works that benefit the community with theoretical analysis on efficient RL~\cite{dkwy20,yw20,cyjw20}.

\paragraph{Speedup Cost Per Iteration}
Recently, there have been many works discussing how to improve the cost per iteration for optimization problems (e.g., linear programming, cutting plane method, maximum matching, training neural networks) while maintaining the total number of iterations in achieving the same final error guarantees. However, all of these algorithms are built on sketching \cite{lsz19,jlsw20,jswz21,sy21,bpsw21}, sampling \cite{cls19,blss20,dly21}, vector-maintenance \cite{b20,jswz21}, sparse recovery \cite{blss20,bln+20} techniques, none of them have used {\lsh}. We hope that our novel combination of data structures and iterative algorithms will open the door for further study into cost reduction in optimization.

\section{Background}\label{sec:background}
\subsection{Locality Sensitive Hashing}\label{sec:intro_lsh}

We present a well-known data structure called locality sensitive hashing \cite{im98} for approximate nearest neighbor search and approximate maximum inner produce search.

\begin{definition}[Locality Sensitive Hashing]
Let $\ov{c}$ denote a parameter such that $\ov{c} >1$. Let $r$ denote a parameter. Let $p_1, p_2$ denote two parameters such that $0 < p_2 < p_1 < 1$. A family $\mathcal{H}$ is called $(r, \ov{c} \cdot r,p_1,p_2)$-sensitive if and only if, for any two point $x,y \in \R^d$, $h$ chosen uniformly from $\mathcal{H}$ satisfies the following: 
(1) if $ \| x - y \|_2 \leq r$, then $\Pr_{h \sim \mathcal{H}} [ h(x)=h(y) ] \geq p_1$, (2)
 if $ \| x - y \|_2 \geq \ov{c} \cdot r$, then $\Pr_{h \sim \mathcal{H}} [ h(x)=h(y) ] \leq p_2$.
\end{definition}
We want to remark that the original {\lsh} definition supports more general distance function than $\ell_2$ distance. In our application, $\ell_2$ distance is sufficient, therefore we only define {\lsh} based on $\ell_2$ distance. It is well-known that an efficient {\lsh} family implies data structure ($\ov{c},r$)-$\ann$ which can be defined as
\begin{definition}[Approximate Near Neighbor ($\ann$)]\label{def:ann}
Let $\ov{c} >1$ and $r \in (0,2)$. Given an $n$-point dataset $P \subset \mathbb{S}^{d-1}$ on the sphere, the goal of the $(\ov{c},r)$-Approximate Near Neighbor ($\ann$) problem is to build a data structure that, given a query $q \in \mathbb{S}^{d-1}$ with the promise that there exists a datapoint $p \in P$ with $\| p - q \|_2 \leq r$ reports a datapoint $p' \in P$ within distance $\ov{c} \cdot r$ from $q$.
\end{definition}

In the iterative-type reinforcement learning algorithm, we care about the dual version of the problem (Definition~\ref{def:amaxip}),
\begin{definition}[Approximate $\maxip$]\label{def:amaxip}
Let $c \in (0,1)$ and $\tau \in (0,1)$.
Given an $n$-point dataset $P \subset \mathbb{S}^{d-1}$ on the sphere, the goal of the $(c,\tau)$-Maximum Inner Product Search ({$\maxip$}) is to build a data structure that, given a query $q \in \mathbb{S}^{d-1}$ with the promise that there exists a datapoint $p \in P$ with $\langle p , q \rangle \geq \tau$, it reports a datapoint $p' \in P$ with similarity $\langle p' , q \rangle \geq c \cdot \tau$.
\end{definition}

We briefly discuss the connection. Let us consider the distance function as Euclidean distance and similarity function as inner product. We also assume all the points are from unit sphere. In this setting, the relationship between two problems are primal vs dual. For any two points $x,y$ with $\| x \|_2 = \|y \|_2=1$, we have $\|x-y\|_2^2= 2 - 2\langle x , y\rangle$. This implies that $r^2 = 2 - 2 \tau$. Further, if we have a data structure for $(\ov{c}, r)$-$\ann$, it automatically becomes a data structure for $(c, \tau)$-$\maxip$  with parameters $\tau = 1-0.5 r^2$ and $c = \frac{1-0.5 \ov{c}^2 r^2}{1 - 0.5 r^2}$.  This implies that 
$
\ov{c}^2= \frac{ 1 - c(1-0.5 r^2) }{0.5r^2} = \frac{1 - c \tau }{1-\tau} 
$.

Our algorithmic result is mainly built on this data structure.

\begin{theorem}[Andoni and Razenshteyn \cite{ar15}]\label{thm:ar15:informal}
Let $\ov{c} > 1$ and $r \in (0,2)$. Let $\rho = \frac{1}{2\ov{c}^2-1} +o(1)$. The $(\ov{c},r)$-$\ann$ (see Definition~\ref{def:ann}) on a unit sphere $\mathbb{S}^{d-1}$ can be solved in space $O(n^{1+\rho} + d n)$ and query time $O(d \cdot n^{\rho})$.
\end{theorem}

Using the standard reduction, we can derive the following.

\begin{corollary}\label{cor:c_tau_maxip:informal}
Let $c \in (0,1)$ and $\tau \in(0,1)$. The $(c,\tau)$-$\maxip$  (see Definition~\ref{def:amaxip}) on a unit sphere $\mathbb{S}^{d-1}$ can be solved in preprocessing time/space $O(n^{1+\rho} + d n)$ and query time $O(d \cdot n^{\rho})$, where
$
    \rho =
     \frac{1-\tau}{1-2c\tau+\tau} + o(1).
$
\end{corollary}
Using \cite{alrw17}, we can improve the preprocessing time and space to $n^{1+o(1)} + dn$ while having a slightly weaker $\rho$ in query. We provide a detailed and formal version of Corollary~\ref{cor:c_tau_maxip:informal} in Theorem~\ref{thm:maxip_lsh}. We present our main result based on that. Moreover, it is reasonable for us to regard $d=n^{o(1)}$  using Johnson-Lindenstrauss Lemma~\cite{jl84}.

Finally, to combine the maximum inner product search with reinforcement algorithm to get sublinear time cost, we still need to deal with many issues, such as the inner product can be negative, $\tau$ is arbitrarily close to $0$, and $\tau$ can arbitrarily close to $1$. We will explain how to handle these challenges in later section. 

\subsection{Reinforcement Learning}
In this section, we introduce some backgrounds about reinforcement learning. We start with defining the episodic Markov decision process.  Let $\mdp({\cal S}, {\cal A}, H, \P, r)$ denotes the episodic Markov decision process, where ${\cal S}$ denotes the set of available states, ${\cal A}$ denotes the set of available actions, $H \in \mathbb{N}$ denotes the total number of steps in each episode, $\P = \{ \P_h \}_{h=1}^H $ with $\P_h [ s' | s, a] $  denotes the probability of transition from state $s\in{\cal S}$ to state $s'\in{\cal S}$ when take actions $a\in{\cal A}$ at step $h$, $ r = \{ r_h \}_{h=1}^H $ denotes the reward obtained at each step. Here the reward $r_h$ is a function that maps ${\cal S} \times {\cal A} $ to $ [0.55,1]$\footnote{Note that in standard reinforcement learning, we assume reward is $[0,1]$, but it is completely reasonable to do a shift. We will provide more discussion in Section~\ref{sec:prevent_negative_zero_ip}.} In practice, we build an agent in $\mdp({\cal S}, {\cal A}, H, \P, r)$ and play $K$ episodes.

In this work, we focus on the linear Markov decision process (linear MDP). In this setting, each pair of state and action is represented as an embedding vector $\phi(s,a)$, where $\phi: {\cal S} \times {\cal A} \rightarrow \R^d$. Moreover, the probability $\P_h [ s' | s, a]$ for state transition and function $r_h$ for reward are linear in this embedding vector. 

 In the MDP framework, a policy $\pi=\{\pi_1,\cdots,\pi_H\}$ is defined as sequence such that $\pi_h:{\cal S}\rightarrow {\cal A}$ for each step $h$. $\pi_h(s)=a$ represents the action taken when we are at step $h$ and state $s$. Next, we represent the Bellman equation with policy $\pi$ as
\begin{align*}
    Q^{\pi}_{h}(s, a) = [r_h + \P_h V^{\pi}_{h+1}](s, a) , \qquad   V^{\pi}_{h}(s) = Q^{\pi}_{h}(s, \pi_h(s)),    
    \qquad V^{\pi}_{H+1}(s) = 0.  
\end{align*}
where $Q^{\pi}(s,a)$  denotes the Q function for policy $\pi$ when taking action $a$ at state $s$ and step $h$ and $V^{\pi}(s)$ denotes the value function of state $s$ at step $h$. We use $[\P_hV_{h+1}](s, a)$ to represent the expect value functions when aking action $a$ at state $s$ at step $h$.  For more detailed definitions, please refer to Section~\ref{sec:preli}.

\section{Our Results}\label{sec:results}
We present the results in this section. 
We start with summarizing all of our main results in Table~\ref{tab:main_compare}. According to Table~\ref{tab:main_compare}, we reduce the value iteration complexity of LSVI~\cite{bb96}, LSVI-UCB~\cite{jywj20}, LSVI-UCB under policy switch limitation~\cite{gxdy21} and model-free version of LSVI-UCB~\cite{wdys20} from linear to sublinear in action space. Meanwhile, the total regret is preserved as same as before.  To achieve this, we pay tolerable time to preprocess pairs of state-action into {\lsh} type approximate $\maxip$ data structure. In the following section, we would elaborate on the details for these main results.

\subsection{Sublinear Least-Squares Value Iteration}

In LSVI~\cite{bb96} with large action space, the runtime in each value iteration step is dominated by computing the estimated value function as below:
\begin{align}\label{eq:LSVI_problem_mips:informal}
    \hat{V}_h(s) = \max_{a\in {\cal A}_{\core}} \langle \hat{w}_h , \phi(s,a) \rangle
\end{align}
where $\hat{w}_h$ is computed by solving the least-squares problem and $\phi(s,a)$ is the embedding for state-action pair.  Eq.~\eqref{eq:LSVI_problem_mips:informal} is a standard $\maxip$ problem and thus, takes $O(Ad)$ to obtain the exact solution. In this work, we relax Eq.~\eqref{eq:LSVI_problem_mips:informal} into an  $(c,\tau)$-$\maxip$ problem, where $c\in (0,1)$ is the approximation parameter and $\tau$ is close to the $\max_{a\in {\cal A}_{\core}} \langle \hat{w}_h , \phi(s,a) \rangle$. Then, we apply {\lsh} type data structure to retrieve $\hat{V}_h(s)\geq c\cdot \max_{a\in {\cal A}_{\core}} \langle \hat{w}_h , \phi(s,a) \rangle$ in $o(A) \cdot O(d)$ time complexity. 
\begin{table}[H]
    \centering
    \begin{tabular}{|l|l|l|l|l|} \hline
        &  {\bf Statement} & {\bf Preprocess} & {\bf \#Regret} & {\bf V. Iter. C.}  \\ \hline \hline
        LSVI  & \cite{bb96} & 0 & $H^2\sqrt{\iota/n}$ & $HSdA$   \\ \hline
        Ours& Theorem~\ref{thm:convergence_SLSVI:informal}  & $SA^{1+o(1)}+Sd A$ & $H^2\sqrt{\iota/n}$ & $HS dA^\rho$ \\ \hline \hline
        LSVI-UCB &  \cite{jywj20} & 0  & $\sqrt{H^4 K d^3 \iota^2}$ &$HKd^2A$   \\ \hline
        Ours  & Theorem~\ref{thm:convergence_SLSVI-UCB:informal} &$KA^{1+o(1)}+Kd^2 A$   &$\sqrt{ H^4 K d^3 \iota^2}$ & $H Kd^2A^\rho$ \\ \hline \hline
        LGSC &  \cite{gxdy21} & 0  & $\sqrt{H^4 K d^3 \iota^2}$ &$HKd^2A$   \\ \hline
        Ours  & Corollary~\ref{coro:convergence_SMFLSVI-UCB:informal} &$KA^{1+o(1)}+Kd^2 A$   &$\sqrt{ H^4 K d^3 \iota^2}$ & $H Kd^2A^\rho$ \\ \hline \hline
        MF &  \cite{wdys20} & 0  & $\sqrt{H^4 K d^3 \iota^2}$ &$HKd^2A$   \\ \hline
        Ours  & Corollary~\ref{coro:convergence_SMFLSVI-UCB:informal} &$KA^{1+o(1)}+Kd^2 A$   &$\sqrt{ H^4 K d^3 \iota^2}$ & $H Kd^2A^\rho$ \\ \hline \hline
    \end{tabular}
    \caption{\small Comparison between our algorithms with previous results such as LSVI, LSVI-UCB, LGSC and MF. We compare our algorithm with: (1) LSVI denotes the Least-Square Value Iteration algorithm~\cite{bb96} (2) LSVI-UCB denotes the Least-Square Value Iteration algorithm with UCB in ~\cite{jywj20}.  (3) LGSC denotes the LSVI-UCB with low global switching cost~\cite{gxdy21}. (4) MF denotes the model free LSVI-UCB presented in~\cite{wdys20}.   Note that ``{\bf V. Iter. C.}'' denotes the Value iteration complexity. Let $S$ denotes the number of available states. Let $A$ denote the number of available actions. Let $d$ denotes the dimension of $\phi(s,a)$. Let $H$ denotes the number of steps per episode. Let $K$ denotes the total number of episodes. Let $n$ be the quantity of times played for each core pair of state-action. Let $\iota=\log(Hd/p)$ and $p$ is the failure probability.  We ignore the big-Oh notation ``$O$'' in the table. Let $\rho \in (0,1)$  denote a parameter determined by data structure.
    In fact, the preprocessing time for Sublinear LSVI-UCB is $O(SA^{1+o(1)}+Sd^2A)$. Since $K>S$, we write the preprocessing time as $O(KA^{1+o(1)}+Kd^2A)$. This table is a union of simplified version of Table~\ref{tab:LSVI_compare} (both our algorithm and LSVI have the exact dependence on another $L$, we omit here and discuss this dependence in Section~\ref{sec:SLSVI}.), Table~\ref{tab:LSVI-UCB_compare} and Table~\ref{tab:UCB_extend_compare}.}
    \label{tab:main_compare}
\end{table} 
Next, we present our main theorem for Sublinear LSVI in Theorem~\ref{thm:convergence_SLSVI:informal}, which gives the same $O(LH^2\sqrt{\iota/n})$ regret as LSVI~\cite{bb96} and reduce the value iteration complexity from $O(HS dA)$ to  $O(H S d) \cdot o(A)$.

\begin{theorem}[Main result, convergence result of Sublinear Least-Squares Value Iteration
 (Sublinear LSVI), an informal version of Theorem~\ref{thm:convergence_SLSVI:formal}]\label{thm:convergence_SLSVI:informal}
Let $\mdp({\cal S}, {\cal A}, H, \P, r)$ denotes a linear MDP. Let $p$ denotes a fixed probability. Let $\iota=\log ( {Hd} / {p} )$. If we set approximate $\maxip$ parameter $c= 1- \Theta(  \sqrt{ \iota / n} )$, then Sublinear LSVI  has regret at most $O(H^2\sqrt{\iota/n})$ with probability at least $1-p$. Moreover, with $SA^{1+o(1)}+Sd A$ preprocessing time and space, the value iteration complexity of Sublinear LSVI is $O(HS dA^{\rho})$ where $\rho= 1 - \Theta(\iota/n)$.
\end{theorem}

Note that we could improve the value iteration complexity to  with $\rho=1-\Theta( \sqrt{ \iota / n} )$ by increasing the preprocessing time and space to $O(SA^{1+\rho}+Sd A)$ using Theorem~\ref{thm:ar15:formal}.  We provide a detailed and formal version of  Theorem~\ref{thm:convergence_SLSVI:informal} in Theorem~\ref{thm:convergence_SLSVI:formal}.
    
\subsection{Sublinear Least-Squares Value Iteration with UCB}\label{sec:results_LSVI-UCB}

We extend the Sublinear LSVI with UCB exploration in this section. In LSVI-UCB~\cite{jywj20} with large action space, the runtime in each value iteration step is dominated by by computing the estimated value function as below:
\begin{align}\label{eq:LSVI-UCB_problem_original:informal}
    \hat{V}_h(s^{\tau}_{h+1}) = \max_{a\in {\cal A}}~\min\{\langle w^k_h,\phi(s^{\tau}_{h+1}, a)\rangle  + \beta \cdot \|\phi(s^{\tau}_{h+1},a)\|_{\Lambda_h^{-1}}, H\}
\end{align} 
where $w^k_h$ is computed by solving the least-squares problem, $\phi(s^{\tau}_{h+1},a)$ is the embedding for state-action pair and $\Lambda_h=\sum_{\tau =1}^{k-1}  \phi(s^{\tau}_h,  a^{\tau}_h)\phi(s^{\tau}_h, a^{\tau}_h)^{\top} + \lambda \cdot  \I_d$. The complexity for Eq.~\eqref{eq:LSVI-UCB_problem_original:informal} is $O(Ad^2)$

The key challenge of Sublinear LSVI-UCB here is that Eq.~\eqref{eq:LSVI-UCB_problem_original:informal} cannot be formulated as a $\maxip$ problem. First, to deal with this issue, we propose a value function estimation approach as below:
\begin{align}\label{eq:LSVI-UCB_problem_mips:informal}
    \hat{V}_h(s^{\tau}_{h+1}) = \max_{a\in {\cal A}}~ \min\{  \|\phi(s^{\tau}_{h+1},a)\|_{2\beta^2 \Lambda_h^{-1}+2w_h^k w_h^{k\top}}, H\}
\end{align}
where $\|\phi(s^{\tau}_{h+1},a)\|_{2\beta^2 \Lambda_h^{-1}+2w_h^k w_h^{k\top}}$ is the upper bound of $\langle w^k_h,\phi(s^{\tau}_{h+1}, a)\rangle  + \beta \cdot \|\phi(s^{\tau}_{h+1},a)\|_{\Lambda_h^{-1}}$.  

Next, we relax this maximum matrix norm search as a $(c,\tau)$-$\maxip$ problem, where $c\in (0,1)$ is the approximation parameter and $\tau$ is the maximum inner product for Eq.~\eqref{eq:LSVI-UCB_problem_mips:informal}.  Then, we apply {\lsh} type data structure to retrieve $\hat{V}_h(s^{\tau}_{h+1})\geq c\cdot \max_{a\in {\cal A}}~ \min\{  \|\phi(s^{\tau}_{h+1},a)\|_{2\beta^2 \Lambda_h^{-1}+2w_h^k w_h^{k\top}}, H\}$ in $o(A) \cdot O( d^2)$ time complexity.

Using {\lsh} data structure for maximum matrix norm search, we present our main theorem for Sublinear LSVI-UCB in Theorem~\ref{thm:convergence_SLSVI-UCB:informal}, which gives the same $O(\sqrt{d^3 H^4 K \iota^2})$ regret as LSVI-UCB~\cite{jywj20} and reduce the value iteration complexity from $O(H Kd^2A)$ to  $O(H Kd^2A) \cdot o(A)$. We start with the setting up the parameters for our algorithm.

\begin{definition}[Sublinear LSVI-UCB Parameteres]\label{def:LSVI-UCB_param}
Let $\mdp({\cal S}, {\cal A}, H, \P, r)$ denotes a linear MDP. For this MDP, we set LSVI-UCB parameter $\lambda = 1$. Let $c= 1-\frac{1}{\sqrt{K}}$ denotes the approximate $\maxip$ parameter. Let $p$ denotes a fixed probability. Let $ \iota = \log (2dT/p)$.
\end{definition}
Then, we present the Theorem.
\begin{theorem}[Main result, convergence result of Sublinear Least-Squares Value Iteration with UCB (Sublinear LSVI-UCB), an informal version of Theorem~\ref{thm:convergence_SLSVI-UCB:formal}] \label{thm:convergence_SLSVI-UCB:informal}
With parameters defined in Definition~\ref{def:LSVI-UCB_param}, Sublinear LSVI-UCB (Algorithm \ref{alg:sublinearLSVI-UCB}) has total regret at most $O(\sqrt{d^3 H^4 K \iota^2})$ with probability at least $1-p$.  Moreover, with $O(KA^{1+o(1)}+Kd^2A)$ preprocessing time and space, the value iteration complexity of Sublinear LSVI-UCB is $O(H Kd^2A^\rho)$, where $\rho = 1-1/K$. 
\end{theorem}

Similarly,  we could improve the value iteration complexity to  with $\rho=1-\frac{1}{\sqrt{K}} $ by increasing the preprocessing time and space to $O(KA^{1+\rho}+Kd A)$ using Theorem~\ref{thm:ar15:formal}.  We provide a detailed and formal version of  Theorem~\ref{thm:convergence_SLSVI-UCB:informal} in Theorem~\ref{thm:convergence_SLSVI-UCB:formal}.

Next, we extend the results in Theorem~\ref{thm:convergence_SLSVI-UCB:informal} to two LSVI-UCB variations. The first algorithm is the LSVI-UCB  under constraints on the switch of the policy~\cite{gxdy21}. We denote this algorithm as LGSC. The second algorithm is the model-free version of LSVI-UCB~\cite{wdys20}. We denote this algorithm as MF\footnote{We discuss the policy switch cost of LGSC in Section~\ref{sec:LGSC} and number of explorations of MF in Section~\ref{sec:MF}}. We propose sublinear version of two algorithms with statement as:

\begin{corollary}
[Main result, informal versions of Corollary~\ref{coro:convergence_SMFLSVI-UCB:formal} and Corollary~\ref{coro:convergence_SLSVI-UCB_LSGSC:formal}]\label{coro:convergence_SMFLSVI-UCB:informal}
With parameters defined in Definition~\ref{def:LSVI-UCB_param}, LGSC and MF have total regret at most $O(\sqrt{d^3 H^4 K \iota^2})$ with probability at least $1-p$. Further more, with $O(KA^{1+o(1)}+Kd^2A)$ preprocessing time and space, the value iteration complexity of LGSC and MF is $O(H Kd^2A^\rho)$, where $\rho = 1-1/K$. 
\end{corollary}

\section{Our Techniques}\label{sec:tech}

As mentioned in Section~\ref{sec:intro_lsh}, we need to tackle five major issues to use {\lsh} based approximate $\maxip$ algorithm for sublinear runtime time LSVI and LSVI-UCB in RL.
\begin{itemize}
    \item How to prevent the maximum inner product between query and data from being negative or arbitrary close to $0$? If the maximum inner product is negative, $\maxip$ data structures cannot be applied to solve this problem with theoretical guarantee. If the maximum inner product is arbitrary close to $0$, the query time of $(c,  \tau)$-$\maxip$ would be close to $O(dn)$.
    \item  How to prevent the maximum inner product between query and data from being close to one? If $\tau$ is close to one, the time cost would also be $O(dn)$ so that $(c,   \tau)$-$\maxip$ cannot reduce the time cost from linear to sublinear.
    \item How to apply $(c,\tau)$-$\maxip$ for LSVI with UCB exploration? The estimated value function with an additional UCB bonus term could not be written as an inner product, which prevents $\maxip$ techniques from accelerating the runtime efficiency. 
    \item How to generalize the $\maxip$ data structure to support maximum matrix norm search? Is $\maxip$ equivalent to maximum matrix norm search?
    \item How to improve the running time while preserving the regret? Although approximate $\maxip$ could accelerate the computation for estimated value function, it brings errors to the value function estimation and thus, affects the total regret. Therefore, a key challenge is quantifying the relationship between regret and the approximation factor $c$ in $(c,\tau)$-$\maxip$. 
    \item How to handle the adaptive queries? The weight $\hat{w}_h$ in Eq.~\eqref{eq:LSVI_problem_mips:informal} and $w_h^k$ in Eq.~\eqref{eq:LSVI-UCB_problem_original:informal} are dependent to $h-1$ step. Therefore, the queries for $(c,\tau)$-$\maxip$ during the Q-learning are adaptive  but not arbitrary. Thus, we could not union bound the failure probability of {\lsh} for $(c,\tau)$-$\maxip$.
    
\end{itemize}

Next, we provide details on how we handle these problems.

\subsection{Avoid Negative Inner Product or Inner Product Close to $0$}\label{sec:prevent_negative_zero_ip}

In our setting, we assume the reward function $r$ lies in $[0.55,1]$\footnote{Note that for any reward range $[a,b]$, there exists a shift $c$ and scaling $\alpha$ so that $(a+c)/\alpha =0.55$ and $(b+c)/\alpha =1$.}.  This shift on the reward function would not affect the convergence results of our Sublinear LSVI and Sublinear LSVI-UCB. Moreover, it would benefits the $\maxip$ by generating acceptable maximum inner product. For Sublinear LSVI, as $r_h(s,a)\in [0.55,1]$, the optimal value function $V_h^*(s)\geq 0.55$. Then according to Theorem~\ref{thm:convergence_SLSVI:informal} the estimated $\hat{V}_h(s)=\max_{a\in {\cal A}}\langle w_h, \psi(s,a)\rangle$ satisfies  $|V_h^*(s)-V_h^*(s)|\leq \epsilon$ if we query each pair of state-action  from span matrix for $n=O(\epsilon^{-2} L^2 H^4 \iota)$ times. In this way, we could assure the maximum inner product is greater than $0.5$ if we set $\epsilon\leq 0.05$. For Sublinear LSVI-UCB, the $\maxip$ is applied on $\hat{V}_h(s)=\max_{a\in {\cal A}}Q^k_h(s,a)$, where $Q^k_h(s,a)$ is a Q function with additional UCB term. From~\cite{jywj20}, we know that for all pair of state-action ,$Q^k_h(s,a)\geq Q^*_h(s,a)$. Therefore, the maximum inner product for Sublinear LSVI-UCB is always greater than $0.5$.

\subsection{Avoid Inner Product Close to $1$}

In the optimization problem that could be accelerated by $\maxip$, the query and data vectors are usually not unit vectors. To apply results in Section~\ref{sec:intro_lsh}, we demonstrate how to transform both query and data vectors into unit vectors. Moreover, we also modify the transformation to avoid the inner product from being too close to $1$. 

Given two vector $x,y\in \R^{d}$ with $\|y\|_2\leq 1 $ and $\|x\|_2\leq D_x$, we apply the following transformations 
\begin{align}\label{eq:trans_intro}
    P(y) = \begin{bmatrix} y^\top & \sqrt{1-\|y\|_2^2}  & 0 \end{bmatrix}^\top ~~
Q(x) = \begin{bmatrix}\frac{0.8\cdot x^\top}{D_x}  & 0 & \sqrt{1-\frac{0.64\cdot\|x\|_2^2}{D_x^2}} \end{bmatrix}^\top
\end{align}

Using this transformations, we transform $x,y$ into unit vectors $P(y)$ and $Q(x)$. Therefore, the $\maxip$ of $Q(x)$ with respect to $P(Y)$ is equivalent to the $\ann$ problem of $Q(x)$ with respect to $P(Y)$, which could be solved via {\lsh}. Moreover, we show that  $
Q(x)^\top P(y)=\frac{0.8x^\top y}{D_x}\leq \frac{0.8\cdot \|x\|_2\|y\|_2}{D_x}=0.8$. Further more, it is sufficient to show that  $  \arg\max_{y} Q(x)^\top P(y)=\arg\max_{y}\frac{0.8\cdot x^\top y}{D_x}=\arg\max_{y} x^\top y$.

If we perform maximum inner product search on $Q(x)$ and $P(y)$ using the {\lsh} data structures described in Section~\ref{sec:intro_lsh}, we have $\tau = \max_{y} Q(x)^\top P(y)\leq 0.8$.  In this way, we could assure $\tau$ is not close to $1$ so that we could reduce the runtime complexity of value function estimation to be sublinear over actions.

\subsection{Approximate $\maxip$ Data Structure for LSVI-UCB}
As shown in Section~\ref{sec:results_LSVI-UCB}, Eq.~\eqref{eq:LSVI-UCB_problem_original:informal} cannot be formulated as a $\maxip$ problem. To overcome this barrier, we bound the term $ Q_h(s^{\tau}_{h+1}, a)= \{w_h^\top\phi(s^{\tau}_{h+1}, a)  + \beta \cdot \|\phi(s^{\tau}_{h+1},a)\|_{\Lambda_h^{-1}}, H\}$ by matrix norms.  Then, we perform the maximum matrix norm search for value function estimation.

We start with the upper bound of $ w_h^\top\phi(s^{\tau}_{h+1}, a)  + \beta \cdot \|\phi(s^{\tau}_{h+1},a)\|_{\Lambda_h^{-1}}$. As both $\langle w_h^k,\phi(s^{\tau}_{h+1}, a)\rangle$ and $\|\phi(s^{\tau}_{h+1},a)\|_{\Lambda_h^{-1}}$ are non-negative, we have
\begin{align*}
    \langle w_h^k,\phi(s^{\tau}_{h+1}, a)\rangle  + \beta \cdot \|\phi(s^{\tau}_{h+1},a)\|_{\Lambda_h^{-1}}
    \leq &~ \sqrt{2(w_h^\top\phi(s^{\tau}_{h+1}, a))^2+2\beta^2 \cdot \|\phi(s^{\tau}_{h+1},a)\|_{\Lambda_h^{-1}}^2}\\
     = &~ \|\phi(s^{\tau}_{h+1},a)\|_{2\beta^2 \Lambda_h^{-1}+2w_h^k (w_h^k)^\top}
\end{align*}
where the first step follows from $a+b\leq \sqrt{2a^2+2b^2}$.

Next, we lower bound the $ w_h^\top\phi(s^{\tau}_{h+1}, a)  + \beta \cdot \|\phi(s^{\tau}_{h+1},a)\|_{\Lambda_h^{-1}}$ as
\begin{align*}
    w_h^\top\phi(s^{\tau}_{h+1}, a)  + \beta \cdot \|\phi(s^{\tau}_{h+1},a)\|_{\Lambda_h^{-1}} 
    \geq &~ \sqrt{(w_h^\top\phi(s^{\tau}_{h+1}, a))^2+\beta^2 \cdot \|\phi(s^{\tau}_{h+1},a)\|_{\Lambda_h^{-1}}^2}\\
     = &~ \|\phi(s^{\tau}_{h+1},a)\|_{\beta^2 \Lambda_h^{-1}+w_h^k (w_h^k)^\top}
\end{align*}
where the first step follows from the fact that both $w_h^\top\phi(s^{\tau}_{h+1}, a)$ and $\|\phi(s^{\tau}_{h+1},a)\|_{\Lambda_h^{-1}}$ are non-negative and  $a+b\geq \sqrt{a^2+b^2}$ if $a,b\geq 0$, the second step is an reorganization.

After we obtain the lower and upper bound of $w_h^\top\phi(s^{\tau}_{h+1}, a)  + \beta \cdot \|\phi(s^{\tau}_{h+1},a)\|_{\Lambda_h^{-1}}$, we could also lower bound the term $\{w_h^\top\phi(s^{\tau}_{h+1}, a)  + \beta \cdot \|\phi(s^{\tau}_{h+1},a)\|_{\Lambda_h^{-1}}, H\}$ with $\min\{  \|\phi(s^{\tau}_{h+1},a)\|_{\beta^2 \Lambda_h^{-1}+w_h^k (w_h^k)^\top}, H\}$  and upper bound it with $ \min\{  \|\phi(s^{\tau}_{h+1},a)\|_{2\beta^2 \Lambda_h^{-1}+2w_h^k (w_h^k)^\top}, H\}$.

Next we use this lower and upper bound and propose a modified value function estimation shown in Eq.~\eqref{eq:LSVI-UCB_problem_mips:informal}.  Therefore, our problem becomes designing an approximate maximum matrix norm search data structure. We will discuss this in the following section and propose our Sublinear LSVI-UCB algorithm.

\subsection{Generalize the Approximate $\maxip$ Data Structure for $\maxmatnorm$}

We demonstrate how to extend $\maxip$ to maximum matrix norm search for Sublinear LSVI-UCB in this section. We first define the approximate Maximum Matrix Norm. Let $c \in (0,1)$ and $\tau \in (0,1)$.
Given an $n$-point dataset $Y \subset \R^{d}$, the goal of the $(c,\tau)$-Maximum Matrix Norm ($\maxmatnorm$) is to construct a data structure that, given a query matrix $x \in \R^{d\times d}$ with the promise that there exists a datapoint  $y \in Y$ with $\|y\|_x\geq \tau$, it reports a datapoint $z \in Y$ with $ \|z\|_{x}\geq c \cdot \tau$.

We solve the approximate maximum matrix norm by transform it into a $\maxip$ problem. We start with showing the relationship between $\maxmatnorm$ and $\maxip$ as
$$\maxmatnorm(X,Y)^2 =\max_{y \in Y} y^\top x y=\max_{y \in Y} \langle \vect(x),\vect(y y^\top)$$
where $\vect$ vectorizes $d\times d$ matrix $x$ into a $d^2$ vector. 

Next, we show that if we obtain $z\in Y$ by $(c^2,\tau^2)$-$\maxip$ so that $\langle \vect(x),\vect(z z^\top) \rangle \geq   c^2 \tau^2$,
we use $z$ and obtain $\|z\|_{x}=\sqrt{\langle \vect(x),\vect(z z^\top) \rangle}\geq c\tau$. In other words, $z$ is the candidate for $(c,\tau)$-$\maxmatnorm$. In this way, we could build an efficient data-structure for $(c^2,\tau^2)$-$\maxip$ to solve $(c,\tau)$-$\maxmatnorm$. In this way, we summarize our approach for $\maxmatnorm$ as three steps: (1) transform matrix $x$ into $\vect(x)$ and $y$ into $\vect(yy^\top)$, (2) transform $\vect(x)$ and $\vect(yy^\top)$ into unit vectors following Eq.~\eqref{eq:trans_intro}, (3) use {\lsh} to solve the $\maxip$ with respect to dataset on the unit sphere.

\subsection{Preserving Regret While Reducing the Runtime}
In our work, we maintain the same regret with LSVI~\cite{bb96} and LSVI-UCB~\cite{jywj20} by carefully setting the approximation parameter $c\in(0,1)$ in $\maxip$. For Sublinear LSVI, we set $c=1-\Theta(\sqrt{\iota/n})$ so that the final regret is as same as LSVI~\cite{bb96}. In Sublinear LSVI-UCB, we set $c=1-\frac{1}{\sqrt{K}}$ so that the final regret is as same as LSVI-UCB~\cite{jywj20}. Because $K$, $\iota$ and $n$ are global parameter, we could set $c$ in the preprocessing step before value iteration. In this way, we show that our two algorithms are novel demonstration of combining {\lsh} with reinforcement learning without losing on the regret.

\subsection{Handle Adaptive Queries in $(c,\tau)$-$\maxip$}

We use a quantization method to handle adaptive queries. We denote $Q$ as the convex hull of all queries for $(c,\tau)$-$\maxip$. Our method contains two steps: (1) Preprocessing: we quantize $Q$ to a lattice $\hat{Q}$ with quantization error ${\lambda}/{d}$. In this way, each coordinate would be quantized into the multiples of ${\lambda}/{d}$.  (2) Query: given a query $q$ in the adaptive sequence $X\subset Q$, we first quantize it to the nearest $\hat{q}\in \hat{Q}$ and perform $(c,\tau)$-$\maxip$. As each $\hat{q}\in \hat{Q}$ is independent, we could union bound the failure probability of adaptive queries. On the other hand, this would generate an $\lambda$ additive error in the returned inner product. Our analysis indicates that the additive error $\lambda$ could be handled without breaking the regret. 

\section{Future Data Structure Design}

In this paper, we show that {\lsh} type data structures could accelerate Q-learning with linear function approximation~\cite{jywj20,gxdy21}. In this setting, the Q function is formulated as the inner product of state-action embedding and weight. However, in the literature of theoretical RL, the Q function is in a more general form.  \cite{jazbj18,bxjw19,wzd+20,zzj20,dkl+21} present the theoretical analysis on general Q function. \cite{ss19} assumes the Lipschitz continuous of optimal $Q$ function and show how to upper bound it. \cite{yw20} assumes the existence of a non-linear mapping in the computation of Q function. 

To extend our results in general Q functions, an efficient similarity search data structure is required. We provide its form as: Let $f(x,y):\R^d\times\R^d\rightarrow \R$ denotes a similarity measure. Let $c\in (0,1)$. Given a $n$-point dataset $P \subset \R^{d}$,  a data structure is an sublinear similarity search data structure if given a query $q\in \R^{d}$, it reports a datapoint $p' \in P$ with similarity $f(q,p')\geq c\cdot \max_{p\in P} f(q,p)$ with query complexity sublinear in $n$.

It is clear to extend {\lsh} for general norms~\cite{ann+17,ann+18a,ann+18b}. However, it remains unclear to extend {\lsh} for more general functions that are widely used in RL. We believe that our work gives a strong motivation on designing novel data structures for more general similarity measures and apply it to solve efficiency bottlenecks in RL and other ML tasks.

\ifdefined\isarxiv

\else
\newpage
\bibliographystyle{alpha}
\bibliography{ref}
\fi

\newpage
\ifdefined\isarxiv
\else
{\hypersetup{linkcolor=black}
\tableofcontents
}
\newpage
\fi

{\hypersetup{linkcolor=black}
\tableofcontents
}

\newpage

\paragraph{Roadmap.}  Section~\ref{sec:preli} introduces the preliminary notations and definitions, Section~\ref{sec:data_structure} introduces the {\lsh} data structure in detail, Section~\ref{sec:SLSVI} presents the results for Sublinear LSVI, Section~\ref{sec:SLSVI_UCB} presents the results for Sublinear LSVI-UCB, Section~\ref{sec:extend} presents the extension of Sublinear LSVI-UCB to different RL settings, Section~\ref{sec:adaptive} shows how to handle adaptive queries in $\maxip$.
\section{Preliminaries}\label{sec:preli}

This section introduces the preliminaries for our work. 
\begin{itemize}
    \item In Section~\ref{sec:basic_not}, we present the basic notations used in our work.
    \item In Section~\ref{sec:definitions}, we introduce several reinforcement learning. 
    \item In Section~\ref{sec:properties}, we list the standard proprieties of linear MDP.
    \item In Section~\ref{sec:def_lsh}, we introduces the definitions of locality sensitive hashing data structures and their applications in nearest neighbor search.
    \item In Section~\ref{sec:prob_tools}, we list the probabilistic tools used in our work.
    \item In Section~\ref{sec:ineq}, we list the inequalities to help the proof.
\end{itemize}
\subsection{Basic Notations}\label{sec:basic_not}

We use $\Pr[]$ to denote probability and $\E[]$ to denote expectation if it exists.

For a matrix $A$, we use $\| A \|_F := (\sum_{i,j} A_{i,j}^2 )^{1/2}$ to denote the Frobenius norm of $A$, we use $\| A \|_1:=\sum_{i,j} |A_{i,j}|$ to denote the entry-wise $\ell_1$ norm of $A$, we use $\| A \|$ to denote the spectral norm of $A$. We say matrix $A\in \R^{d\times d}$ is a positive semidefinite matrix if for all $x\in \R^d$, $x^\top A x \geq 0$.   We say matrix $A\in \R^{d\times d}$ is a positive definite matrix if for all $x\in \R^d$, $x^\top A x > 0$.

For a vector $x$, we use $\| x \|_2:= (\sum_{i} x_i^2)^{1/2}$ to denote the $\ell_2$ norm of $x$,  we use $\| x \|_1:=\sum_{i} |x_i|$ to denote the $\ell_1$ norm of $x$, we use $\| x \|_{\infty}$ to denote the $\ell_{\infty}$ norm.

For a vector $x \in \R^d$ and a psd matrix $A \in \R^{d \times d}$, we use $\| x \|_{A}:=(x^\top A x)^{1/2}$ to denote the matrix norm of $x$ over $A$.

We use $\mathbb{S}^{d-1}$ to denote the unit sphere.

\subsection{Notations and Definitions}\label{sec:definitions}
In this section, we present the notation and definitions for reinforcement learning. We summarize our notations in Table~\ref{tab:notations}.
\begin{table}[H]
\begin{center}
\begin{tabular}{|l|l|}
\hline  
{\bf Notation} & {\bf Meaning} \\
\hline  
${\cal S}$& states space\\ \hline
${\cal A}$& action space\\ \hline
${\cal S}_{\core}$&  core state set \\
\hline 
${\cal A}_{\core}$&  core action set \\
\hline 
$S$&  \# states \\
\hline 
$A$&  \# actions \\
\hline 
$H$& number of steps per episode\\ \hline
$K$& length of episode\\ \hline
$s'$& next state of state $s$\\ \hline
$\P$& state transition probability\\ \hline
$\P_h [s'|s,a]$& transition probability when we take action $a\in{\cal A}$ at step $h\in [H]$ from state $s\in{\cal S}$.\\ \hline
$r_h(s,a)$&  reward at step $h$ given state $s$ and action $a$ \\ \hline
$r$&  $\{ r_h \}_{h=1}^H$ \\ \hline
$\phi(s,a)$&  feature map $\phi(s,a)\in \R^{d}$\\\hline
$\mu_h(s)$&  unknown measure that $\P_h [s'|s,a]=\langle\phi(s,a), \mu_h(s')\rangle$ \\\hline
$\theta_h$&  unknown measure that $r_h(s,a)=\langle\phi(s,a) ,\theta_h\rangle$ \\
\hline 
$\Phi$ &$\Phi \in \R^{d \times M}$\\\hline 
$n$ & number of samples played given from each $\phi_j$.\\\hline 
\end{tabular}    
\end{center}\caption{Notations related to reinforcement learning.}\label{tab:notations}
\end{table}

We start with the definition of the Episodic Markov decision process. 

\begin{definition}[Episodic Markov decision process (episodic MDP)]\label{def:mdp}
Let $\mdp({\cal S}, {\cal A}, H, \P, r)$ denotes the episodic Markov decision process, where ${\cal S}$ denotes the set of available states, ${\cal A}$ denotes the set of available actions, $H \in \mathbb{N}$ denotes the total number of steps in each episode, $\P = \{ \P_h \}_{h=1}^H $ with $\P_h [ s' | s, a] $  denotes the probability of transition from state $s\in{\cal S}$ to state $s'\in{\cal S}$ when take actions $a\in{\cal A}$ at step $h$, $ r = \{ r_h \}_{h=1}^H $ denotes the reward obtained at each step. Here the reward $r_h$ is a function that maps ${\cal S} \times {\cal A} $ to $ [0.55,1]$\footnote{Note that in standard reinforcement learning, we assume reward is $[0,1]$, but it is completely reasonable to do a shift. We will provide more discussion in Section~\ref{sec:prevent_negative_zero_ip}.}
\end{definition}

Note that for any reward range $[a,b]$, there exists a shift $c$ and scaling $\alpha$ so that $(a+c)/\alpha =0.55$ and $(b+c)/\alpha =1$ The shift in reward is designed for sublinear runtime in maximum inner product search. We will provide more discussion in Section~\ref{sec:true_sublinear_query}.

In this work, we focus on linear Markov decision process (linear MDP). In this setting, each pair of state-action is represented as an embedding vector. Moreover,  the transition probability $\P_h [ s' | s, a] $ and reward function $r_h$ are linear in this embedding vector.

\begin{definition}[Linear MDP~\cite{bb96, mr07}]\label{def:linear_mdp}
The $\mdp({\cal S}, {\cal A}, H, \P, r)$ becomes a linear MDP
if there exists a function $\phi: {\cal S} \times {\cal A} \rightarrow \R^d$ and an unknown signed measure set $\mu_h = (\mu_h^{(1)}, \ldots, \mu_h^{(d)})$ over ${\cal S}$ such that the transition probability $\P_h [s'|s, a] = \langle\phi(s, a), \mu_h(s')\rangle$ at any step any $h\in [H]$. Here we assume $\max_{(s,a ) \in {\cal S} \times {\cal A}}\|\phi(s, a)\|_2 \leq 1$.  Moreover, there exists a hidden vector $\theta_h \in \R^d$ so that $r_h(s, a) = \langle\phi(s, a), \theta_h\rangle$. Here we assume $\max_{h \in [H]}\{\|\mu_h({\cal S})\|_2, \|\theta_h\|_2\} \le     \sqrt{d}$.
\end{definition}

In the MDP framework, we define the policy $\pi$ as a sequence of functions that map state to actions. 
\begin{definition}[Policy]\label{def:policy:formal}
Given a MDP with form $\mdp({\cal S}, {\cal A}, H, \P, r)$, a policy $\pi=\{\pi_1,\cdots,\pi_H\}$ is defined as sequence such that $\pi_h:{\cal S}\rightarrow {\cal A}$ for each step $h$. $\pi_h(s)=a$ represents the action taken when we are at state $s$ and step $h$.
\end{definition}

Moreover, we use $V^{\pi}_{h}(s):{\cal S}\rightarrow \R$ to define the value of cumulative rewards in expectation if the agent follows received under a given policy $\pi$ when the start state is $s$ and the start step is $h$. 
\begin{definition}[Value function]\label{def:value_function:formal}
Given a MDP with form $\mdp({\cal S}, {\cal A}, H, \P, r)$, we let the value function be: 
\begin{align*}
    V^{\pi}_{h}(s) := \E\left[\sum_{h' = h}^H r_h(s_{h}', \pi_h(s_{h}'))  ~\bigg |~ s_h = s\right], \qquad \forall s\in {\cal S}, h \in [H].
\end{align*}

\end{definition}

Further more, we define the Q function $Q^{\pi}_{h}(s, a):{\cal S}\times {\cal A}\rightarrow \R$ as the expected cumulative rewards if a agent follows policy $\pi$ and starts from takeing action $a$ at state $s$ and step $h$. This representation of $Q^{\pi}_{h}(s, a)$ is also associated with the well-known Bellman equation~\cite{s18}.
\begin{definition}[Q-Learning]\label{def:q_learning:formal}
Let $\mdp({\cal S}, {\cal A}, H, \P, r)$ denotes an episodic MDP. We use a simplified notation $[\P_hV_{h+1}](s, a) := \E_{s' \sim \P_h [s'|s, a]} [V_{h+1}(s')]$. Then, we represent the Bellman equation with policy $\pi$ as
\begin{align*}
    Q^{\pi}_{h}(s, a) = [r_h + \P_h V^{\pi}_{h+1}](s, a) , \qquad   V^{\pi}_{h}(s) = Q^{\pi}_{h}(s, \pi_h(s)),    
    \qquad V^{\pi}_{H+1}(s) = 0.  
\end{align*}
Similarly, for optimal policy $\pi^*$, we have
\begin{align}\label{eq:bellman:formal}
    Q^{*}_{h}(s, a) = [r_h + \P_h V^{*}_{h+1}](s, a) , \qquad   V^{*}_{h}(s) = \max_{a\in{\cal A}} Q^{*}_{h}(s, a),    
    \qquad V^{*}_{H+1}(s) = 0.  
\end{align}
Note that as $r_h\in[0,1]$. All $Q_h^\pi$ and $V_h^\pi$ are upper bounded by $H+1-h$.

\end{definition}

After formulate the MDP and its value functions, we start listing conditions on the space of state and action for the convenience of our Sublinear LSVI and Sublinear LSVI-UCB. We first present the definition for the convex hull.
\begin{definition}[Convex hull]\label{def:cvx_hull:formal}
Given a set $\{x_1, x_2, \cdots, x_n\} \subset \R^d$ that denotes as a matrix $A\in\R^{d\times n}$, we define its convex hull $\mathcal{B}(A)$ to be the collection of all finite linear combinations $y$ that satisfies
$y = \sum_{i=1}^n a_i\cdot x_i$, where $a_i\in [0,1]$ for all $i\in[n]$ and $\sum_{i\in [n]}a_i=1$.
\end{definition}

In this work, we focus on the Sublinear LSVI under continuous state  and action space. Given the action space ${\cal A}$ and state space ${\cal S}$, we formulate $\phi(({\cal S}\times {\cal A}))$ as the convex hull of $\phi({\cal S}_{\core}\times {\cal A}_{\core})$, where ${\cal S}_{\core}$ is core state set and ${\cal A}_{\core}$ is core action set.

\begin{definition}[Core state and core action sets]\label{def:coresets:formal}
Given a linear MDP with form $\mdp({\cal S}, {\cal A}, H, \P, r)$, we define set ${\cal S}_{\core}\subset {\cal S}$ as the core states set and ${\cal A}_{\core}\subset {\cal A}$ as the core action set. We denotes cardinality of ${\cal S}_{\core}$ and ${\cal A}_{\core}$  as $S$ and $A$. Specifically, we have ${\cal B}(\phi({\cal S}_{\core}\times {\cal A}_{\core}))=\phi({\cal S}\times {\cal A})$.  Without loss of generality, we let $A\geq d$.

\end{definition}

In LSVI~\cite{bb96}, the value iteration procedure requires a span matrix that contains state-action  embeddings. Moreover, there also exists a series of assumptions on the span matrix. We provide these assumptions as below:
\begin{definition}[Span matrix]\label{def:spansets:formal}
Given a linear MDP with form $\mdp({\cal S}, {\cal A}, H, \P, r)$, we define the span matrix $\Phi\in \R^{d \times M}$ as follows: in total $M\leq d$ columns, the $j$th column is denoted as $\phi_j=\phi(s_j,a_j)$, where $(s_j,a_j)\in {\cal S} \times {\cal A}$. Moreover, $\{\phi_1, \phi_2, \cdots,\phi_M\}$ is the linear span of $\phi(S \times A)$. Specifically, $\Phi$ satisfies:
\begin{itemize}
\item $\phi(s,a)=\sum_{j=1}^{M}w_j\phi_j$, $w_j\in \R$ for all $(s,a)\in {\cal S} \times {\cal A}$,
\item $\rank(\Phi)=M$,
\item $\max_{(s,a) \in {\cal S} \times {\cal A}}\|\Phi^{-1} \phi(s,a)\|_1\leq L$.
\end{itemize}

\end{definition}

Next, we follow \cite{jywj20} and making assumptions for Sublinear LSVI-UCB. Given a linear MDP with form $\mdp({\cal S}, {\cal A}, H, \P, r)$, we assume ${\cal S}$ is finite with cardinally $S$ and ${\cal A}$ is finite with cardinally $A$.

\subsection{Standard Proprieties of Linear MDP} \label{sec:properties}

We list the tools for analyzing linear MDPs properties from \cite{jswz21} in this section.

\begin{lemma}[Proposition 2.3 \cite{jywj20}] \label{prop:realizable}
The Q function with form $Q_h^\pi(s, a)$ in linear MDP could be represented it as a inner product $Q_h^\pi(s, a) = \langle\phi (s, a),  w^{\pi}_h\rangle$, where $w^{\pi}_h\in \R^d$ is a weight vector.
\end{lemma}

Next, we show the upper bound of weight $w^\pi_h$ for any policy $\pi$.

\begin{lemma}[Lemma B.2 \cite{jywj20}] \label{lem:w_h_pi_bound}
Given a linear MDP, let $w^{\pi}_h$ denotes the weight that achieves $Q_h^\pi(s, a) = \langle\phi (s, a),  w^{\pi}_h\rangle$  for all $(s, a) \in {\cal S}\times{\cal A}$ at step $h\in [H]$. We show that for $\|w^\pi_h\|_2 \leq 2H\sqrt{d}$ for any $h\in [H]$, 
\end{lemma}

\subsection{Locality Sensitive Hashing}\label{sec:def_lsh}
We define locality sensitive hashing ({\lsh}). These definitions are very standard, e.g., see Indyk and Motwani~\cite{im98}. 
\begin{definition}[Locality Sensitive Hashing]
Let $\dist$ denotes a metric distance.  Let $\ov{c}$ denote a parameter such that $\ov{c} >1$. Let $p_1, p_2$ denote two parameters such that $0 < p_2 < p_1 < 1$. A family $\mathcal{H}$ is called $(r,\ov{c} \cdot r,p_1,p_2)$-sensitive if and only if, for any two point $x,y \in \R^d$, $h$ chosen uniformly from $\mathcal{H}$ satisfies the following:
\begin{itemize}
    \item if $ \dist (x,y) \leq r$, then $\Pr_{h \sim \mathcal{H}} [ h(x)=h(y) ] \geq p_1$,
    \item if $ \dist (x,y) \geq \ov{c} \cdot r$, then $\Pr_{h \sim \mathcal{H}} [ h(x)=h(y) ] \leq p_2$.
\end{itemize}
\end{definition}

We focus on situations where $\dist$ is $\ell_2$ or cosine distance.

{\lsh} is designed to accelerate the runtime of the Approximate  Nearest Neighbor ($\ann$) problem. We start with define the exact $\nn$ problem as:
\begin{definition}[Exact Nearest  Neighbor ($\nn$)]\label{def:nn:formal}
 Given an $n$-point dataset $Y \subset \mathbb{S}^{d-1}$ on the sphere, the goal of the  Nearest Neighbor ($\nn$) problem  is
to find a datapoint $y \in Y$ for a query $x \in \mathbb{S}^{d-1}$ such that 
\begin{align*}
    \nn (x,Y) := \min_{y \in Y} \|x-y\|_2.
\end{align*}
\end{definition}

\cite{im98} relax the $\nn$ problem in Definition~\ref{def:nn:formal} as with approximation and define the Approximate Nearest Neighbor ($\ann$) problem.  
\begin{definition}[Approximate Nearest  Neighbor ($\ann$)]\label{def:ann:formal}
Let $\ov{c} >1$ and $r \in (0,2)$. Given an $n$-point dataset $P \subset \mathbb{S}^{d-1}$ on the sphere, the goal of the $(\ov{c},r)$-Approximate Near Neighbor problem ($\ann$) is to build a data structure that, given a query $q \in \mathbb{S}^{d-1}$ with the promise that there exists a datapoint $p \in P$ with $\| p - q \|_2 \leq r$ reports a datapoint $p' \in P$ within distance $\ov{c} \cdot r$ from $q$.
\end{definition}

Then, the query complexity of $\ann$ is reduced to sublinear by {\lsh} following Theorem~\ref{thm:ar15:formal} and Theorem~\ref{thm:ar17:formal}. Note that here we write $O(1/\sqrt{\log n})$ as $o(1)$.

\begin{theorem}[Andoni and Razenshteyn \cite{ar15}]\label{thm:ar15:formal}
Let $\ov{c} > 1$ and $r \in (0,2)$. The $(\ov{c},r)$-$\ann$ on a unit sphere $\mathbb{S}^{d-1}$ can be solved  with query time $O(d \cdot n^{\rho})$, space $O(n^{1+\rho} + d n)$ and preprocessing time $O(dn^{1+\rho})$, where $\rho = \frac{1}{2\ov{c}^2-1} +o(1)$.
\end{theorem}

\begin{theorem}[Andoni, Laarhoven, Razenshteyn and Waingarten \cite{alrw17}]\label{thm:ar17:formal}
Let $\ov{c} > 1$ and $r \in (0,2)$. The $(\ov{c},r)$-$\ann$ on a unit sphere $\mathbb{S}^{d-1}$ can be solved  with query time $O(d \cdot n^{\rho})$, space $O(n^{1+o(1)} + d n)$ and preprocessing time $O(dn^{1+o(1)})$, where $\rho = \frac{2}{\ov{c}^2} -\frac{1}{\ov{c}^4}+o(1)$.
\end{theorem}

In this work, we focus on the $\maxip$, which is a well-known problem in the field of computational complexity, we follow the standard notation in this work \cite{c18}. We define the exact and approximate $\maxip$ problem as follows:
\begin{definition}[Exact $\maxip$]\label{def:exact_maxip} 
Given a data set $Y\subseteq \R^d$, we define $\maxip$ for a query point $x\in \R^d$ with respect to $Y$ as follows:
\begin{align*}
    \maxip (x,Y) := \max_{y \in Y} \langle x,y \rangle .
\end{align*}
\end{definition}

\begin{definition}[Approximate $\maxip$]\label{def:approximate_maxip}
Let $c \in (0,1)$ and $\tau \in (0,1)$.
Given an $n$-point dataset $Y \subset \mathbb{S}^{d-1}$, the goal of the $(c,\tau)$-{$\maxip$} is to build a data structure that, given a query $x \in \mathbb{S}^{d-1}$ with the promise that there exists a datapoint $y \in Y$ with $\langle x , y \rangle \geq \tau$, it reports a datapoint $z \in Y$ with similarity $\langle x , z \rangle \geq c \cdot \maxip(x,Y)$.
\end{definition}

To solve $(c,\tau)$-{$\maxip$}, we define a dual version of {\lsh} data structure (Shrivastava and Li~\cite{sl14} call it asymmetric {\lsh}):
\begin{definition}[Asymmetric  Locality Sensitive Hashing]\label{def:alsh}
Let $c$ denotes a parameter such that $c \in (0,1)$.  Let $\tau$ denotes a parameter such that $\tau>0$. Let $p_1, p_2$ denote two parameters such that $0 < p_2 < p_1 < 1$. Let $\mathrm{sim}(x,y)$ denote a binary similarity function between $x,y\in \R^d$.
A family $\mathcal{H}$ is called $(\tau,c \cdot \tau,p_1,p_2)$-sensitive if and only if, for any query point $x \in \R^d$ and a data point $y \in \R^d$, $h$ chosen uniformly from $\mathcal{H}$ satisfies the following:

\begin{itemize}
    \item if $\mathrm{sim}(x,y) \geq \tau$ then $\Pr_{h \sim \mathcal{H}} [ h(  x  ) = h(  y  ) ]\geq p_1$,
    \item if $\mathrm{sim}(x,y)\leq c \cdot \tau$ then $\Pr_{h \sim \mathcal{H}} [ h(x) = h(y) ]\leq p_2$.
\end{itemize}
\end{definition}

It is shown from~\cite{sl14} that {\lsh} type data structure with asymmetric transformations could achieve sublinear runtime complexity of $(c,\tau)$-{$\maxip$}.

\subsection{Probabilistic Tools}\label{sec:prob_tools}
\begin{lemma}[Hoeffding bound \cite{h63}]\label{lem:hoeffding}
Let $x_1,\cdots,x_n$ be n independent bounded variables in $[a_i,b_i]$. Let , then we show the Hoeffding bound over $x=\sum_{i=1}^{n} x_i$ as:
\begin{align*}
    \Pr[|x-\E[x]|\geq t]\leq 2\exp\Big(-\frac{2t^2}{\sum_{i=1}^{n}(b_i-a_i)^2}\Big).
\end{align*}
\end{lemma}


\subsection{Inequalities}\label{sec:ineq}

In this sections, we present the supporting inequalities for our work.
\begin{fact}[Lemma D.1 in \cite{jywj20}]\label{lem:lamdba_ineq} Given a matrix $\Lambda_t = \lambda \I_d + \sum_{i=1}^t \phi_i \phi_i^{\top}$ with $ \phi_i \in \R^d$ and $\lambda > 0$, we show that:
\begin{align*}
\sum_{i=1}^t \phi_i^{\top} (\Lambda_t)^{-1} \phi_i \le d.
\end{align*}
\end{fact}

\begin{lemma}[Lemma D.4 in \cite{jywj20}]\label{lem:self_norm_covering}
Let $\mathcal{V}$ denotes a function family that $\max_{V\in \mathcal{V},x\in {\cal S}} |V(x)| \leq H$. Let $G$ denotes the $\epsilon$-covering number of $\mathcal{V}$. Let ${\cal S}$ denotes a state space. Let $\{\F_\tau\}_{\tau = 0}^\infty$ denotes the filtration of ${\cal S}$. Let $\{x_\tau\}_{\tau = 1}^\infty$ denotes a random process defined on ${\cal S}$. Let $\{\phi_\tau\}_{\tau = 0}^\infty$ denotes a real valued random process in $\R^d$. Moreover, $\phi_\tau \in \F_{\tau-1}$ and we have upper bound $\| \phi_\tau \|_2 \le 1$. Given a matrix $\Lambda_k\in \R^{d\times d}$ so that $\Lambda_k = \lambda I_d + \sum_{\tau=1}^k \phi_\tau \phi_\tau^{\top}$, for any $\delta >0$, for any $k \geq 0$, for any  $V \in \mathcal{V}$, we have 

\begin{align*} 
\Big\| \sum_{\tau = 1}^k \phi_\tau (  V(x_\tau) - \E[V(x_\tau) ~|~ \F_{\tau-1}] )  \Big\|^2_{\Lambda_k^{-1}}
\le 4H^2 \big( d\log ( 1+ k/\lambda  )  + \log({G_{\epsilon}}/{\delta}) \big)  + {8k^2\epsilon^2}/{\lambda}, 
\end{align*}
\end{lemma}

\section{Data Structures}\label{sec:data_structure}
This section presents the data Structures for our work. 
\begin{itemize}
    \item In Section~\ref{sec:primal_dual}, we introduce the transformations that build primal-dual connections between approximate $\maxip$ and $\ann$. 
    \item In Section~\ref{sec:sublinear_ds}, we present our  data structure that achieves sublinear query time in approximate $\maxip$.
    \item In Section~\ref{sec:sublinear_matnorm}, we show how to perform approximate $\maxmatnorm$ via approximate $\maxip$ data structure.
    \item In Section~\ref{sec:efficient_trans}, we present our efficient transformations for $\maxip$ in optimization.
    \item In Section~\ref{sec:query_part1}, we formally provide the theoretical results of sublinear approximate $\maxip$ using one {\lsh} data structure.
    \item In Section~\ref{sec:query_part2}, we provide the theoretical results of sublinear approximate $\maxip$ using another {\lsh} data structure.
\end{itemize}

\subsection{Existing Transformation from Primal to Dual}\label{sec:primal_dual}
In this section, we show a transformation that builds the connection between $\maxip$ and $\nn$. Under this asymmetric transformation, $\nn$  is formulated as a dual problem of $\maxip$.

We start with presenting the asymmetric transformation.
\begin{definition}[Asymmetric  transformation \cite{ns15}]\label{def:asym_pq_transform}
Let $Y\in \R^d$ and $\|y\|_2\leq 1$ for all $y\in Y$. Let $x\in \R^d$ and $\|x\|_2\leq D_{x}$. We define the following asymmetric transform:
\begin{eqnarray}
  \label{eq:simpleP}
  P(y) &=& \begin{bmatrix} y^\top & \sqrt{1-\|y\|_2^2} & 0 \end{bmatrix}^\top \\
  \nonumber
  Q(x) &=& \begin{bmatrix} (x D_{x}^{-1})^\top & 0 & \sqrt{1-\|x D_{x}^{-1}\|_2^2} \end{bmatrix}^\top
  \nonumber
\end{eqnarray}
Therefore, we have
\begin{align*}
\|Q(x)-P(y)\|_2^2=2-2 D_{x}^{-1}\langle x,y\rangle,  ~~~\arg\max_{y\in Y}\langle x,y \rangle =\arg\min_{y\in Y} \|Q(x)-P(y)\|^2.
\end{align*}
\end{definition}

In this way, we regard $\maxip$ as the primal problem and $\nn$ as a dual problem.

\subsection{Sublinear $\maxip$ Data Structure}\label{sec:sublinear_ds}
In this section, we show the theorem that provides sublinear query time for  $\maxip$ problem using {\lsh} type data structure.

\begin{theorem}[Formal statement of Corollary~\ref{cor:c_tau_maxip:informal}]\label{thm:maxip_lsh}
Let $c \in (0,1)$ and $\tau \in(0,1)$. Given a set of $n$-points $Y \subset {\cal S}^{d-1}$ on the sphere, one can construct a data structure with ${\cal T}_{\mathsf{init}}$ preprocessing time and ${\cal S}_{\mathsf{space}}$ space so that for any query $x \in {\cal S}^{d-1}$, we take query time complexity $O(d\cdot n^{\rho})$:
\begin{itemize}
    \item if $ \maxip(x,Y)\geq\tau $, then we output a vector in $Y$ which is a $(c,\tau)$-$\maxip$ with respect to $(x,Y)$ with probability at least $0.9$\footnote{It is obvious to boost probability from constant to $\delta$ by repeating the data structure $\log(1/\delta)$ times.}, where $\rho:= f(c,\tau)+o(1)$.
    \item otherwise, we output {\fail}.
\end{itemize}
Further, 
\begin{itemize}
    \item If ${\cal T}_{\mathsf{init}} = O( d n^{1+\rho})$ and ${\cal S}_{\mathsf{space}} = O(n^{1+\rho} + d n)$, then $f(c,\tau) = \frac{1-\tau}{1-2c\tau+\tau}$.
    \item If ${\cal T}_{\mathsf{init}} = O( d n^{1+o(1)})$ and ${\cal S}_{\mathsf{space}} = O( n^{1+o(1)} + d n)$, then $f(c,\tau) = \frac{2(1-\tau)^2}{(1-c\tau)^2}-\frac{(1-\tau)^4}{(1-c\tau)^4}$.
\end{itemize}

\end{theorem}

\begin{proof}
 We start with showing that for any two points $x,y$ with $\| x \|_2 = \|y \|_2=1$, we have $\|x-y\|_2^2= 2 - 2\langle x , y\rangle$. This implies that $r^2 = 2 - 2 \tau$ for a $(\ov{c}, r)$-{$\ann$} and a $(c, \tau)$-$\maxip$ on $x,Y$.

 Further, if we have a data structure for $(\ov{c}, r)$-$\ann$, it automatically becomes a data structure for $(c, \tau)$-$\maxip$  with parameters $\tau = 1-0.5 r^2$ and $c = \frac{1-0.5 \ov{c}^2 r^2}{1 - 0.5 r^2}$.  This implies that 
\begin{align*}
\ov{c}^2= \frac{ 1 - c(1-0.5 r^2) }{0.5r^2} = \frac{1 - c \tau }{1-\tau} .
\end{align*}

Next, we show how to solve $(c,\tau)$-$\maxip$ by solving  $(\ov{c}, r)$-$\ann$ using two different data structures.

{\bf Part 1.} If we initialize the data-structure following Theorem~\ref{thm:ar15:formal}, we show that the $(c,\tau)$-$\maxip$ on a unit sphere ${\cal S}^{d-1}$ can be solved by solving  $(\ov{c}, r)$-$\ann$ with query time $O(d \cdot n^{\rho})$, space $O(n^{1+\rho} + d n)$ and preprocessing time $O(dn^{1+\rho})$, where
\begin{align*}
    \rho = \frac{1}{2\ov{c}^2-1} +o(1) = \frac{1}{ 2 \frac{1-c\tau}{1-\tau} -1 } +o(1) = \frac{1-\tau}{1-2c\tau+\tau} + o(1).
\end{align*}
Thus, $f(c,\tau)=\frac{1-\tau}{1-2c\tau+\tau}$.

{\bf Part 2.} If we initialize the data-structure following Theorem~\ref{thm:ar17:formal}, we show that the $(c,\tau)$-$\maxip$ on a unit sphere ${\cal S}^{d-1}$ can be solved by solving  $(\ov{c}, r)$-$\ann$ with query time $O(d \cdot n^{\rho})$, space $O(n^{1+o(1)} + d n)$ and preprocessing time $O(dn^{1+o(1)})$, where
\begin{align*}
    \rho =  \frac{2}{\ov{c}^2} -\frac{1}{\ov{c}^4} +o(1) = \frac{2(1-\tau)^2}{(1-c\tau)^2}-\frac{(1-\tau)^4}{(1-c\tau)^4}+o(1). 
\end{align*}
Thus, $f(c,\tau)=\frac{2(1-\tau)^2}{(1-c\tau)^2}-\frac{(1-\tau)^4}{(1-c\tau)^4}$.

\end{proof}

In practice, we tune parameter $\tau$ close to $\maxip(x,Y)$ to achieve higher $c$. Moreover, Theorem~\ref{thm:maxip_lsh} could be applied to general $\maxip$ problem. To do this, we first apply asymmetric transformation in Definition~\ref{def:asym_pq_transform} and transfer it to a $(c,\tau)$-{$\maxip$} problem over $Q(x)$ and $Q(Y)$. Then, we solve this $(c,\tau)$-{$\maxip$} problem by solving its dual problem, which is $(\ov{c},r)$-{$\ann$}. Finally, the solution to the $(\ov{c},r)$-{$\ann$} would be the approximate solution to the original $\maxip(x,Y)$. Meanwhile, it is reasonable for us to regard $d=n^{o(1)}$  using Johnson-Lindenstrauss Lemma~\cite{jl84}.

\subsection{Sublinear $\maxip$ Data Structure for Maximum Matrix Norm Search}\label{sec:sublinear_matnorm}

In this section, we extend {\lsh} type $\maxip$ data structure for maximum matrix norm search.
\begin{definition}[Exact Maximum Matrix Norm ($\maxmatnorm$)]\label{def:exact_maxmatnorm:formal} 
Given a data set $Y\subseteq \R^d$ and a query matrix $x\in\R^{d\times d}$,  we define Maximum Matrix Norm as follows:
\begin{align*}
     \maxmatnorm(x,Y):= \max_{y \in Y} \|y\|_{x}.
\end{align*}
\end{definition}

Next, we define the approximate version of the Maximum Matrix Norm.

\begin{definition}[Approximate $\maxmatnorm$]\label{def:approximate_maxmatnorm:formal}

Let $c \in (0,1)$ and $\tau \in (0,1)$. Let $vec$ denotes the vectorization of ${d\times d}$ matrix into a $d^2$ vector.
Given an $n$-point dataset $Y \subset \R^d$ and $yy^\top \in \mathbb{S}^{d^2-1}$ for all $y\in Y$, the goal of the $(c,\tau)$-$\maxmatnorm$ is to cosntruct a data structure that, given a query matrix $x \in \R^{d\times d}$ and $\vect(x)\in \mathbb{S}^{d^2-1}$ with the promise that there exists a datapoint  $y \in Y$ with $\|y\|_x\geq \tau$, it reports a datapoint $z \in Y$ with $ \|z\|_{x}\geq c \cdot \maxmatnorm(x,Y)$. 
\end{definition}

Next, we show the relationship between $\maxmatnorm$ and $\maxip$

\begin{lemma}[Relation between $\maxmatnorm$ and $\maxip$]\label{lem:maxmatnorm_maxip:formal}
We show that
\begin{align*}
    \maxmatnorm(X,Y)^2 =\max_{y \in Y} \langle \vect(x),\vect(y y^\top)
\end{align*}
where $\vect$ vectorizes $d\times d$ matrix $x$ into a $d^2$ vector.

\end{lemma}
\begin{proof}
We show that
\begin{align*}
    \maxmatnorm(x,Y)^2
    =&~ \max_{y \in Y} \|y\|_{x}^2\\
     =&~ \max_{y \in Y} y^\top x y\\
     =&~ \max_{y \in Y} \langle \vect(x),\vect(y y^\top) \rangle
\end{align*}
where the first step follows the definition of $\maxmatnorm$, the second step follows from the definition of $\|y\|_{x}^2$, the third step decomposes the quadratic form into a inner product.

\end{proof}

Next, we present our main theorem for $\maxmatnorm(x,Y)$.
\begin{theorem}[]\label{thm:maxmatnorm_lsh}
Let $c$ denote a parameter such that $c \in (0,1)$.  Let $\tau$ denote a parameter such that $\tau \in (0,1)$. Let $vec$ denotes the vectorization of ${d\times d}$ matrix into a $d^2$ vector. Given a $n$-points set $Y\subseteq \R^d$ and $yy^\top \in \mathbb{S}^{d^2-1}$ for all $y\in Y$, one can construct a data structure with ${\cal T}_{\mathsf{init}}$  preprocessing time and ${\cal S}_{\mathsf{space}}$ so that for any query matrix $x \in \R^{d\times d}$ with $\vect(x)\in \mathbb{S}^{d^2-1}$, we take query time complexity $O( d^2 n^{\rho} \cdot\log(1/\delta))$:
\begin{itemize}
    \item if $\maxmatnorm (x,Y)\geq\tau $, then we output a vector in $Y$ which is a $(c,\tau)$-$\maxmatnorm$ with respect to $(x,Y)$ with probability at least $1-\delta$, where $\rho:= f(c,\tau)  + o(1)$.
    \item otherwise, we output {\fail}.
\end{itemize}
Further, 
\begin{itemize}
    \item If ${\cal T}_{\mathsf{init}} = O( d^2n^{1+\rho}\cdot\log(1/\delta))$ and ${\cal S}_{\mathsf{space}} = O((n^{1+\rho} + d^2 n)\cdot\log(1/\delta))$, then $f(c,\tau) = \frac{1-\tau^2}{1-c^2\tau^2+\tau^2}$.
    \item If ${\cal T}_{\mathsf{init}} = O( d^2n^{1+o(1)}\cdot\log(1/\delta))$ and ${\cal S}_{\mathsf{space}} = O( (n^{1+o(1)} + d^2 n)\cdot\log(1/\delta))$, then $f(c,\tau) = \frac{2(1-\tau^2)^2}{(1-c^2\tau^2)^2}-\frac{(1-\tau^2)^4}{(1-c^2\tau^2)^4}$.
\end{itemize}
\end{theorem}

\begin{proof}

 We start with showing that if we have a $(c^2,\tau^2)$-$\maxip$ data structure over $\vect(x)$ and every $\vect(yy^\top)$, $y\in Y$, we would obtain a $z\in Y$ such that

\begin{align}\label{eq:matmaxnorm_c}
    \langle \vect(x),\vect(z z^\top) \rangle \geq  c^2 \max_{y \in Y} \langle \vect(x),\vect(y y^\top) \rangle,
\end{align}
 we could use it and derive the following propriety for $z$:

\begin{align*}
    \|z\|_{x}
    =&~\sqrt{\langle \vect(x),\vect(z z^\top) \rangle}\\
    \geq &~\sqrt{c^2 \max_{y \in Y} \langle \vect(x),\vect(y y^\top) \rangle} \\
    = &~c \max_{y \in Y}\sqrt{ \langle \vect(x),\vect(y y^\top) \rangle}\\
    =&~ c\max_{y \in Y}  \|y\|_{x}
\end{align*}
where the second step follows from Eq.~\eqref{eq:matmaxnorm_c}.

Therefore, $z$ is the solution for $(c,\tau)$-$\maxmatnorm(x,Y)$.

Next, we show how to retrieve $z$ via two data structures used for $(c,\tau)$-$\maxip(x,Y)$  in Theorem~\ref{thm:maxip_lsh}.

{\bf Part 1.} 
If we initialize the data structure following Theorem~\ref{thm:ar15:formal}, we can construct a data structure with $O( (n^{1+\rho} + d^2 n)\cdot \log(1/\delta))$ preprocessing time and $O((n^{1+\rho} + d^2 n)\cdot\log(1/\delta))$ space so that for any query matrix $x \in \R^{d\times d}$ with $\vect(x)\in \mathbb{S}^{d^2-1}$, we take query time complexity $O(d^2 n^{\rho} \cdot \log(1/\delta))$ to retrieve $z$. Here $\rho=\frac{1-\tau^2}{1-c^2\tau^2+\tau^2} + o(1)$ and we are able to improve the failure probability to $\delta$ by repeating the {\lsh} for $\log(1/\delta)$ times. 

{\bf Part 2.}
If we initialize the data structure following Theorem~\ref{thm:ar17:formal}, we can construct a data structure with $O( (n^{1+o(1)} + d^2 n)\cdot \log(1/\delta))$ preprocessing time and $O( (n^{1+o(1)} + d n)\cdot\log(1/\delta))$ space so that for any query matrix $x \in \R^{d\times d}$ with $\vect(x)\in \mathbb{S}^{d^2-1}$, we take query time complexity $O(d^2 n^{\rho} \cdot \log(1/\delta))$ to retrieve $z$. Here $\rho=\frac{2(1-\tau^2)^2}{(1-c^2\tau^2)^2}-\frac{(1-\tau^2)^4}{(1-c^2\tau^2)^4} + o(1)$ and we also improve the failure probability to $\delta$ by repeating the {\lsh} for $\log(1/\delta)$ times.

\end{proof}

Moreover, Theorem~\ref{thm:maxmatnorm_lsh} could be applied to general $\maxmatnorm$ problem. To do this, we first apply transform $(c,\tau)$-$\maxmatnorm$ problem into a $(c^2,\tau^2)$-$\maxip$ problem using Lemma~\ref{lem:maxmatnorm_maxip:formal}. Next, we apply transformations in Definition~\ref{def:asym_pq_transform} and transfer the $(c^2,\tau^2)$-$\maxip$ problem to a $(c^2,\tau^2)$-{$\maxip$} problem over $Q(x)$ and $Q(Y)$. Then, we solve this $(c^2,\tau^2)$-{$\maxip$} problem by solving its dual problem, which is $(\ov{c},r)$-{$\ann$}. Finally, the solution to the $(\ov{c},r)$-{$\ann$} would be the approximate solution to the original $\maxmatnorm(x,Y)$.

\subsection{Transformation for Efficient Query}\label{sec:efficient_trans}
In the optimization problem that could be accelerated by  $(c,\tau)$-$\maxip$, the query and data vectors are usually not unit vectors so that we apply transformations in Definition~\ref{def:asym_pq_transform} to map both query and data vectors into unit vectors. However, if the mapped inner product is too close to $1$. The formulation of $\rho$ would break and the time complexity would be linear. To avoid this, we propose a new set of asymmetric transformations:

\begin{definition}[Efficient asymmetric  transformation]\label{def:efficient_asym_pq_transform}
Let $Y\in \R^d$ and $\|y\|_2\leq 1$ for all $y\in Y$. Let $x\in \R^d$ and $\|x\|_2\leq D_{x}$. We define the following asymmetric transform:
\begin{align*}
    P(y) = \begin{bmatrix} y^\top & \sqrt{1-\|y\|_2^2}  & 0 \end{bmatrix}^\top, ~~~
  Q(x) = \begin{bmatrix}\frac{0.8\cdot x^\top}{D_x}  & 0 & \sqrt{1-\frac{0.64\cdot \|x\|_2^2}{D_x^2}} \end{bmatrix}^\top.
\end{align*}
\end{definition}

Next, we use Lemma~\ref{lem:avoid_tau_1} to show how to enforce $\tau$ to be away from $1$ via our efficient asymmetric transformation.

\begin{lemma}\label{lem:avoid_tau_1}
Given the transformation $P$ and $Q$ defined in Definition~\ref{def:efficient_asym_pq_transform}, we show that both $\maxip(Q(x),P(Y))$ and $\nn(Q(x),P(Y))$  are equivalent to $\maxip(x,Y)$. Moreover,
\begin{align*}
\maxip(Q(x),P(Y))\leq 0.8.
\end{align*}
\end{lemma}

\begin{proof}
Using transformations in Definition~\ref{def:efficient_asym_pq_transform}, for all $y\in Y$,  we have
\begin{align*}
    Q(x)^\top P(y)=\frac{0.8\cdot x^\top y}{D_x}\leq 0.8\cdot\frac{\|x\|_2\|y\|_2}{D_x}\leq 0.8
\end{align*}
where the third step follows from $\|x\|_2\leq D_x$ and $\|y\|_2\leq 1$.

Next, we show that $\maxip(Q(x),P(Y))$ is equivalent to $\maxip(x,Y)$.
\begin{align*}
    \arg\max_{y\in Y} Q(x)^\top P(y)=\arg\max_{y\in Y}\frac{0.8\cdot \langle x,y \rangle }{D_x}=\arg\max_{y\in Y} \langle x,y \rangle. 
\end{align*}

Further more, $\nn(Q(x),P(Y))$ (see Definition~\ref{def:nn:formal}) is equivalent to $\maxip(x,Y)$.
\begin{align*}
\|Q(x)-P(y)\|_2^2=2-1.6 D_{x}^{-1}\langle x,y\rangle,  ~~~\arg\min_{y\in Y}\|Q(x)-P(y)\|^2 =\arg\max_{y\in Y} \langle x,y \rangle .
\end{align*}
\end{proof}

\subsection{Sublinear Query Time: Part 1}\label{sec:true_sublinear_query}\label{sec:query_part1}

In this section, we show that $\rho$ is strictly less than $1$ using  {\lsh} in \cite{ar15}.

\begin{lemma}\label{lem:ar15_rho_bound}
 If {\lsh} data structure's parameters $c$ and $\tau$ satisfy that $c\in [0.5,1)$ and $\tau\in [0.5,1)$
then, we could upper bound $\rho$ as:
\begin{align*}
    \rho< 1-\frac{\gamma}{2}+ O(1/\sqrt{\log n})
\end{align*}
where $\gamma=1-c$.
\end{lemma}

\begin{proof}

We can upper bound $\rho$ as follows: 
\begin{align*}
    \rho =&~\frac{1-\tau}{1-2c\tau+\tau} +  O(1/\sqrt{\log n})\\
    = & ~ 1 - \frac{  2\tau - 2 c \tau }{1-2c\tau + \tau}+  O(1/\sqrt{\log n}) \\
    = & ~ 1 - (1-c) \cdot \frac{2\tau}{1-2c\tau + \tau}+  O(1/\sqrt{\log n}) \\
    \leq & ~ 1 - (1-c) \cdot \frac{1}{1-2c\tau + \tau}+  O(1/\sqrt{\log n})  & \text{~by~} \tau \geq 0.5\\
    < & ~ 1 - (1-c) \cdot \frac{1}{2}+  O(1/\sqrt{\log n}) & \text{~by~} \tau < 1 \\
    = & ~ 1-\frac{\gamma}{2}+ O(1/\sqrt{\log n})
\end{align*}
where the second and third steps are reorganizations, the forth step follows from $\tau\geq 0.5$, the fifth step follows from $\tau< 1$ and $c\geq 0.5$, the last step is a reorganization.

Therefore, we complete the proof.
\end{proof}

For Sublinear LSVI, we set $c= 1-C_0 L\sqrt{\iota/n}$ and and $\tau\geq 0.5$ by shifting the reward function.
In this way, we have
\begin{align}
    \rho
    <&~1-\frac{C_0 L\sqrt{\iota/n}}{2}+ O(\frac{1}{\sqrt{\log A}})<1- \frac{1}{4} C_0L\sqrt{\iota/n}
\end{align}
where the first step follows from $\gamma=1-c=C_0 L\sqrt{\iota/n}$, the second step follows from $\frac{1}{4} C_0L\sqrt{\iota/n}>\Omega(\frac{1}{\sqrt{\log A}})$.

For Sublinear LSVI-UCB, we set $c=1-\frac{1}{\sqrt{K}}$ and $\tau\geq 0.5$ by shifting the reward function. In this way, we have
\begin{align}
    \rho
    <&~1-\frac{1}{2\sqrt{K}}+ O(\frac{1}{\sqrt{\log A}})<1-\frac{1}{4\sqrt{K}}
\end{align}
where the first step follows from $\gamma=1-c=\frac{1}{\sqrt{K}}$, the second step follows from $\frac{1}{4\sqrt{K}}>\Omega(\frac{1}{\sqrt{\log A}})$.

Therefore, we show that sublinear value iteration can be achieved while preserving the same regret.

\subsection{Sublinear Query Time: Part 2}\label{sec:query_part2}

In this section, we show that $\rho$ is strictly less than $1$ using  {\lsh} in \cite{alrw17}.

Using \cite{alrw17}, the $\rho$ for {\lsh} based $\maxip$ data structure with parameters $c$ and $\tau$ becomes
\begin{align*}
    \rho=\frac{2(1-\tau)^2}{(1-c\tau)^2}-\frac{(1-\tau)^4}{(1-c\tau)^4}+o(1)
\end{align*}
where is a function over $c$ and $\tau$.

To upper bound the $\rho$, we start with showing that it is decreasing as $\tau$ increase when $c\in [0.5,1)$ and $\tau\in [0.5,1)$.

\begin{lemma}\label{lem:ar17_rho_bound}
 Let $c\in [0.5,1)$ and $\tau\in [0.5,1)$. We show that function 
 \begin{align*}
    f(c,\tau):= \frac{2(1-\tau)^2}{(1-c\tau)^2}-\frac{(1-\tau)^4}{(1-c\tau)^4}
 \end{align*} 
 is decreasing as $\tau$ increase.
\end{lemma}

\begin{proof}
We take the derivative of $f(c,\tau)$ in $\tau$ and get
\begin{align*}
     \frac{\partial}{\partial \tau} f(c,\tau)=-\frac{4(c-1)^2(\tau-1)\tau(c\tau+\tau-2)}{(1-c\tau)^5}<0
\end{align*}
where the second step follows from $c\in [0.5,1)$ and $\tau\in [0.5,1)$.

Thus, $f(c,\tau)$ is decreasing as $\tau$ increase when $c\in [0.5,1)$ and $\tau\in [0.5,1)$.

\end{proof}

Next, we have our results in upper bounding $\rho$.

\begin{lemma}
 If {\lsh} data structure's parameters $c$ and $\tau$ satisfy that $c\in [0.5,1)$ and $\tau\in [0.5,1)$
then, we could upper bound $\rho$ as:
\begin{align*}
    \rho< 1- \frac{   \gamma^2  }{ 4 }+ O(1/\sqrt{\log n})
\end{align*}
where $\gamma=1-c$.
\end{lemma}

\begin{proof}

Let $\gamma=1-c$, we have

\begin{align*}
    \rho
    =&~\frac{2}{\ov{c}^2}-\frac{1}{\ov{c}^4}+O(1/\sqrt{\log n})\\
    =&~ \frac{2(1-\tau)^2}{(1-c\tau)^2}-\frac{(1-\tau)^4}{(1-c\tau)^4}+  O(1/\sqrt{\log n})\\
     \leq &~ \frac{0.5}{(1-0.5c)^2}-\frac{0.0625}{(1-0.5c)^4}+  O(1/\sqrt{\log n})\\
    =&~\frac{0.5}{(0.5+0.5\gamma)^2}-\frac{0.0625}{(0.5+0.5\gamma)^4}+  O(1/\sqrt{\log n})\\
    =&~\frac{2}{(1+\gamma)^2}-\frac{1}{(1+\gamma)^4}+  O(1/\sqrt{\log n})\\
    = & ~ \frac{ 2 + 4 \gamma + 2\gamma^2 -1 }{ (1+\gamma)^4 } +  O(1/\sqrt{\log n}) \\
    = & ~ \frac{ 1+ 4\gamma + 2 \gamma^2 }{ (1+\gamma)^4 } +  O(1/\sqrt{\log n})\\
    = & ~ 1 - \frac{   4 \gamma^2 + 4 \gamma^3 + \gamma^4 }{ (1+\gamma)^4 } +  O(1/\sqrt{\log n}) \\
    < & ~ 1 - \frac{   4 \gamma^2  }{ (1+\gamma)^4 } +  O(1/\sqrt{\log n}) & \text{~by~}\gamma > 0 \\
    <  & ~ 1 - \frac{   \gamma^2  }{ 4 } + O(1/\sqrt{\log n}) & \text{~by~} \gamma < 1
\end{align*}
where the second step follows from $\ov{c}^2=\frac{1-c\tau}{1-\tau}$, the third step follows from that $\rho$ is monotonic decrease as $\tau$ increase and $\tau\geq 0.5$, the forth to eighth steps are reorganizations, the ninth step follows from $\gamma=1-c> 0$, the tenth step follows from $\gamma=1-c<1$.

\end{proof}

For Sublinear LSVI, we set $c= 1-C_0 L\sqrt{\iota/n}$ and and $\tau\geq 0.5$ by shifting the reward function.
In this way, we have
\begin{align}
    \rho
    <&~1-\frac{C_0^2 L^2\iota}{4n}+ O(\frac{1}{\sqrt{\log A}})<1-\frac{1}{8} C_0^2L^2\iota/n
\end{align}
where the first step follows from $\gamma=1-c=C_0 L\sqrt{\iota/n}$, the second step follows from $\frac{1}{8} C_0^2L^2\iota/n>\Omega(\frac{1}{\sqrt{\log A}})$.

For Sublinear LSVI-UCB, we set $c=1-\frac{1}{\sqrt{K}}$ and $\tau\geq 0.5$ by shifting the reward function. In this way, we have
\begin{align}
    \rho
    <&~1-\frac{1}{4K}+ O(\frac{1}{\sqrt{\log A}})<1-\frac{1}{8K}
\end{align}
where the first step follows from $\gamma=1-c=\frac{1}{\sqrt{K}}$, the second step follows from $\frac{1}{8K}>\Omega(\frac{1}{\sqrt{\log A}})$.

Therefore, we show that sublinear value iteration can be achieved while preserving the same regret.


\newpage
\section{Sublinear Least-Squares Value Iteration}\label{sec:SLSVI}

This section presents the Sublinear Least-Squares Value Iteration (Sublinear LSVI)
\begin{itemize}
    \item In Section~\ref{sec:lsvi_alg}, we introduce the Sublinear LSVI algorithm.
    \item In Section~\ref{sec:value_diff}, we provide the upper bound of the difference between the optimal value function and the estimated value function.
    \item In Section~\ref{sec:lsvi_regret}, we present the regret analysis of Sublinear LSVI.
    \item In Section~\ref{sec:runtime}, we perform a runtime analysis on the building blocks of Sublinear LSVI to analyze its efficiency.
    \item In Section~\ref{sec:lsvi_compare}, we compare Sublinear LSVI with LSVI~\cite{bb96} in regret and value iteration complexity.
\end{itemize}

\subsection{Algorithm}\label{sec:lsvi_alg}

We present our Sublinear LSVI algorithm in Algorithm~\ref{alg:sublinear_LSVI}. We summarize our algorithm as several steps: (1) sample collection: we query a pair of state and action in the span matrix for $n$ times at each step and observe its reward and next state, (2) data structure construction, we preprocess the embeddings for state and action pairs and build a nearest neighbor data structure, (3) we perform least-squares solver to estimate the weight in the linear MDP model, (4) we use {\lsh} for value function estimation, (5) we construct policy based on the estimated value function.

\begin{algorithm}[!ht]\caption{Sublinear LSVI}\label{alg:sublinear_LSVI}
\begin{algorithmic}[1]
\State {\bf data structure} \textsc{LSH} \Comment{Theorem~\ref{thm:maxip_lsh}}
    \State \hspace{4mm} \textsc{Init}($S \subset \R^d$, $n \in \mathbb{N}$, $d \in \mathbb{N}$, $c\in [0.5,0.8)$, $\tau \in [0.5,0.8)$)
    \State\Comment{$|S|=n$, $c,\tau$ is the approximate $\maxip$ parameter and $d$ is the dimension of data}
    \State \hspace{4mm} \textsc{Query}($x \in \R^d$)
\State {\bf end data structure}
\State

\Procedure{SublinearLSVI}{${\cal S}_{\core}$, ${\cal A}_{\core}$, $N \in \mathbb{N}$, $H \in \mathbb{N},c_{\textit{\lsh}}\in [0.5,0.8],\tau_{\textit{\lsh}}\in [0.5,0.8]$} 
\State \Comment{${\cal S}_{\core}$ and ${\cal A}_{\core}$ are in Definition~\ref{def:coresets:formal}}
\State {\color{blue}/*Collect Samples*/}
\For{step $h \in [H]$}
\State ${\cal D}_h \leftarrow \emptyset$
\For {$j=1,\cdots,M$} \Comment{For each column in the span matrix defined in Definition~\ref{def:spansets:formal}}
\For{$l=1,\cdots,n$} \Comment{Play $n$ times}
\State Query $(s_j,a_j)$ at step $h$, observe the next state $s_{jl}'$. 
\State \Comment{$s_j,a_j $ defined in Definition~\ref{def:spansets:formal}}
\State ${\cal D}_h \leftarrow {\cal D}_h\cup \{(s_j,a_j,s_{jl}')\}$ \Comment{$|{\cal D}_h|=Mn$}
\EndFor
\EndFor
\EndFor
\State {\color{blue}/*Preprocess data and build a nearest neighbor data structure*/} 
\State \Comment{This step takes $O(S\cdot( A^{1+\rho}+dA))$}
\For{$s\in {\cal S}_{\core}$}
\State $\Phi_s\leftarrow \{\phi(s,a)|\ \forall  a\in {\cal A}_{\core}\}$ 
\State {\bf static} \textsc{LSH} $\textsc{lsh}_s$
\State $\textsc{lsh}_s$.\textsc{Init}($\Phi_s,A,d,c_{\textit{\lsh}},\tau_{\textit{\lsh}}$)
\EndFor
\State {\color{blue}/*Precompute $\Lambda$ matrix*/} \Comment{This step takes $O(Md^2+d^\omega)$}
\State $\Lambda\leftarrow n\sum_{j=1}^{M}  \phi(s_j,a_j)\phi(s_j,a_j)^\top$
\State Compute $\Lambda_h^{-1}$ 

\State {\color{blue}/*Update value function*/} 
\Comment{This step takes $O(H(d^2+Md+Mn+SdA^{\rho}))$}

\For{step $h = H, \ldots, 1$}
\State $\hat{w}_h \leftarrow \Lambda^{-1}\sum_{(\dot{s},\dot{a},\dot{s_l}')\in{\cal D}_h} \phi(\dot{s},\dot{a})\left(r_h(\dot{s},\dot{a})+\hat{V}_{h+1}(\dot{s_l}')\right)$
\For{all $s\in {\cal S}_{\core}$}
\State $a\leftarrow \textsc{lsh}_s.\textsc{Query}(\hat{w}_h)$ 
\State $\hat{V}_h(s) \leftarrow \langle \hat{w}_h , \phi(s,a) \rangle$
\EndFor
\EndFor
\State {\color{blue}/*Construct policy*/}\Comment{This step takes $O(HSdA)$}
\State policy $\hat{\pi} \leftarrow \emptyset$
\For{step $h = 1, \ldots, H$}
\State $\hat{\pi}_h(s) \leftarrow \argmax_{a\in {\cal A}_{\core}} \langle \hat{w}_h, \phi(s,a) \rangle$ for all $s\in {\cal S}_{\core}$
\EndFor
\State \Return $\hat{\pi}$
\EndProcedure
\end{algorithmic}
\end{algorithm}

\subsection{Value Difference}\label{sec:value_diff}

In this section, we provide the tools for regret analysis. The goal of this section is to prove Lemma~\ref{lem:value_diff_lsh}.
\begin{lemma}\label{lem:value_diff_lsh} Let $\mdp({\cal S}, {\cal A}, H, \P, r)$ denotes a linear MDP. Let $V_1^{*}(s)$ be the optimal value function defined in Definition~\ref{def:q_learning:formal}. Let $\hat{V}_1(s)$ be the estimated value function defined in Definition~\ref{def:q_learning:formal}. We show that via Algorithm~\ref{alg:sublinear_LSVI}, the difference $V_1^{*}(s)-\hat{V}_1(s)$ is upper bounded by:

\begin{align}
    V_1^{*}(s)-\hat{V}_1(s) \leq  \E_{\pi^*}\Big[\sum_{h=1}^{H}[(\P_h-\hat{\P}_h)\hat{V}_{h+1}](s_h,a_h)|s_1=s\Big]+ \frac{1-c}{2}\cdot H(H+1)
\end{align}
where $c$ is the parameter for $\maxip$.
\end{lemma}
\begin{proof}

We start with lower bounding $\hat{V}_h(s)$ as
\begin{align}\label{eq:thm_bound_vh}
    \hat{V}_h(s)
    \geq&~ c\cdot \max_{a\in {\cal A}_{core}} \langle \hat{w}_h, \phi(s,a)\rangle\notag\\
    =&~ c \max_{a\in {\cal A}} \hat{Q}_{h}(s,a)
\end{align}
where the first step follows from Theorem~\ref{thm:maxip_lsh}, the second step follows from the definition of $\hat{Q}_{h}(s,a)$ in Definition~\ref{def:q_learning:formal} and the definition of convex hull.

Next, we upper bound $V_h^{*}(s)-\hat{V}_h(s)$ as
\begin{align}\label{eq:lm1_v_diff_lsh}
 V_h^{*}(s)-\hat{V}_h(s)
 = & ~ \max_{a\in {\cal A}} Q_{h}^*(s,a)-\hat{V}_h(s) \notag\\
 \leq & ~ \max_{a\in {\cal A}} Q_{h}^*(s,a)-c\max_{a\in {\cal A}} \hat{Q}_{h}(s,a) \notag\\
 \leq & ~ Q_{h}^*(s,\pi^*(s))- c\max_{a\in {\cal A}} \hat{Q}_{h}(s,a)
 \notag\\
 \leq & ~ Q_{h}^*(s,\pi^*(s))- c\hat{Q}_{h}(s,\pi^*(s))\notag\\
 =&~c\Big(Q_{h}^*(s,\pi^*(s))-\hat{Q}_{h}(s,\pi^*(s))\Big) +(1-c) Q_{h}^*(s,\pi^*(s))\notag \\
 \leq&~c\Big(Q_{h}^*(s,\pi^*(s))-\hat{Q}_{h}(s,\pi^*(s))\Big) +(1-c) (H+1-h)\notag\\
   \leq&~\Big(Q_{h}^*(s,\pi^*(s))-\hat{Q}_{h}(s,\pi^*(s))\Big) +(1-c) (H+1-h)
\end{align}
where the first step follows from $V_h^{*}(s)=\max_{a\in {\cal A}} Q_{h}^*(s,a)$, the second step follows from Eq.~\eqref{eq:thm_bound_vh}, the third step follows from $\max_{a\in {\cal A}} Q_{h}^*(s,a)=Q_{h}^*(s,\pi^*(s))$ and the forth step follows from\\ $\max_{a\in{\cal A}} \hat{Q}_{h}(s,a)\geq \hat{Q}_{h}(s,\pi^*(s))$, the fifth step is an reorganization, the sixth step follows the upper bound for $Q^*_h$ in Definition~\ref{def:q_learning:formal}, the seventh step follows from $c\in (0,1)$ and $c$ is close to $1$.

Next, we can write the difference $Q^{*}_{h}(s, a)-\hat{Q}_{h}(s, a)$ as,
\begin{align}\label{eq:lm1_q_diff_lsh}
    Q^{*}_{h}(s, a)-\hat{Q}_{h}(s, a)
    =&~[r_h + \P_h V^{*}_{h+1}](s, a)-[r_h + \hat{\P}_h \hat{V}_{h+1}](s, a)\notag\\
    =&~ [\P_h V^{*}_{h+1}](s,a)-[\hat{\P}_h\hat{V}_{h+1}](s,a)\notag\\
    =&~ [\P_h V^{*}_{h+1}](s,a)-[\P_h \hat{V}_{h+1}](s,a)+[\P_h \hat{V}_{h+1}](s,a)-[\hat{\P}_h\hat{V}_{h+1}](s,a)\notag\\
    =&~[\P_h(V^*_{h+1}-\hat{V}_{h+1})](s,a)+[(\P_h-\hat{\P}_h)\hat{V}_{h+1}](s,a)
\end{align}
where the first step follows from the definition of $Q_h(s,a)$ in Definition~\ref{def:q_learning:formal}, the second step follows from eliminating the common term $r_h(s,a)$, the third step follows from inserting an additional term $[\P_h \hat{V}_{h+1}](s,a)$, and the last step is a reorganization.

Combining Eq.~\eqref{eq:lm1_v_diff_lsh} and Eq.~\eqref{eq:lm1_q_diff_lsh}, we have
\begin{align*}
     & ~ V_h^{*}(s)-\hat{V}_h(s) \\
     \leq&~ \Big( Q_{h}^*(s,\pi^*(s))- \hat{Q}_{h}(s,\pi^*(s))\Big)+(1-c) (H+1-h)\\
     =&~[\P_h(V^*_{h+1}-\hat{V}_{h+1})](s,\pi^*(s))+[(\P_h-\hat{\P}_h)\hat{V}_{h+1}](s,\pi^*(s))+(1-c) (H+1-h)\\
     =&~ \E_{\pi^*}\Big[(V^*_{h+1}-\hat{V}_{h+1})(s_{h+1})~\Big|~s_h=s\Big]+ \E_{\pi^*}\Big[[(\P_h-\hat{\P}_h)\hat{V}_{h+1}](s_h,a_h)~\Big|~s_h=s\Big] \\ 
     & ~ +(1-c) (H+1-h)\\
     =&~ (V^*_{h+1}-\hat{V}_{h+1})+ \E_{\pi^*}\Big[[(\P_h-\hat{\P}_h)\hat{V}_{h+1}](s_h,a_h)~\Big|~s_h=s\Big]\\
     & ~ +(1-c) (H+1-h)
\end{align*}
 where the first step follows the Eq.~\eqref{eq:lm1_v_diff_lsh}, the second step follows the Eq.~\eqref{eq:lm1_q_diff_lsh}, the third step rewrites both terms into an expectation over $\pi^*$, and the last step follows the definition of $ V^*_{h+1}$ and $\hat{V}_{h+1}$. 

Using induction from $1$ to $H$, we have
\begin{align*}
    V_1^{*}(s)-\hat{V}_1(s)
    \leq&~ \E_{\pi^*}\Big[\sum_{h=1}^{H}[(\P_h-\hat{\P}_h)\hat{V}_{h+1}](s_h,a_h) ~\Big|~s_1=s\Big]+(1-c) \sum_{h=1}^{H}(H+1-h)\\
    =&~ \E_{\pi^*}\Big[\sum_{h=1}^{H}[(\P_h-\hat{\P}_h)\hat{V}_{h+1}](s_h,a_h) ~\Big|~s_1=s\Big]+ \frac{1-c}{2}\cdot H(H+1)
\end{align*}
where the second step is a reorganization.
\end{proof}

\subsection{Regret Analysis}\label{sec:lsvi_regret}
The goal of this section is to prove Theorem~\ref{thm:convergence_SLSVI:formal}.
\begin{theorem}[Convergence Result of Sublinear Least-Squares Value Iteration
 (Sublinear LSVI), a formal version of Theorem~\ref{thm:convergence_SLSVI:informal}]\label{thm:convergence_SLSVI:formal}
Given a linear MDP with form $\mdp({\cal S}, {\cal A}, H, \P, r)$ with core sets ${\cal S}_{\core}$, ${\cal A}_{\core}$ defined in Definition~\ref{def:coresets:formal}, if we chose $n=O( C_0^2 \cdot \epsilon^{-2} L^2 H^4 \iota)$, where $\iota=\log ( {Hd} / {p} )$ and $C_0$ is a constant,  the Sublinear LSVI (Algorithm~\ref{alg:sublinear_LSVI}) with approximate $\maxip$ parameter $c= 1- \Theta( L\cdot \sqrt{ \iota / n} )$ has regret at most $O(LH^2\sqrt{\iota/n})$ with probability at least $1-p$.
\end{theorem}
\begin{proof}

We have two definitions for $\hat{Q}_{h}(s, a)$. The first definition is given by  Definition~\ref{def:mdp}, it says
\begin{align}\label{eq:thm_q_def1_lsh}
    \hat{Q}_{h}(s, a)
     = & ~ r_h(s,a)+[\hat{\P}_h\cdot  \hat{V}_{h+1}](s,a).
\end{align}

The second definition is given by Definition~\ref{def:linear_mdp}, it says
\begin{align}
    \hat{Q}_{h}(s, a)= & ~ \phi(s,a)^\top \hat{w}_h.
\end{align}

Given the second definition, our goal is to derive $\hat{\P}_h$.

To do this, we write $\hat{Q}_{h}(s, a)$ as
\begin{align}\label{eq:lm_q_lsh}
    & ~ \hat{Q}_{h}(s, a) \notag \\
    = & ~ \phi(s,a)^\top \hat{w}_h\notag\\
    = & ~ \phi(s,a)^\top\Lambda^{-1}\sum_{(\dot{s},\dot{a},\dot{s_l}')\in{\cal D}_h} \phi(\dot{s},\dot{a})\left(r_h(\dot{s},\dot{a})+\hat{V}_{h+1}(\dot{s_l}')\right)\notag \\
    = & ~ \phi(s,a)^\top\Lambda^{-1}\sum_{(\dot{s},\dot{a},\dot{s_l}')\in{\cal D}_h} \phi(\dot{s},\dot{a})\left(\phi(\dot{s},\dot{a})^\top\theta_h+\hat{V}_{h+1}(\dot{s_l}')\right)\notag \\
    = & ~ \phi(s,a)^\top\Lambda^{-1}\sum_{(\dot{s},\dot{a},\dot{s_l}')\in {\cal D}_h} \phi(\dot{s},\dot{a})\phi(\dot{s},\dot{a})^\top\theta_h+\phi(s,a)^\top\Lambda^{-1}\sum_{(\dot{s},\dot{a},\dot{s_l}')\in{\cal D}_h} \phi(\dot{s},\dot{a})\hat{V}_{h+1}(\dot{s_l}')\notag\\
    = & ~ \phi(s,a)^\top\theta_h+\phi(s,a)^\top\Lambda^{-1}\sum_{(\dot{s},\dot{a},\dot{s_l}')\in{\cal D}_h} \phi(\dot{s},\dot{a})\hat{V}_{h+1}(\dot{s_l}')\notag\\
    = & ~ \phi(s,a)^\top\theta_h+\int\Big(\phi(s,a)^\top\Lambda^{-1}\sum_{(\dot{s},\dot{a},\dot{s_l}')\in {\cal D}_h} \phi(\dot{s},\dot{a})\delta(s',\dot{s_l}')\Big)\hat{V}_{h+1}(s') \d s'\notag\\
    = & ~ r_h(s,a)+\int\Big(\phi(s,a)^\top\Lambda^{-1}\sum_{(\dot{s},\dot{a},\dot{s_l}')\in {\cal D}_h} \phi(\dot{s},\dot{a})\delta(s',\dot{s_l}')\Big)\hat{V}_{h+1}(s') \d s'
\end{align}
where the first step follows the definition of $\hat{Q}_{h}(s, a)$ in Definition~\ref{def:q_learning:formal}, the second step follows the definition of $\hat{w}_h$ in Algorithm~\ref{alg:sublinearLSVI-UCB}, the third step follows the definition of reward $r_h$ in Definition~\ref{def:linear_mdp}, the forth step is an reorganization, the fifth step follows from $\Lambda^{-1}\sum_{(\dot{s},\dot{a},\dot{s_l}')\in {\cal D}_h} \phi(\dot{s},\dot{a})\phi(\dot{s},\dot{a})^\top=\I_d$, the sixth step rewrites the second term in a integral format, where $\delta(x,y)$ is a Dirichlet function, the last step follows the definition of reward $r_h$ in Definition~\ref{def:linear_mdp}.

By comparing Eq.~\eqref{eq:lm_q_lsh} with Eq.~\eqref{eq:thm_q_def1_lsh}, we should define $\hat{\P}_h(s'|s,a)$ as
\begin{align}\label{eq:thm_ph_lsh}
    \hat{\P}_h(s'|s,a)=\phi(s,a)\Lambda^{-1}\sum_{(\dot{s},\dot{a},\dot{s_l}')\in {\cal D}_h} \phi(\dot{s},\dot{a})\delta(s',\dot{s_l}') .
\end{align}
Combining Eq.~\eqref{eq:thm_ph_lsh} with the definition of $[\hat{\P}_h  \hat{V}_{h+1}](s,a)$ in Definition~\ref{def:q_learning:formal}.
\begin{align}\label{eq:thm1_pv_lsh}
    [\hat{\P}_h  \hat{V}_{h+1}](s,a)=\phi(s,a)^\top\Lambda^{-1}\sum_{(\dot{s},\dot{a},\dot{s_l}')\in{\cal D}_h} \phi(\dot{s},\dot{a})\hat{V}_{h+1}(\dot{s_l}').
\end{align}

In the next a few paragraphs, we will explain how to rewrite $[(\P_h-\hat{\P}_h)\hat{V}_{h+1}](s,a)$.
\begin{align}\label{eq:thm_p_error_bound_lsh}
    & ~  [(\P_h-\hat{\P}_h)\hat{V}_{h+1}](s,a) \notag \\
    =&~\phi(s,a)^\top\int \hat{V}_{h+1}(s') \d \mu(s')-[\hat{\P}_h\hat{V}_{h+1}](s,a)\notag\\
    =&~\phi(s,a)^\top\int \hat{V}_{h+1}(s') \d \mu(s')-\phi(s,a)^\top\Lambda^{-1}\sum_{(\dot{s},\dot{a},\dot{s_l}')\in{\cal D}_h}\phi(\dot{s},\dot{a})\hat{V}_{h+1}(\dot{s_l}')\notag\\
    =&~\phi(s,a)^\top\Lambda^{-1}\sum_{(\dot{s},\dot{a},\dot{s_l}')\in{\cal D}_h}\phi(\dot{s},\dot{a})\phi(\dot{s},\dot{a})^\top\int \hat{V}_{h+1}(s') \d \mu(s')\notag\\
    &~-\phi(s,a)^\top\Lambda^{-1}\sum_{(\dot{s},\dot{a},\dot{s_l}')\in{\cal D}_h}\phi(\dot{s},\dot{a})\hat{V}_{h+1}(\dot{s_l}')\notag\\
    =&~\phi(s,a)^\top\Lambda^{-1}\sum_{(\dot{s},\dot{a},\dot{s_l}')\in{\cal D}_h}\phi(\dot{s},\dot{a})\Big(\phi(\dot{s},\dot{a})^\top\int \hat{V}_{h+1}(s') \d \mu(s')-\hat{V}_{h+1}(\dot{s_l}')\Big)\notag\\
    =&~\phi(s,a)^\top\Lambda^{-1}\sum_{(\dot{s},\dot{a},\dot{s_l}')\in{\cal D}_h}\phi(\dot{s},\dot{a})\Big(\int \hat{V}_{h+1}(s')\phi(\dot{s},\dot{a})^\top \d \mu(s')-\hat{V}_{h+1}(\dot{s_l}')\Big)\notag\\
    =&~\phi(s,a)^\top\Lambda^{-1}\sum_{(\dot{s},\dot{a},\dot{s_l}')\in{\cal D}_h}\phi(\dot{s},\dot{a})\Big(\int \hat{V}_{h+1}(s')\P_h[s'|\dot{s},\dot{a}] \d s'-\hat{V}_{h+1}(\dot{s_l}')\Big)\notag\\
    =&~\phi(s,a)^\top\Lambda^{-1}\sum_{(\dot{s},\dot{a},\dot{s_l}')\in{\cal D}_h}\phi(\dot{s},\dot{a})\Big(\E[ \hat{V}_{h+1}(s')|\dot{s},\dot{a} ] - \hat{V}_{h+1}(s_i')\Big)
\end{align}
where the first step follows the definition of  $\P_h[s'|s_i,a_i]=\phi(s_i,a_i)\mu_h(s')$, the second step follows Eq.~\eqref{eq:thm1_pv_lsh}, the third step adds the $\Lambda^{-1}\sum_{(\dot{s},\dot{a},\dot{s_l}')\in {\cal D}_h} \phi(\dot{s},\dot{a})\phi(\dot{s},\dot{a})^\top=\I_d$  to the left term, the forth and fifth steps are reorganizations, the sixth step follows the definition of $\P_h$ in Definition~\ref{def:linear_mdp}, the last step follows the definition of expectation.

Next, we rewrite $ \sum_{(\dot{s},\dot{a},\dot{s_l}')\in{\cal D}_h}\phi(\dot{s},\dot{a})$ as
\begin{align}~\label{eq:thm_phi_sum_lsh}
    \sum_{(\dot{s},\dot{a},\dot{s_l}')\in{\cal D}_h}\phi(\dot{s},\dot{a})
    =~n\sum_{j=1}^{M}\phi_(s_j,a_j)
    =~n\sum_{j=1}^{M}\phi_j
\end{align}
where the first steps follows from Algorithm~\ref{alg:sublinear_LSVI} that for each 
$\phi(s_j,a_j)$, we query it $n$ times and put all $\{(s_j,a_j,s_{j1}'),\cdots,(s_j,a_j,s_{jn}')\}$ in ${\cal D}_h$, the second step follows by $\phi_j=\phi(s_j,a_j)$ in Definition~\ref{def:coresets:formal}.

Next, we rewrite $\Lambda$ as
\begin{align}~\label{eq:thm_Lambda_lsh}
    \Lambda
    =&~\sum_{(\dot{s},\dot{a},\dot{s_l}')\in{\cal D}_h}\phi(\dot{s},\dot{a})\phi(\dot{s},\dot{a})^\top\notag\\
    =&~n\sum_{j=1}^{M}\phi_(s_j,a_j)\phi_(s_j,a_j)^\top\notag\\
    =&~n\Phi\Phi^\top
\end{align}
where the first steps follows by the definition of $\Lambda$ in Algorithm~\ref{alg:sublinear_LSVI},the second steps follows from Algorithm~\ref{alg:sublinear_LSVI} that for each 
$\phi(s_j,a_j)$, we query it $n$ times and put all $\{(s_j,a_j,s_{j1}'),\cdots,(s_j,a_j,s_{jn}')\}$ in ${\cal D}_h$, the third step follows from the definition of $\Phi$ in Definition~\ref{def:coresets:formal}.

Combining Eq.~\eqref{eq:thm_Lambda_lsh} with Eq.~\eqref{eq:thm_p_error_bound_lsh}, we get
\begin{align}~\label{eq:thm_p_error_bound_replace_lambda_lsh}
    [(\P_h-\hat{\P}_h)\hat{V}_{h+1}](s,a)
    =&~\phi(s,a)^\top (n\Phi\Phi^\top)^{-1}\sum_{(\dot{s},\dot{a},\dot{s_l}')\in{\cal D}_h}\phi(\dot{s},\dot{a})\Big(\E[ \hat{V}_{h+1}(s')|\dot{s},\dot{a} ] - \hat{V}_{h+1}(s_i')\Big)
\end{align}

Next, we further bound $[(\P_h-\hat{\P}_h)\hat{V}_{h+1}](s,a)$ as:

\begin{align*}
    & ~ [(\P_h-\hat{\P}_h)\hat{V}_{h+1}](s,a) \\
    =&~\phi(s,a)^\top (n\Phi\Phi^\top)^{-1}\sum_{(\dot{s},\dot{a},\dot{s_l}')\in{\cal D}_h}\phi(\dot{s},\dot{a})\Big(\E[ \hat{V}_{h+1}(s')|\dot{s},\dot{a} ] - \hat{V}_{h+1}(s_i')\Big)\\
    =&~\phi(s,a)^\top (n\Phi\Phi^\top)^{-1} n\sum_{j=1}^{M}\phi(s_j,a_j)\sum_{l=1}^{n}\Big(\E[ \hat{V}_{h+1}(s')|s_j,a_j] - \hat{V}_{h+1}(s_{jl}')\Big)\\
    =&~\phi(s,a)^\top (\Phi\Phi^\top)^{-1}  \sum_{j=1}^{M}\phi(s_j,a_j) \Big(\E[ \hat{V}_{h+1}(s')|s_j,a_j] - \frac{1}{n}\sum_{l=1}^{n} \hat{V}_{h+1}(s_{jl}')\Big)\\
    =&~\phi(s,a)^\top (\Phi\Phi^\top)^{-1}  \sum_{j=1}^{M}\phi_j \Big(\E[ \hat{V}_{h+1}(s')|s_j,a_j] - \frac{1}{n}\sum_{l=1}^{n} \hat{V}_{h+1}(s_{jl}')\Big)\\
\end{align*}
where the first step follows from Eq.~\eqref{eq:thm_p_error_bound_replace_lambda_lsh}, the second steps follows from Algorithm~\ref{alg:sublinear_LSVI} that for each 
$\phi(s_j,a_j)$, we query it $n$ times and put all $\{(s_j,a_j,s_{j1}'),\cdots,(s_j,a_j,s_{jn}')\}$ in ${\cal D}_h$, the third step is an reorganization, the last step follows the definition of $\phi_j$ in Definition~\ref{def:coresets:formal}.

For each $j \in [M]$, we define random variable 
\begin{align*}
z_j := \E[ \hat{V}_{h+1}(s')|\phi_j ] - \frac{1}{n}\sum_{l=1}^{n} \hat{V}_{h+1}(s_{jl}')
\end{align*}

By Hoefding Inequality in Lemma~\ref{lem:hoeffding}, we can show 

\begin{align*}
|z_j| \leq C_0\cdot H \cdot \sqrt{ {\iota} / {n} }
\end{align*}

For convenient, we define vector $z \in \R^M$ to be $z:=[z_1,\cdots,z_{M}]$.

Now, we can upper bound $[(\P_h-\hat{\P}_h)\hat{V}_{h+1}](s,a)$ as follows:
\begin{align}\label{eq:thm1_pp_h_lsh}
    [(\P_h-\hat{\P}_h)\hat{V}_{h+1}](s,a)
    =&~\phi(s,a)^\top(\Phi\Phi^\top)^{-1}\Phi z\notag\\
    =&~\phi(s,a)^\top(\Phi^{\dagger})^\top z\notag\\
    =&~\Big(\Phi^{\dagger}\phi(s,a)\Big)^\top z\notag\\
    \leq &~\|\Phi^{\dagger}\phi(s,a)\|_1 \cdot\| z \|_{\infty}\notag\\
    \leq&~ L\cdot C_0\cdot H \cdot \sqrt{{\iota}/{n}}
\end{align}
where the first step follows the $\sum_{j=1}^{M}\phi_jz_j=\Phi z$, the second step is an reorganization, the third step follows the holders inequality, the last step uses the bound for $\|\Phi^{-1}\phi(s,a)\|_1$ in Definition~\ref{def:coresets:formal} and $\|z_j\|_2$.

Combining Eq.~\eqref{eq:thm1_pp_h_lsh} with Lemma~\ref{lem:value_diff_lsh}, we could upper bound $V_1^{*}(s)-\hat{V}_1(s)$
\begin{align*}
    V_1^{*}(s)-\hat{V}_1(s)
    \leq &~ 
    \E_{\pi^*}\Big[\sum_{h=1}^{H}[(\P_h-\hat{\P}_h)\hat{V}_{h+1}](s_h,a_h)|s_1=s\Big]+\frac{1-c}{2}\cdot H(H+1)\\
    \leq&~H \cdot L\cdot C_0\cdot H \cdot \sqrt{ \iota / n}+\frac{1-c}{2}\cdot H(H+1) \\
    =&~ L\cdot C_0\cdot H^2 \cdot \sqrt{ \iota / n} +\frac{1-c}{2}\cdot H(H+1)\\
     \leq &~ L\cdot C_0\cdot H^2 \cdot \sqrt{ \iota / n} +(1-c)H^2\\
    \leq &~ 2C_0LH^2\sqrt{\iota/n}\\
    \leq & ~ \epsilon
\end{align*}
where the first step follows from Lemma~\ref{lem:value_diff_lsh}, the second step  follows the upper bound of $[(\P_h-\hat{\P}_h)\hat{V}_{h+1}](s,a)$ in Eq.~\ref{eq:thm1_pp_h_lsh}, the third step is an reorganization, the forth step follows from $H\geq 1$ so that $H^2\geq H$, the fifth step follows from $1-c= C_0 L\sqrt{\iota/n}$, the sixth step follows from $n= O( C_0^2 \cdot \epsilon^{-2} L^2 H^4 \iota)$.

\end{proof}

\begin{algorithm}[h]
\caption{LSVI~\cite{bb96}}\label{alg:LSVI} 
\begin{algorithmic}[1]
\Procedure{LSVI}{${\cal S}$, ${\cal A}$, $N \in \mathbb{N}$, $H \in \mathbb{N}$}  \Comment{${\cal S}$ and ${\cal A}$ are in Definition~\ref{def:linear_mdp}}
\State {\color{blue}/*Collect Samples*/}
\For{$h \in [H]$}
\State ${\cal D}_h \leftarrow \emptyset$
\For {$j=1,\cdots,M$} \Comment{For each element in the span set defined in Definition~\ref{def:spansets:formal}}
\For{$l=1,\cdots,n$} \Comment{Play $n$ times}
\State Query $(s_j,a_j)$ at step $h$, observe the next state $s_{jl}'$. 
\State \Comment{$s_j,a_j $ defined in Definition~\ref{def:spansets:formal}}
\State ${\cal D}_h \leftarrow {\cal D}_h\cup \{(s_j,a_j,s_{jl}')\}$ \Comment{$|{\cal D}_h|=Mn$}
\EndFor
\EndFor
\EndFor
\State {\color{blue}/*Precompute $\Lambda$ matrix*/} \Comment{This step takes $O(Md^2 + d^{\omega})$}
\State $\Lambda \leftarrow n \sum_{j=1}^{M}  \phi(s_j,a_j)\phi(s_j,a_j)^\top$\Comment{$\Lambda\in \R^{d\times d}$}
\State Compute $\Lambda^{-1}$

\State {\color{blue}/*Update value function*/} \Comment{This step takes $O(H(d^2+Md+Mn+SAd))$}
\For{$h = H, \ldots, 1$}
\State $\hat{w}_h \leftarrow \Lambda^{-1}\sum_{(\dot{s},\dot{a},\dot{s_l}')\in{\cal D}_h} \phi(\dot{s},\dot{a})\left(r_h(\dot{s},\dot{a})+\hat{V}_{h+1}(\dot{s_l}')\right)$
\For{all $s\in {\cal S}_{\core}$}
\State $\hat{V}_h(s) \leftarrow \max_{a\in {\cal A}_{\core}} \langle \hat{w}_h , \phi(s,a) \rangle$
\EndFor
\EndFor

\State {\color{blue}/*Construct policy*/} \Comment{This step takes $O(HSAd)$}
\State policy $\hat{\pi}\leftarrow \emptyset$
\For{$h = 1, \ldots, H$}
\State $\hat{\pi}_h(s) \leftarrow \argmax_{a\in {\cal A}_{\core}} \langle \hat{w}_h, \phi(s,a) \rangle$ for all $s\in {\cal S}$
\EndFor
\State \Return $\hat{\pi}$
\EndProcedure
\end{algorithmic}
\end{algorithm}
\subsection{Running Time Analysis}\label{sec:runtime}

\begin{lemma}
The running time of pre-computing $\Lambda^{-1}$ takes
\begin{align*}
    O(M d^2 + d^{\omega})
\end{align*}
\end{lemma}
\begin{proof}
It takes $O(Md^2)$ to sum up every $\phi(s_j,a_j)\phi(s_j,a_j)^\top$. It takes $O(d)$ constant to multiply the sum results by $n$. Computing the inverse matrix of $\Lambda$ takes $O(d^\omega)$. Combining the complexity together, we obtain the pre-computing complexity $O(M d^2 + d^{\omega})$.
\end{proof}

\begin{lemma}\label{lem:lsvi_runtime_vi_time}
The running time of updating value takes
\begin{align*}
    O( H \cdot (d^2 + M d + M n + SdA^\rho ) )
\end{align*}

Further more,
\begin{itemize}
    \item If initialize the {\lsh} data-structure using Theorem~\ref{thm:ar15:formal}, $\rho=1- \frac{1}{4} C_0L\sqrt{\iota/n}$.
    \item If initialize the {\lsh} data-structure using Theorem~\ref{thm:ar17:formal}, $\rho=1- \frac{1}{8} C_0^2L^2\iota/n$.
\end{itemize}
\end{lemma}
\begin{proof}
We can rewrite $\wh{w}_h$ as follows:
\begin{align*}
    \hat{w}_h 
    =&~ \Lambda^{-1}\sum_{(\dot{s},\dot{a},\dot{s_l}')\in{\cal D}_h} \phi(\dot{s},\dot{a})\left(r_h(\dot{s},\dot{a})+\hat{V}_{h+1}(\dot{s_l}')\right)\\
    =&~ \Lambda^{-1}n\sum_{j=1}^{M} \phi(s_j,a_j)(r_h(s_j,a_j)+\frac{1}{n}\sum_{l=1}^{n}\hat{V}_{h+1}(s_{jl}'))
\end{align*}
where the second step follows the definition of ${\cal D}_h$.

For each of the $H$ step,
\begin{itemize}
\item It takes $O(SdA^\rho )$ to compute $\hat{V}_h(s_{jl}')$ for each state $s_j\in {\cal S}_{\core}$. If we initialize the {\lsh} data-structure using Theorem~\ref{thm:ar15:formal}, we determine $\rho=1- \frac{1}{4} C_0L\sqrt{\iota/n}$ using Lemma~\ref{lem:ar15_rho_bound}. If we initialize the {\lsh} data-structure using Theorem~\ref{thm:ar17:formal}, we determine $\rho=1- \frac{1}{8} C_0^2L^2\iota/n$ using Lemma~\ref{lem:ar17_rho_bound}. 
\item It takes $O(Mn)$ to compute $r_h(s_j,a_j)+\frac{1}{n}\sum_{l=1}^{n}\hat{V}_{h+1}(s_{jl}')$ for the total $n$ number of $s_{jl}'$ observed by $(s_j,a_j)$.
\item It takes $O(Md)$ to sum up the $M$ dimensional vector $\phi(s_j,a_j)(r_h(s_j,a_j)+\frac{1}{n}\sum_{l=1}^{n}\hat{V}_{h+1}(s_{jl}'))$.
\item It takes $O(d^2)$ to multiply $\Lambda$ with the sum of vectors. 
\item All other operations take $O(d)$.
\end{itemize}

Combining the complexity together and multiply by $H$ steps, we finish the proof.
\end{proof}

\begin{lemma}
The running time of constructing policy takes
\begin{align*}
    O(HSdA)
\end{align*}
\end{lemma}

\begin{proof}
For each step, it takes $O(SdA)$ to find the optimal action. Thus, it takes $O(HSdA)$ for inference.
\end{proof}

\subsection{Comparison}\label{sec:lsvi_compare}

In this section, we show the comparison between our Sublinear LSVI with LSVI~\cite{bb96}.

We start with presenting the LSVI algorithm in Algorithm~\ref{alg:LSVI}.

Next, we show the comparison results in Table~\ref{tab:LSVI_compare}.
\begin{table}[H]
    \centering
    \begin{tabular}{|l|l|l|l|} \hline
        {\bf Algorithm} & {\bf Preprocess} & {\bf \#Value Iteration} &  {\bf Regret} \\ \hline
        Ours & $O(SdA^{1+\rho_1})$ & $O(HSdA^{\rho_1})$ &  $O(C_0LH^2\sqrt{\iota/n})$ \\ \hline
        Ours & $O(SdA^{1+o(1)})$ & $O(HSdA^{\rho_2})$ &  $O(C_0LH^2\sqrt{\iota/n})$ \\ \hline
         LSVI & 0 & $O(HSdA)$ & $O(C_0LH^2\sqrt{\iota/n})$\\ \hline 
    \end{tabular}
    \caption{Comparison between Our Sublinear LSVI with LSVI. Let $S$ and $A$ denotes the cardinality of ${\cal S}_{core}$ and ${\cal A}_{core}$. Let $d$ denotes the dimension of $\phi(s,a)$. Let $H$ be the number of steps played in each episode.  Let $n$ denotes the quantity of times played for each pair of core state-action. Let $L$ denotes the constant in Definition~\ref{def:spansets:formal}. Let $\iota=\log(Hd/p)$ and $p$ is the failure probability. Let $\rho_1= 1- \frac{1}{4} C_0L\sqrt{\iota/n} $ be the parameter of data structures in Theorem~\ref{thm:ar15:formal} and  $\rho_2=1- \frac{1}{8} C_0^2L^2\iota/n$  be the parameter of data structure Theorem~\ref{thm:ar17:formal}. This table is a detailed version of corresponding part of Table~\ref{tab:main_compare}.}
    \label{tab:LSVI_compare}
\end{table}
\newpage
\newpage
\section{Sublinear Least-Squares Value Iteration with UCB}\label{sec:SLSVI_UCB}

This section extend the Sublinear LSVI with UCB exploration.
\begin{itemize}
    \item In Section~\ref{sec:ucb_alg}, we present the Sublinear LSVI-UCB algorithm.
    \item In Section~\ref{sec:ucb_not}, we define several simplified notations for the convenience of proof.
    \item In Section~\ref{sec:ucb_w_bound}, we provide the upper bound of weight estimated by Sublinear LSVI-UCB. 
    \item In Section~\ref{sec:ucb_net}, we introduce a modified version of net argument for Sublinear LSVI-UCB.
    \item In Section~\ref{sec:bound_fluc}, we upper bound the fluctuation on the value function when performing Sublinear LSVI-UCB Algorithm. 
    \item In Section~\ref{sec:ucb_bound_q_diff}, we provide the upper bound on the difference between the estimated Q function and the actual Q function.
    \item In Section~\ref{sec:ucb_induct}, we given the upper bound on the difference between the estimated Q function and the actual Q function at the first step using induction.
    \item In Section~\ref{sec:ucb_recur}, we introduce the recursion formula for the regret analysis.
    \item In Section~\ref{sec:ucb_regret}, we formally provide the regret analysis of LSVI-UCB.
    \item In Section~\ref{sec:ucb_runtime}, we analyze the runtime Sublinear LSVI-UCB by calculating the time complexity for each block.
    \item In Section~\ref{sec:ucb_compare}, we compare Sublinear LSVI-UCB with LSVI-UCB~\cite{jywj20} in terms of regret and value iteration complexity.
\end{itemize}

 In the following sections we show how to tackle the problem and provide our Sublinear LSVI-UCB. Moreover, we provide the regret analysis of our Sublinear LSVI-UCB.
 
\subsection{Algorithm}\label{sec:ucb_alg}
In LSVI-UCB~\cite{jywj20} with large action space, the runtime in each value iteration step is dominated by by computing the estimated value function as below:

\begin{align}\label{eq:LSVI-UCB_problem_original:formal}
    \hat{V}_h(s^{\tau}_{h+1}) = \max_{a\in {\cal A}}~\min\{\langle w^k_h,\phi(s^{\tau}_{h+1}, a)\rangle  + \beta \cdot \|\phi(s^{\tau}_{h+1},a)\|_{\Lambda_h^{-1}}, H\}
\end{align} 
where $w^k_h$ is computed by solving the least-squares problem and $\phi(s^{\tau}_{h+1},a)$ is the embedding for a pair of state-action. The complexity for Eq.~\eqref{eq:LSVI-UCB_problem_original:formal} is $O(d^2A)$

The key challenge of Sublinear LSVI-UCB here is that Eq.~\eqref{eq:LSVI-UCB_problem_original:informal} cannot be formulated as a $\maxip$ problem.

To handle this, we demonstrate how to develop Sublinear LSVI-UCB algorithm. We start with bounding the Q function in \cite{jywj20} as

\begin{lemma}\label{lem:bound_q_func_chi}
We show that

\begin{align*}
    \min\{  \|\phi(s^{\tau}_{h+1},a)\|_{\beta^2 \Lambda_h^{-1}+w_h^k (w_h^k)^\top}, H\} \leq Q_h(s^{\tau}_{h+1}, a) \leq \min\{  \|\phi(s^{\tau}_{h+1},a)\|_{2\beta^2 \Lambda_h^{-1}+2w_h^k (w_h^k)^\top}, H\}.
\end{align*}

\end{lemma}
\begin{proof}
We start with rewriting $Q_h(s^{\tau}_{h+1}, a)$,
\begin{align*}
    Q_h(s^{\tau}_{h+1}, a) = \min\{ w_h^\top\phi(s^{\tau}_{h+1}, a)  + \beta \cdot \|\phi(s^{\tau}_{h+1},a)\|_{\Lambda_h^{-1}}, H\} .
\end{align*}

Next, we show that
\begin{align*}
    w_h^\top\phi(s^{\tau}_{h+1}, a)  + \beta \cdot \|\phi(s^{\tau}_{h+1},a)\|_{\Lambda_h^{-1}} 
    \leq &~ \sqrt{2(w_h^\top\phi(s^{\tau}_{h+1}, a))^2+2\beta^2 \cdot \|\phi(s^{\tau}_{h+1},a)\|_{\Lambda_h^{-1}}^2}\\
     = &~ \|\phi(s^{\tau}_{h+1},a)\|_{2\beta^2 \Lambda_h^{-1}+2w_h^k (w_h^k)^\top}
\end{align*}
where the first step follows from Cauchy-Schwartz inequality, the second step is an reorganization.

Next, we show that
\begin{align*}
    w_h^\top\phi(s^{\tau}_{h+1}, a)  + \beta \cdot \|\phi(s^{\tau}_{h+1},a)\|_{\Lambda_h^{-1}} 
    \geq &~ \sqrt{(w_h^\top\phi(s^{\tau}_{h+1}, a))^2+\beta^2 \cdot \|\phi(s^{\tau}_{h+1},a)\|_{\Lambda_h^{-1}}^2}\\
     = &~ \|\phi(s^{\tau}_{h+1},a)\|_{\beta^2 \Lambda_h^{-1}+w_h^k (w_h^k)^\top}
\end{align*}
where the first step follows from the fact that both $w_h^\top\phi(s^{\tau}_{h+1}, a)$ and $\|\phi(s^{\tau}_{h+1},a)\|_{\Lambda_h^{-1}}$ are non-negative, the second step is an reorganization.

Finally, consider the propriety of $\min$ function, we finish the proof of the lemma.

\end{proof}

\begin{algorithm}[!ht]
\caption{Modified LSVI-UCB}\label{alg:MLSVI-UCB}
\begin{algorithmic}[1]
\For{$k = 1, \ldots, K$}
\State Initialize the state to $s^k_1$.
\For{$h = H, \ldots, 1$}

\State {\color{blue}/*Compute $\Lambda_h^{-1}$*/}\Comment{This step takes $O(Kd^2+d^{\omega})$}
\State $\Lambda_h \leftarrow \sum_{\tau =1}^{k-1}  \phi(s^{\tau}_h,  a^{\tau}_h)\phi(s^{\tau}_h, a^{\tau}_h)^{\top} + \lambda \cdot  \I_d$.  \label{line:Lambda} 
\State Compute $\Lambda_h^{-1}$

\State {\color{blue}/* Value Iteration*/} \Comment{This takes $O(AKd^2)$}
\State $w_h^k \leftarrow \Lambda_h^{-1} \sum_{\tau=1}^{k-1} \phi(s^{\tau}_h,  a^{\tau}_h) \cdot ( r_h(s^{\tau}_h, a^{\tau}_h) + \hat{V}_{h+1} (s^{\tau}_{h+1})) $  

\For{$\tau=1,\cdots, k-1$}
\For{$a\in {\cal A}$}

\State $Q_h(s^{\tau}_{h+1}, a) \leftarrow \min\{  \|\phi(s^{\tau}_{h+1},a)\|_{2\beta^2 \Lambda_h^{-1}+2w_h^k w_h^{k\top}}, H\}$.\label{line:mlsvi-ucb_Q_h}
\EndFor
\State $\hat{V}_h (s^{\tau}_{h}) \leftarrow \max_{a\in {\cal A}} Q_h(s^{\tau}_{h}, a)$\label{line:maxip_q_function}
\State $a^{\tau}_h\leftarrow \arg \max_{a\in {\cal A}} Q_h(s, a) $\label{line:maxip_policy} \Comment{$a^{\tau}_h$ is the maximum value action taken at state $s^{\tau}_{h}$.}

\EndFor

\label{line:ucb}
\EndFor
\State {\color{blue}/* Construct Policy*/}
\For{$h = 1, \ldots, H$} 
\State Given state $s^k_{h}$, take action $a^k_h$, and observe $s^k_{h+1}$. 
\EndFor
\EndFor
\end{algorithmic}
\end{algorithm}

Next, we present a modified version of  LSVI-UCB in Algorithm~\ref{alg:MLSVI-UCB}. The major difference between our modified version of LSVI-UCB and \cite{jywj20} lies in in Line~\ref{line:mlsvi-ucb_Q_h} of Algorithm~\ref{alg:MLSVI-UCB}. Here we choose $Q_h(s^{\tau}_{h+1}, a) \leftarrow \min\{  \|\phi(s^{\tau}_{h+1},a)\|_{2\beta^2 \Lambda_h^{-1}+2w_h^k (w_h^k)^\top}, H\}$, which is the upper bound of  $ \min\{ w_h^\top\phi(s^{\tau}_{h+1}, a)  + \beta \cdot \|\phi(s^{\tau}_{h+1},a)\|_{\Lambda_h^{-1}}, H\}$ according to Lemma~\ref{lem:bound_q_func_chi}. 

Based on Algorithm~\ref{alg:MLSVI-UCB}, we propose our Sublinear LSVI-UCB in Algorithm~\ref{alg:sublinearLSVI-UCB}, which reduce the value iteration complexity to sublinear in actions. Note that to let $\rho$ strict less than $1$, we set $c^2\in [0.5,0.8]$ and $\tau^2\in [0.5,0.8])$ following Lemma~\ref{lem:ar15_rho_bound}.

\begin{algorithm}[!ht]
\caption{Sublinear LSVI-UCB}\label{alg:sublinearLSVI-UCB}
\begin{algorithmic}[1]
\State {\bf data structure} \textsc{MatrixLSH} \Comment{Theorem~\ref{thm:maxmatnorm_lsh}}
    \State \hspace{4mm} \textsc{Init}($S \subset \R^d$, $n \in \mathbb{N}$, $d \in \mathbb{N}$, $c\in (0.72,0.9)$, $\tau \in (0.72,0.9)$)
    \State\Comment{$|S|=n$, $c,\tau$ is the approximate $\maxmatnorm$ parameter and $d$ is the dimension of data}
    \State \hspace{4mm} \textsc{Query}($x \in \R^d$)
\State {\bf end data structure}
\State
\Procedure{SublinearLSVI-UCB}{${\cal S}$, ${\cal A}$, $N \in \mathbb{N}$, $H \in \mathbb{N},c_{\mathsf{MatLSH}} \in (0.72,0.9), \tau_{\mathsf{MatLSH}} \in (0.72,0.9)$}

\State {\color{blue}/*Preprocess $\phi(s,a)$ and build a {\lsh} data structure*/} \Comment{This step takes $O( S \cdot (A^{1+\rho}+d^2A))$}
\For{$s\in {\cal S}$}
\State $\Phi_s\leftarrow \{\phi(s,a)|\ \forall  a\in {\cal A}\}$ 
\State {\bf static} \textsc{MatrixLSH} $\textsc{matlsh}_s$
\State $\textsc{matlsh}_s$.\textsc{Init}($\Phi_s,A,d,c_{\mathsf{MatLSH}}, \tau_{\mathsf{MatLSH}}$)
\EndFor
\State
\For{$k = 1, \ldots, K$}
\State Initialize state to $s^k_1$.
\For{$h = H, \ldots, 1$}

\State {\color{blue}/*Compute $\Lambda_h^{-1}$*/}\Comment{This step takes $O(Kd^2+d^{\omega})$}
\State $\Lambda_h \leftarrow \sum_{\tau =1}^{k-1}  \phi(s^{\tau}_h,  a^{\tau}_h)\phi(s^{\tau}_h, a^{\tau}_h)^{\top} + \lambda \cdot  \I_d$.   

\State Compute $\Lambda_h^{-1}$

\State {\color{blue}/* Value Iteration*/} \Comment{This takes $O(Kd^2A^{\rho})$}
\State $w_h^k \leftarrow \Lambda_h^{-1} \sum_{\tau=1}^{k-1} \phi(s^{\tau}_h,  a^{\tau}_h) \cdot ( r_h(s^{\tau}_h, a^{\tau}_h) + \hat{V}_{h+1} (s^{\tau}_{h+1})) $  

\For{$\tau=1,\cdots, k-1$}

\State $a^{\tau}_h\leftarrow \textsc{matlsh}_s.\textsc{Query}( 2\beta^2\Lambda_h^{-1}+2w_h^kw_h^{k\top} )$ 
\State $\hat{V}_h (s^{\tau}_{h}) \leftarrow   \min\{  \|\phi(s^{\tau}_{h+1},a^{\tau}_h)\|_{2\beta^2 \Lambda_h^{-1}+2w_h^k w_h^{k\top}}, H\}$

\EndFor

\EndFor
\State {\color{blue}/* Construct Policy*/}
\For{step $h = 1, \ldots, H$} 
\State Take action $a^k_h$ at $s^k_{h}$, and observe $s^k_{h+1}$. 
\EndFor
\EndFor
\EndProcedure
\end{algorithmic}
\end{algorithm}

\subsection{Notations for Proof of Convergence}\label{sec:ucb_not}

Next, we start the regret analysis of our Sublinear LSVI-UCB. We first define a series of notations. At episode $k$, we first estimate the weight $w^k_h$ and matrix $\Lambda_h^k$. Next, we use them to estimate Q function $Q_h^k$. Then, using our {\lsh} data structures, we obtain the value function $V_h^k(s)$ following line~\ref{line:maxip_q_function} of Algorithm~\ref{alg:sublinear_LSVI}. We also obtain the corresponding action associated with the value function and form the polity $\pi_k$ following Line~\ref{line:maxip_policy} of Algorithm~\ref{alg:sublinear_LSVI}. We also simplify
$\phi(s^{k}_h, a^{k}_h)$ as $\phi^k_h$.

\subsection{Upper Bound on Weights in Sublinear LSVI-UCB}\label{sec:ucb_w_bound}
In this section, we show how to bound the weights $w^k_h$ in Algorithm \ref{alg:sublinearLSVI-UCB} using Lemma~\ref{lem:wn_estimate}. The weight we would like to bound is different from~\cite{jywj20}. But the bound inequalities is very standard and similar to the proof in \cite{jywj20}.
\begin{lemma}[]\label{lem:wn_estimate}
The weight $w^k_h$ in Algorithm \ref{alg:sublinearLSVI-UCB} at episode $k\in [K] $ and step $h\in [H]$ satisfies:
\begin{align*}
\|w^k_h\|_2 \leq 2H \sqrt{dk/\lambda}.
\end{align*}
\end{lemma}
\begin{proof}

If we perform $v^\top w^k_h$ where $v \in \R^d$ could be any vector in $\R^d$, we could bound $|v^\top w^k_h|$ as
\begin{align*}
|v^\top w^k_h| 
= & ~ \Big|v^\top (\Lambda^k_h)^{-1} \sum_{\tau=1}^{k-1} \phi^\tau_h \Big(r(s^{\tau}_h, a^{\tau}_h) +  \hat{V}_{h+1}(s^{\tau}_{h+1})\Big)\Big|\\
\leq & ~ \Big|v^\top (\Lambda^k_h)^{-1} \sum_{\tau=1}^{k-1} \phi^\tau_h \Big(r(s^{\tau}_h, a^{\tau}_h) + \max_{a\in {\cal A}} Q_{h+1}(s^{\tau}_{h+1}, a)\Big)\Big|\\
\leq & ~  2H\cdot \Big|v^\top (\Lambda^k_h)^{-1} \sum_{\tau=1}^{k-1} \phi^\tau_h \Big| \\
= & ~  2H\cdot \sum_{\tau = 1}^{k-1}  \Big|v^\top (\Lambda^k_h)^{-1} \phi^\tau_h\Big|  \\
\leq & ~ 2H\cdot \Big( \big( \sum_{\tau = 1}^{k-1}  v^\top (\Lambda^k_h)^{-1}v \big)  \cdot \big( \sum_{\tau = 1}^{k-1}  (\phi^\tau_h)^\top (\Lambda^k_h)^{-1}\phi^\tau_h \big) \Big)^{1/2} \\
\leq & ~ 2H \|v\|_2\sqrt{dk/\lambda}, 
\end{align*}
where the first step follows from the definition of $w_h^k$ in Algorithm~\ref{alg:sublinearLSVI-UCB}, the second step follows from the definition of $\hat{V}_{h+1}$ in Algorithm~\ref{alg:sublinearLSVI-UCB}, the third step follows from Definition~\ref{def:linear_mdp} that $r(s, a) +  \hat{V}_{h+1}(s) \leq 2H$ for all $s\in {\cal S}$ and $a\in {\cal A}$, the forth step is a reorganization, the fifth step follows Cauchy–Schwarz inequality, the last step follows from Lemma \ref{lem:lamdba_ineq}.

Next, we rewrite $\|w^k_h\|_2 = \max_{v:\|v\|_2 = 1} |v^\top w^k_h|$, in this way, 

\begin{align*}
    \|w^k_h\|_2 = \max_{v:\|v\|_2 = 1} |v^\top w^k_h|\leq 2H \sqrt{dk/\lambda}
\end{align*}
where the last step follows from $|v^\top w^k_h|\leq 2H \sqrt{dk/\lambda}$.

\end{proof}

\subsection{Our Net Argument}\label{sec:ucb_net}
We present our net argument to support the proof in the this section. We start with defining the covering number of euclidean ball.

\begin{lemma} [] \label{lem:euclid}
    Let ${\cal B}$ denotes a Euclidean ball in $\R^d$. ${\cal B}$ has radius greater than $0$. For any  $\epsilon > 0$, we upper bound the $\epsilon$-covering number of ${\cal B}$ by $(1 + 2 R/ \epsilon )^d$. 
\end{lemma}

This is a standard statement.  We reder readers to \cite{v10} for more details.

Next, we upper bound the covering number of a function ${\cal V}(s) = \min \Big\{\| \phi(s,a) \|_{ \beta^2 \Lambda^{-1}+w w^\top }  , H \Big\}$. The $\mathcal{V}$ we would like to bound is is different from~\cite{jywj20}. But the net argument is very standard and similar to proof in \cite{jywj20}.

\begin{lemma}[Our Net Argument] \label{lem:covering_number_new}
Let $\Lambda \in \R^{d\times d}$ denotes a invertible matrix whose minimum eigenvalue is greater than a constant $\lambda$. Let $w$ denotes a vector such that $\|w\|_2\leq L$. Let $\beta \in [0,B]$. Let $\max_{(s,a)\in {\cal S}\times {\cal A}}\| \phi(s, a) \|_2 \leq 1$. Let ${\cal V}$ denotes a famility of functions such that $V:{\cal S} \rightarrow \R $ for any $V\in {\cal V}$  Let $\mathcal{N}_{\epsilon}$ denotes the $\epsilon$-covering number of $\mathcal{V}$. The $\epsilon$-covering number is defined on distance $\mathrm{dist}(V, V') = \max_{s\in {\cal S}} |V(s) - V'(s)|$. If for any $V\in {\cal V}$, we have the form
\begin{align}\label{eq:ourform}
V(s) = \min \Big\{\| \phi(s,a) \|_{ \beta^2 \Lambda^{-1}+w w^\top }  , H \Big\}
\end{align}

Then we have
\begin{align*}
\log \mathcal{N}_{\epsilon} \le d  \log (1+ 4L / \epsilon ) + d^2 \log \Big( 1 +  8 d^{1/2} B^2  / (\lambda\epsilon^2)  \Big) .
\end{align*}

\end{lemma}

\begin{proof}
For given two arbitrary functions $V_1, V_2 \in \mathcal{V}$, we have
\begin{align} \label{eq:cover_upper_new}
 \mathrm{dist}(V_1, V_2)  
 \leq & ~ \sup_{s, a} ~\Big( \| \phi(s,a) \|_{ \beta_1^2 \Lambda_1^{-1} w_1 w_1^\top }  -  \| \phi(s,a) \|_{ \beta_2^2 \Lambda_2^{-1}+w_2 w_2^\top } \Big)   \notag \\
 \leq & ~ \sup_{\phi:\|\phi\|_2\le 1}  \Big( \| \phi \|_{ \beta_1^2 \Lambda_1^{-1}+w_1 w_1^\top }  -  \| \phi \|_{ \beta_2^2 \Lambda_2^{-1} +w_2 w_2^\top } \Big) \notag \\
\leq & \sup_{\phi:\|\phi\|_2\le 1} \sqrt{  | \phi^{\top} (\beta_1^2 \Lambda_1^{-1}+w_1 w_1^\top -\beta_2^2 \Lambda_2^{-1} -w_2 w_2) \phi    |   }  \notag \\
 \leq & ~ \sup_{\phi:\|\phi\|_2\le 1}   \Big(  (w_1 - w_2) ^{\top} \phi \Big)  +  \sup_{\phi:\|\phi\|_2\le 1} \sqrt{  | \phi^{\top} (\beta_1^2 \Lambda_1^{-1}-\beta_2^2 \Lambda_2^{-1} ) \phi    |   } \notag \\
= & ~ \| w_1 - w_2 \| + \sqrt{\| \beta_1^2 \Lambda_1^{-1} - \beta_2^2 \Lambda_2^{-1}  \|_2} \notag \\
\le & ~ \| w_1 - w_2 \| + \sqrt{\| \beta_1^2 \Lambda_1^{-1}- \beta_2^2 \Lambda_2^{-1} \|_F}, 
\end{align}
where the first step follows the definition of $V_1$ and $V_2$ in Eq.~\eqref{eq:ourform}, the second step follows from the fact that $\|\phi(s,a)\|_2\leq 1$ in Definition~\ref{def:linear_mdp}, the third step step follows from the fact that for any $x, y \geq 0$, we have $| \sqrt{ x} - \sqrt{y} |  \leq \sqrt{ | x- y|}$, the forth step follows from $\sqrt{x+y} \leq \sqrt{x} + \sqrt{y}$ for any $x,y \geq 0$, the fifth step follows from the fact that the Frobenius norm of matrix is greater than the $\ell_2$ norm.

Next, we denote $\mathcal{C}_{w}$  as the  $(\epsilon/2)$-cover of a ball $\{w\in\R^d ~|~ \|w\|_2 \leq L\}$. Using Lemma \ref{lem:euclid}, we show that it can be upper bound as: $|\mathcal{C}_w | \leq ( 1+ 4L / \epsilon ) ^d$.

Similarly, we denote $\mathcal{C}_{\Lambda}$ as the $(\epsilon^2/4)$-cover of a ball $\{ \beta^2 \Lambda^{-1} \in \R^{d\times d} ~|~ \|\beta^2 \Lambda^{-1}\|_F \le d^{1/2}B^2\lambda^{-1}\}$. Here we define the ball in $\|\cdot\|_F$. Using Lemma \ref{lem:euclid}, we show that it can be upper bound as: $|\mathcal{C}_\Lambda | \leq ( 1 +  8 d^{1/2} B^2 / (\lambda\epsilon^2)  ) ^{d^2}$.

Using, Eq.~\eqref{eq:cover_upper_new}, we know that given any $V_1 \in \mathcal{V}$, we could find a $V_2 \in \mathcal{V}$ with form $V_2(s) = \min \Big\{\| \phi(s,a) \|_{ \beta_2^2 \Lambda_2^{-1}+w_2 w_2^\top }  , H \Big\}$ where $w_2 \in \mathcal{C}_w$ and  $\beta_2^2 \Lambda_2^{-1} \in \mathcal{C}_\Lambda$, such that $\mathrm{dist}(V_1, V_2)  \leq \epsilon$. Therefore, $\mathcal{N}_{\epsilon} \leq | \mathcal{C}_w | \cdot | \mathcal{C}_\A |$. Using this inequality, we have

\begin{align*}
\log \mathcal{N}_{\epsilon}  
\leq&~ \log | \mathcal{C}_\A | +\log  | \mathcal{C}_w |   \\
\leq&~ d  \log (1+ 4L / \epsilon ) + d^2 \log  (  1 +  8 d^{1/2} B^2  / (\lambda\epsilon^2)   ).
\end{align*}
Thus, we conclude the proof.
\end{proof}

\subsection{Upper Bound on Fluctuations}\label{sec:bound_fluc}
We present a concentration lemma so that the fluctuations in LSVI-UCB is upper bounded in this section. The analysis is very standard and similar to proof in \cite{jywj20}. However, we improve the proof of \cite{jywj20} with more detailed constant dependence.
\begin{lemma} \label{lem:stochastic_term}

Let $C_{\beta} > 1$ denote a fixed constant. Let $\beta = C_{\beta} \cdot dH \sqrt{\iota}$. Let $\iota = \log(2dT/p)$.
We show that for any probability $p\in[0, 1]$ that is fixed, if we have an $\xi$ event satisfying that for all $k\in [K]$ and $h\in[H]$:
\begin{align*}
\quad \Big \|\sum_{\tau = 1}^{k-1} \phi^\tau_h \big(V^{k}_{h+1}(s^\tau_{h+1}) - [\P_h V^{k}_{h+1}](s_h^\tau, a_h^\tau)\big) \Big\|_{(\Lambda^k_h)^{-1}}
\le 30 \cdot d H\sqrt{\iota + \log(5 C_{\beta}) } , 
\end{align*}
Then, we have
\begin{align*}
\Pr[\xi] \ge 1-p/2.
\end{align*}
\end{lemma}
\begin{proof}

We show that any fixed $\epsilon>0$, we have

\begin{align}
    \label{eq:combine1_lsh}
& \Big\| \sum_{\tau = 1}^{k-1} \phi^\tau_h \big(V^{k}_{h+1}(s^\tau_{h+1}) - [\P_h V^{k}_{h+1}](s_h^\tau, a_h^\tau)\big) \Big\|_{(\Lambda^k_h)^{-1}} ^2 \notag  \\
\leq & ~ 4H^2 \left(  d \log ( 1 + k/\lambda  ) + d  \log \bigg(1+ \frac{8H \sqrt{dk }} {\epsilon\sqrt{\lambda} } \bigg) + d^2 \log \bigg( 1 +  \frac{8 d^{1/2}\beta^2 } {\epsilon^2\lambda}  \bigg)  + \log( 2 / p ) \right) + \frac{8k^2\epsilon^2}{\lambda} \notag \\
\leq & ~ 4H^2 (d \log (1+k)+d\log(1+8\sqrt{k^3/d}) + d^2\log(1+8C_{\beta}^2d^{0.5}K^2\iota) + \log(2/p)) +8 d^2 H^2 \notag \\
\leq & ~ 30 \cdot d^2 H^2 \log(10 C_{\beta} d T / p) \notag \\
= & ~ 30 \cdot d^2 H^2 ( \iota + \log(5 C_{\beta})),
\end{align}
where the first step follows from combining Lemmas \ref{lem:self_norm_covering} and \ref{lem:covering_number_new},the second step follows from $\lambda=1$, $\epsilon= d H/K$, and $\beta = C_{\beta} \cdot d H \sqrt{\iota} $, the third step follows from $C_{\beta}\geq 1$ and $\iota = \log (2dT/p)$, the last step follows from $\iota = \log (2dT/p)$.

Thus, we complete the proof. 
\end{proof}

\subsection{Upper Bound of Difference of Q Function}\label{sec:ucb_bound_q_diff}

In this section, we bound like to bound the difference between the $Q$ function $Q^{k}_h$ (see Section~\ref{sec:basic_not}) selected by Algorithm \ref{alg:sublinearLSVI-UCB} and the value function $Q^{\pi}_h$ (see Definition~\ref{def:value_function:formal}) of any policy $\pi$.  We bound the their difference by bounding $\langle\phi(s, a), w^k_h\rangle - Q_h^\pi(s, a)$. The analysis is very standard and similar to proof in \cite{jywj20}. However, we improve the proof of \cite{jywj20} with more detailed constant dependence.

\begin{lemma}\label{lem:basic_relation}
Let $\lambda=1$ in Algorithm~\ref{alg:sublinearLSVI-UCB}. Let $\iota = \log (2dT/p)$.  We show that for any policy $\pi$ that is fixed, for all $s\in {\cal S}$, $a\in {\cal A}$, $h\in [H]$ and $k\in [K]$, on the event $\xi$ defined in Lemma \ref{lem:stochastic_term}, we show that exists an absolute constant $C_\beta \geq 100$ such that

\begin{align*}
\langle\phi(s, a), w^k_h\rangle - Q_h^\pi(s, a)  - [\P_h (V^{k}_{h+1} - V^{\pi}_{h+1})] (s, a)\leq  C_\beta dH\sqrt{\iota} \cdot \| \phi(s,a) \|_{ (\Lambda^k_h)^{-1} }
\end{align*}

\end{lemma}

\begin{proof}

We start with rewriting $Q^\pi_h(s,a)$ as
\begin{align*}
Q^\pi_h(s,a) :=  \langle \phi(s,a), w^\pi_h\rangle = r_h(s,a) + [\P_h V^{\pi}_{h+1}](s,a). 
\end{align*}
where the first step follows from Proposition \ref{prop:realizable}, and the second step follows from Eq.~\eqref{eq:bellman:formal}.

Next, we show that
\begin{align*}
w^k_h -  w^\pi_h
&= (\Lambda^k_h)^{-1} \sum_{\tau = 1}^{k-1} \phi^\tau_h (r^\tau_h + V^{k}_{h+1}(s^\tau_{h+1}))- w^\pi_h  \\
&= (\Lambda^k_h)^{-1} \Big( -\lambda w^\pi_h + \sum_{\tau = 1}^{k-1} \phi^\tau_h \bigl (V^{k}_{h+1}(s^\tau_{h+1}) - [\P_h V_{\pi}^{h+1}](s_h^\tau, a_h^\tau) \bigr )\Big) \\
&=  p_1 + p_2 + p_3. 
\end{align*}
where the first step follows from the definition of $w_h^k$, the second step follows from the definition of $w_h^\pi$. the last step follows from 
\begin{align*}
    p_1 := & ~ -\lambda (\Lambda^k_h)^{-1} w^\pi_h \\
    p_2 := & ~ (\Lambda^k_h)^{-1} \sum_{\tau = 1}^{k-1} \phi^\tau_h \big(V^{k}_{h+1}(s^\tau_{h+1}) - [\P_h V^{k}_{h+1}](s_h^\tau, a_h^\tau)\big ) \\
    p_3 := & ~ (\Lambda^k_h)^{-1}\sum_{\tau = 1}^{k-1} \phi^\tau_h [\P_h (V^{k}_{h+1} - V^\pi_{h+1})](s_h^\tau, a_h^\tau)
\end{align*}

Next, we upper bound $p_1$, $p_2$ and $p_3$ separately.

We upper bound $p_1$ as,
\begin{align}\label{eq:bound_phi_q_1}
|\langle \phi(s,a), q_1\rangle| 
= & ~
\lambda  \cdot | \langle \phi(s,a), (\Lambda^k_h)^{-1} w^\pi_h\rangle| \notag \\
\leq & ~ \lambda \cdot \| w_h^\pi \|_2 \cdot \| \phi(s,a) \|_{ (\Lambda^k_h)^{-1}} \notag \\
\leq & ~  2H \sqrt{d \lambda} \cdot \| \phi(s,a) \|_{ (\Lambda^k_h)^{-1}} \notag \\
\leq & ~2H \sqrt{d } \cdot \| \phi(s,a) \|_{ (\Lambda^k_h)^{-1}},
\end{align}
where the second step follows from $|\langle a , b \rangle| \leq \| a \|_2 \cdot \| b \|_2$, and the third step follows from $\|w_h^\pi\|_2 \leq 2H\sqrt{d/\lambda}$ (see Lemma~\ref{lem:wn_estimate}), and the last step follows from $\lambda = 1$.

We upper bound $p_2$ as,
\begin{align}\label{eq:bound_phi_q_2}
|\langle \phi(s,a), q_2\rangle| \leq 30 \cdot dH\sqrt{\iota + \log(5 C_{\beta})} \| \phi(s,a) \|_{ (\Lambda^k_h)^{-1}} 
\end{align}
where the first step follows from Lemma \ref{lem:stochastic_term} on the event $\xi$.

We upper bound $q_3$ as,
\begin{align*}
\langle \phi(s,a), p_3\rangle &= \bigg \langle \phi(s,a), (\Lambda^k_h)^{-1}\sum_{\tau = 1}^{k-1} \phi^\tau_h [\P_h (V^{k}_{h+1} - V^\pi_{h+1})](s_h^\tau, a_h^\tau) \bigg \rangle\\
&=\bigg \langle \phi(s,a), (\Lambda^k_h)^{-1}\sum_{\tau = 1}^{k-1} \phi^\tau_h (\phi^\tau_h)^\top \int (V^{k}_{h+1} - V^\pi_{h+1})(s') \d \mu_h(s')\bigg\rangle\\
&= q_1 + q_2 
\end{align*}
where the first step is a reorganization, the second step decomposes the right hand side as:
\begin{align*}
    q_1 := & ~ \bigg \langle \phi(s,a), \int (V^{k}_{h+1} - V^\pi_{h+1})(s') \d \mu_h(s')\bigg \rangle \\
    q_2 := & ~ -\lambda \bigg\langle \phi(s,a), (\Lambda^k_h)^{-1}\int (V^{k}_{h+1} - V^\pi_{h+1})(s') \d \mu_h(s') \bigg\rangle
\end{align*}

Then, we rewrite $ q_1 = \P_h (V^{k}_{h+1} - V^\pi_{h+1})(s,a)$ following Definition~\ref{def:linear_mdp}. 

Next, we upper bound $q_2$ as
\begin{align}\label{eq:bound_phi_q_3}
    |q_2|  \leq & ~2 H \sqrt{d} \| \phi(s,a) \|_{  (\Lambda^k_h)^{-1}  }
\end{align}
where the first step follows from Lemma~\ref{lem:wn_estimate}.

Finally, because $\langle\phi(s,a), w^k_h\rangle - Q_h^\pi(s,a) =\langle \phi(s,a), p_1 + p_2 + p_3\rangle$, we have 
\begin{align*}
& ~|\langle\phi(s,a), w^k_h\rangle - Q_h^\pi(s,a)  -  \P_h (V^{k}_{h+1} - V^{\pi}_{h+1})(s,a)|\\ 
=& ~ \langle \phi(s,a), p_1 + p_2+q_2\rangle \\
\leq & ~ (2H\sqrt{d}+ 30 \cdot dH\sqrt{\iota + \log(5 C_{\beta})}+2H\sqrt{d})\cdot \| \phi(s,a) \|_{  (\Lambda^k_h)^{-1}  }\\
\leq &~  dH (30 \sqrt{\iota + \log(5 C_{\beta}) } + 4) \cdot \| \phi(s,a) \|_{  (\Lambda^k_h)^{-1}  },
\end{align*}
where the second step follows from combining Eq.~\eqref{eq:bound_phi_q_1}, Eq.~\eqref{eq:bound_phi_q_2} and Eq.~\eqref{eq:bound_phi_q_3}, the third step follows from $d \geq 1, H \geq 1$.

Finally, we choose an absolute constant $C_\beta$ that satisfies:
\begin{equation} \label{eq:beta_constant}
30 ( \sqrt{\iota + \log(5C_\beta )} + 4) \le C_\beta \sqrt{\iota},
\end{equation}
Note that $\iota = \log (2dT/p) \geq 4$, as long as $C_{\beta} \geq 100$ the above inequality holds

Finally, with this choice of $C_\beta$, we finish the proof.
\end{proof}

\subsection{Q Function Difference by Induction}\label{sec:ucb_induct}

In this section, we build a connection between $Q^k_{1}(s,a)$ selected by Algorithm~\ref{alg:sublinearLSVI-UCB} and $ Q_{1}^*(s,a)$. We show in Lemma~\ref{lem:induction_q_k_q_optimal} that  $ Q_{1}^*(s,a)$ is upper bounded by $Q^k_{1}(s,a)$ plus an error term related to the parameter $c$ for approximate $\maxmatnorm$ in Algorithm~\ref{alg:sublinearLSVI-UCB}.

\begin{lemma}[]\label{lem:induction_q_k_q_optimal}
Let $Q_1^k(s,a)$ denotes the estimated Q function for state $s$ when taking action $a$ at the first step. Let $Q_{1}^*(s,a)$ denotes the optimal Q function for state $s$ when taking action $a$ at the first step. Let $H$ denotes the total steps. Let $c$ is the parameter for approximate $\maxmatnorm$. We show that using Sublinear LSVI-UCB (see Algorithm~\ref{alg:sublinearLSVI-UCB}), we have   

\begin{align*}
 Q_{1}^*(s,a)-Q^k_{1}(s,a)\leq H-c\frac{1-c^{H}}{1-c}
\end{align*}

\end{lemma}

\begin{proof}
We start with bounding on the relationship between $Q^{k}_h(s,a)$ and $Q^*_h(s,a)$.
\begin{align}\label{eq:bound_Q_Qstar}
\langle\phi(s,a), w^k_h\rangle +\beta \| \phi(s,a) \|_{ (\Lambda^k_h)^{-1} }  \geq Q_h^*(s,a)+ [\P_h (V^{k}_{h+1}   - V^{*}_{h+1})](s,a)
\end{align}
where the first step follows Lemma~\ref{lem:stochastic_term}.

Next, when $h=H$, as the value functions are all zero in $H+1$ step, we have
\begin{align}\label{eq:bound_Q_Qstar_H}
    Q^k_H(s,a)\geq &~ \langle\phi(s,a), w^k_H\rangle +\beta \| \phi(s,a) \|_{ (\Lambda^k_H)^{-1} }\notag\\
    \geq&~ Q_H^*(s,a) 
\end{align}
where the first step follows from Lemma~\ref{lem:bound_q_func_chi}, the second step follows from Lemma~\ref{lem:basic_relation}.

Next, we have
\begin{align}\label{eq:VH_greate_Vstar}
    \max_{a\in {\cal A}} Q^k_H(s,a)
    \geq &~\max_{a\in {\cal A}} Q^*_H(s,a)\notag\\
    \geq &~V_{H}^*(s)
\end{align}
where the first step follows from Eq.~\eqref{eq:bound_Q_Qstar_H}, the second step follows from the definition of $V_{H}^*(s)$ in Definition~\ref{def:value_function:formal}.

Next, when $h=H-1$, we bound $[\P_h (V^{k}_{H} - V^{*}_{H})](s,a)$ as

\begin{align}\label{eq:bound_pv_vstar}
    [\P_h (V^{k}_{H} - V^{*}_{H})](s,a)
    \geq&~[\P_h (c \max_{a\in {\cal A}} Q^k_H(s,a) - V^{*}_{H})](s,a)\notag\\
    \geq &~ c[\P_h (\max_{a\in {\cal A}} Q^k_H(s,a) - V^{*}_{H})](s,a)-(1-c)[\P_h V^{*}_{H}](s,a)\notag\\
    \geq &~ c[\P_h (\max_{a\in {\cal A}} Q^k_H(s,a) - V^{*}_{H})](s,a)-(1-c)\cdot1\notag\\
    \geq &~-(1-c)\cdot 1
\end{align}
where the first step comes from the property of data structure \textsc{MatrixLSH} in Algorithm~\ref{alg:sublinearLSVI-UCB}, the second step is an reorganization, the third step follows the definition of $V_{H}^*(s)$ in Definition~\ref{def:value_function:formal}, the last step follows the Eq.~\eqref{eq:VH_greate_Vstar}.

Next, we have 
\begin{align}\label{eq:q_recur_h-1}
    Q^k_{H-1}(s,a)\geq&~ \langle\phi(s,a), w^k_{H-1}\rangle +\beta \| \phi(s,a) \|_{ (\Lambda^k_{H-1})^{-1} }\notag\\
    \geq&~ Q_{H-1}^*(s,a) +[\P_h (V^{k}_{H} - V^{*}_{H})](s,a)\notag\\
    \geq&~ Q_{H-1}^*(s,a) -(1-c)\cdot 1
\end{align}
where the first step follows from the Lemma~\ref{lem:bound_q_func_chi}, the second step follows from Eq.~\eqref{eq:bound_Q_Qstar} ,and the third step follows Eq.~\eqref{eq:bound_pv_vstar}.

Next, we have
\begin{align}\label{eq:VH_greate_Vstar_h}
    \max_{a\in {\cal A}} Q^k_{H-1}(s,a)
    \geq &~\max_{a\in {\cal A}} Q^*_{H-1}(s,a)-(1-c)\cdot 1\notag\\
    \geq &~V_{H-1}^*(s)-(1-c)\cdot 1
\end{align}
where the first step follows from Eq.~\eqref{eq:q_recur_h-1}, and the second step follows the definition of $Q^*_{H-1}(s,a)$ in section~\ref{sec:basic_not}.

Next, when $h=H-2$, we lower bound $[\P_h (V^{\hat{k}}_{H-1} - V^{*}_{H-1})](s,a)$ as

\begin{align}\label{eq:bound_pv_vstar_next}
    [\P_h (V^{k}_{H-1} - V^{*}_{H-1})](s,a)
    \geq&~[\P_h (c \max_{a\in {\cal A}} Q^k_{H-1}(s,a) - V^{*}_{H-1})](s,a)\notag\\
    \geq &~ c[\P_h (\max_{a\in {\cal A}} Q^k_{H-1}(s,a) - V^{*}_{H-1})](s,a)-(1-c)\cdot [\P_h V^{*}_{H-1})](s,a)\notag\\
    \geq &~ c[\P_h (\max_{a\in {\cal A}} Q^k_{H-1}(s,a) - V^{*}_{H-1})](s,a)-(1-c)\cdot 2\notag\\
    \geq &~-c(1-c)\cdot 1-(1-c)\cdot 2
\end{align}
where the first step comes from the \textsc{MatrixLSH} in Algoritm~\ref{alg:sublinearLSVI-UCB}, the second step is an reorganization, the third step follows the definition of $V_{H}^*(s)$ in section~\ref{sec:basic_not}, the last step follows the Eq.~\eqref{eq:VH_greate_Vstar_h}.

Next, we have 
\begin{align}
    Q^k_{H-2}(s,a)\geq &~ \langle\phi(s,a), w^k_{H-2}\rangle +\beta \| \phi(s,a) \|_{ (\Lambda^k_{H-2})^{-1} }\notag\\
    \geq&~ Q_{H-2}^*(s,a) +[\P_h (V^{k}_{H-1} - V^{*}_{H-1})](s,a)\notag\\
    \geq&~ Q_{H-2}^*(s,a)-c(1-c)\cdot 1-(1-c)\cdot 2
\end{align}
where the first step follows from the Lemma~\ref{lem:bound_q_func_chi}, the second step follows from Eq.~\eqref{eq:bound_Q_Qstar} ,and the third step follows Eq.~\eqref{eq:bound_pv_vstar_next}.

using induction from $H$ to $1$, we have

\begin{align}
    Q^k_{1}(s,a)
    \geq&~ Q_{1}^*(s,a)-(1-c)\sum_{h=1}^{H}c^{h-1}(H+1-h)\notag\\
    =&~ Q_{1}^*(s,a)-(1-c)\frac{H-cH-c+c^{H+1}}{(1-c)^2}\notag\\
   =&~ Q_{1}^*(s,a)-\frac{H-cH-c+c^{H+1}}{1-c}\notag\\
    =&~ Q_{1}^*(s,a)-\frac{H-cH}{1-c}+\frac{-c+c^{H+1}}{1-c}\notag\\
    =&~ Q_{1}^*(s,a)-(H-\frac{c-c^{H+1}}{1-c})\notag\\
    =&~ Q_{1}^*(s,a)-(H-c\frac{1-c^{H}}{1-c})
\end{align}
where the first step follows the induction rule, the remain steps are reorganizations. 

\end{proof}

We notice from Lemma~\ref{lem:induction_q_k_q_optimal} that there exists a term $H-c\frac{1-c^{H}}{1-c}$. Here we use Fact~\ref{fact:gamma_H_2} to bound this term.

\begin{fact}\label{fact:gamma_H_2}
Let $H\in \mathbb{N}$. Let $c = 1-\gamma$, for any $\gamma \in (0,1/(10 H))$, then we have
\begin{align*}
    H - c\frac{1-c^H}{1-c} \leq 2 \gamma H^2 .
\end{align*}
\end{fact}
\begin{proof}
First, by definition $\gamma = 1 -c \in (0,1)$, then we can rewrite LHS as
\begin{align*}
    H - c\frac{1-c^H}{1-c} = & ~ H - (1-\gamma) (1- (1-\gamma)^H) /\gamma \\
    \leq & ~ H - (1-\gamma) (1- e^{-H \gamma}) / \gamma \\
    \leq & ~ H - (1-\gamma) ( H - 0.5(H^2\gamma) ) \\
    = & ~ H ( 1- (1-\gamma) (1- 0.5(H \gamma)) ) \\
    \leq & ~ H \cdot ( 2 H \gamma  )=2\gamma H^2
\end{align*}
where the second step follows from $(1-\gamma)^{1/\gamma} \leq e^{-1} $, the third step follows from $1-e^{-x} \geq x -0.5 x^2$, $\forall x \in [0,1/10]$.

\end{proof}

\subsection{Recursive Formula}\label{sec:ucb_recur}

In this section, we bound the difference between $Q^k_h(s^{k}_{h}, a) $ and $ Q^{\pi_k}_h(s^{k}_{h}, a)$ in a recursive formula.

\begin{lemma}[Recursion]\label{lem:recursive_q_k_q_pi}
Let $\delta^k_h$ denotes the difference $  Q^k_h(s^{k}_{h}, a) - Q^{\pi_k}_h(s^{k}_{h}, a)$. Let $\zeta^k_{h+1} = 
\E[\delta^k_{h+1}|s^{k}_{h}, a] - \delta_{h+1}^k$ denotes the error between expectation and observed difference. Let $\beta = C_\beta dH\sqrt{\iota}$. Given the event $\xi$ defined in Lemma \ref{lem:stochastic_term}, we bound $\delta^k_h-\delta^k_{h+1}$ for any $k\in [K]$ and $h\in [H]$ as

\begin{align*}
\delta^k_h-\delta^k_{h+1} \leq  \zeta^k_{h+1} + 2\beta \| \phi(s^{k}_{h},a) \|_{ (\Lambda^k_h)^{-1}}. 
\end{align*}
\end{lemma}

\begin{proof}

We bound the $\delta^k_h$ as
\begin{align*}
\delta^k_h
=&~Q^k_h(s,a) - Q^{\pi_k}_h(s,a)\\
\leq &~[\P_h (V^{k}_{h+1} - V^{\pi_k}_{h+1})](s,a)
+ 2C_\beta dH\sqrt{\iota} \| \phi(s,a) \|_{ (\Lambda^k_h)^{-1} }.\\
=&~\zeta^k_{h+1}+\delta_{h+1}^k + 2\beta \| \phi(s,a) \|_{ (\Lambda^k_h)^{-1} }
\end{align*}
where the second step follows from  Lemma \ref{lem:basic_relation}, the third step follows from $[\P_h (V^{k}_{h+1} - V^{\pi_k}_{h+1})](s,a)=\zeta^k_{h+1}+\delta_{h+1}^k$.

Thus, we finish the proof.

\end{proof}

As our algorithm have the same upper bound on recursion with \cite{jywj20}, the upper bound on $\zeta^k_{h+1}$ in \cite{jywj20} could also be used in our analysis. We state the bound as 

\begin{lemma}[\cite{jywj20}]\label{lem:bound_sum_zeta}
Let $\zeta^k_{h+1} = 
\E[\delta^k_{h+1}|s^{k}_{h}, a] - \delta_{h+1}^k$. With probability at least $1 -p / 2$,  we show that 
\begin{align*}
  \sum_{k=1}^{K}\sum_{h=1}^H  \zeta^k_{h} \leq 2 H  \sqrt{ T\iota}, 
\end{align*}
\end{lemma}

We could also upper bound $\sum_{k=1}^{K} \sum_{h=1}^H   \| \phi(s^k_1,a^{k*}_1) \|_{ (\Lambda^k_h)^{-1}}$ following \cite{jywj20}.

\begin{lemma}[\cite{jywj20}]\label{lem:bound_sum_matrix_norm_phi}
Let $a^{k*}_1\in {\cal A}$ denotes the optimal action at state $s^k_1\in {\cal S}$. Given, $\Lambda^k_h$ estimated in each step, we have
\begin{align*}
  \sum_{k=1}^{K} \sum_{h=1}^H\| \phi(s^k_1,a^{k*}_1) \|_{ (\Lambda^k_h)^{-1}}   \leq & ~ H \cdot \sqrt{ 2 dK \iota} 
  \end{align*}
\end{lemma}

\subsection{Regret Analysis}\label{sec:ucb_regret}
In this section, we prove main theorem in Theorem \ref{thm:convergence_SLSVI-UCB:formal}.
\begin{theorem}[Convergence Result of Sublinear Least-Squares Value Iteration with UCB
 (Sublinear LSVI-UCB), a formal version of Theorem~\ref{thm:convergence_SLSVI-UCB:informal}] \label{thm:convergence_SLSVI-UCB:formal}
In a linear MDP in Definition \ref{def:linear_mdp}, we set $\lambda = 1$. Let $C_{\beta} \geq 100$ denotes a fixed constant and $ \iota = \log (2dT/p)$. If we set approximate $\maxmatnorm$ parameter $c= 1-\frac{\iota}{\sqrt{K}}$, then for any $p \in (0, 1)$ that is fixed, with probability $1-p$, Sublinear LSVI-UCB (Algorithm \ref{alg:sublinearLSVI-UCB}) has the cumulative regret at most $O(C_{\beta} \cdot \sqrt{d^3 H^3 T \iota^2})$.

\end{theorem}
\begin{proof}

We start with upper bounding the regret as:
\begin{align}\label{eq:final1}
\text{Regret}(K) 
= & ~ \sum_{k=1}^{K} \left(V^{*}_{1}  (s^k_1) - V^{\pi_k}_{1} (s^{k}_{1})\right) \notag \\
= & ~ \sum_{k=1}^{K} \left(\max_{a\in {\cal A}}Q^{*}_{1}  (s^k_1,a) - \max_{a\in {\cal A}}Q^{\pi_k}_{1}(s^k_1,a) \right) \notag \\
\leq & ~ \sum_{k=1}^{K} \left(\max_{a\in {\cal A}}Q^{*}_{1}  (s^k_1,a^{k*}_1) - Q^{\pi_k}_{1}(s^k_1,a^{k*}_1) \right) \notag \\
\leq & ~ \sum_{k=1}^{K} (Q^{k}_{1}(s^k_1,a^{k*}_1) - Q^{\pi_k}_{1} (s^k_1,a^{k*}_1)+2\gamma H^2) \notag \\
= & ~ 2\gamma K H^2+\sum_{k=1}^{K}  \delta^k_1  \notag\\
\leq & ~ 2\gamma K H^2+\sum_{k=1}^{K}\sum_{h=1}^H  \zeta^k_{h} + 2\beta \sum_{k=1}^{K}\sum_{h=1}^H  \| \phi(s^k_1,a^{k*}_1) \|_{ (\Lambda^k_h)^{-1}}
\end{align}
where the first step follows the definition of regret, the second steps follows from the definition of value function in Definition~\ref{def:value_function:formal}, the third step follows from that $\max_{a\in {\cal A}}Q^{\pi_k}_{1}(s^k_1,a)\geq Q^{\pi_k}_{1}(s^k_1,a^{k*}_1)$, where $a^{k*}_1$ is the optimal action chosen at state $s^k_1$, the forth step follows from Lemma~\ref{lem:induction_q_k_q_optimal}, the fifth step is a follows the definition of $\delta^k_h$ and $\zeta_{h}^k$ as in Lemma \ref{lem:recursive_q_k_q_pi}, the sixth step follows from Lemma \ref{lem:recursive_q_k_q_pi}.

Next, with probability $1-p$, we show that
\begin{align}\label{eq:bound_regret}
  \text{Regret}(K)  
  \leq&~2\gamma K H^2+\sum_{k=1}^{K}\sum_{h=1}^H  \zeta^k_{h} + 2\beta \sum_{k=1}^{K}\sum_{h=1}^H  \| \phi(s^k_1,a^{k*}_1) \|_{ (\Lambda^k_h)^{-1}}\notag \\
  \leq&~2\gamma K H^2+2 H \sqrt{T \iota} + 2\beta \sum_{k=1}^{K}\sum_{h=1}^H  \| \phi(s^k_1,a^{k*}_1) \|_{ (\Lambda^k_h)^{-1}} \notag\\
  \leq&~ 2K\gamma H^2 +2 H \sqrt{T \iota} +  \beta H\sqrt{ 2 dK \iota}\notag\\
  =&~ 2K\gamma H^2 +2 H \sqrt{T \iota} +  C_{\beta} \cdot\sqrt{ 2 d^3H^4 K \iota^2}\notag\\
   \leq &~ 2\sqrt{H^4K\iota^2} +2  \sqrt{H^3K \iota} +  C_{\beta} \cdot\sqrt{ 2 d^3H^4 K \iota^2}\notag\\
  \leq &~ 2C_{\beta}\sqrt{d^3H^4K\iota^2}
  \end{align}
  where the second step follows from Lemma~\ref{lem:bound_sum_zeta}, the third step follows from Lemma~\ref{lem:bound_sum_matrix_norm_phi}, the forth step from $\beta= C_{\beta} \cdot d H \sqrt{\iota}$, the fifth step follows from $\gamma= \frac{1}{\sqrt{K}}$, the sixth step is a reorganizationm the seventh step follows from $C_\beta\geq 100$.
  
  Thus, we finish our proof.

 \end{proof}

 \subsection{Running Time Analysis}\label{sec:ucb_runtime}

We present the running time analysis of our Sublinear LSVI-UCB. We first introduce the running time of each procedure of LSVI-UCB in Section~\ref{sec:lsvi_ucb_time}. Next, we introduce the running time of Sublinear LSVI-UCB in Section~\ref{sec:slsvi_ucb_time}. Therefore, we could compare their efficiency in the next section.

\subsubsection{LSVI-UCB}\label{sec:lsvi_ucb_time}
First, we show the LSVI-UCB algorithm in Algorithm~\ref{alg:LSVI-UCB}

\begin{algorithm}[!ht]
\caption{LSVI-UCB~\cite{jywj20}}\label{alg:LSVI-UCB}
\begin{algorithmic}[1]
\For{$k = 1, \ldots, K$}
\State Initialize the state $s^k_1$.
\For{$h = H, \ldots, 1$}

\State {\color{blue}/*Compute $\Lambda_h^{-1}$*/}\Comment{This step takes $O(Kd^2+d^{\omega})$}
\State $\Lambda_h \leftarrow \sum_{\tau =1}^{k-1}  \phi(s^{\tau}_h,  a^{\tau}_h)\phi(s^{\tau}_h, a^{\tau}_h)^{\top} + \lambda \cdot  \I_d$.

\State Compute $\Lambda_h^{-1}$

\State {\color{blue}/* Value Iteration*/} \Comment{This step takes $O(Kd^2A)$}
\State $w_h^k \leftarrow \Lambda_h^{-1} \sum_{\tau=1}^{k-1} \phi(s^{\tau}_h,  a^{\tau}_h) \cdot ( r_h(s^{\tau}_h, a^{\tau}_h) + \hat{V}_{h+1} (s^{\tau}_{h+1})) $  \label{line:w}

\For{$\tau=1,\cdots, k-1$}
\For{$a\in {\cal A}$}

\State $Q_h(s^{\tau}_{h+1}, a) \leftarrow \min\{\langle w_h^{k},\phi(s^{\tau}_{h+1}, a)\rangle  + \beta \cdot \|\phi(s^{\tau}_{h+1},a)\|_{\Lambda_h^{k-1}}, H\}$.
\EndFor
\State $\hat{V}_h (s^{\tau}_{h}) \leftarrow \max_a  Q_h(s^{\tau}_{h}, a)$
\State $a^{\tau}_h\leftarrow \arg \max_a Q_h(s, a) $ \Comment{$a^{\tau}_h$ is the maximum value action taken at state $s^{\tau}_{h}$.}

\EndFor

\EndFor

\State {\color{blue}/* Construct Policy*/}

\For{$h = 1, \ldots, H$} 
\State Take action $a^k_h$, and observe $s^k_{h+1}$. 
\EndFor
\EndFor
\end{algorithmic}
\end{algorithm}
\begin{lemma}
The running time of pre-computing $\Lambda^{-1}$ in Algorithm~\ref{alg:LSVI-UCB} takes time
\begin{align*}
    O(K d^2 + d^{\omega})
\end{align*}
where $\omega \approx 2.373$ is the exponent of matrix multiplication \cite{w12,l14}.
\end{lemma}
\begin{proof}
It takes $O(Kd^2)$ to compute and sum up every $\phi(s_h^{\tau},a_h^{\tau})\phi(s_h^{\tau},a_h^{\tau})^\top$. Computing the inverse matrix of $\Lambda$ takes $O(d^\omega)$. All other operations take $O(d)$. Combining the complexity together, we obtain the pre-computing complexity $O(K d^2 + d^{\omega})$.
\end{proof}

\begin{lemma}
The running time of value iteration in  Algorithm~\ref{alg:LSVI-UCB} takes
\begin{align*}
    O( H  Kd^2A )
\end{align*}
\end{lemma}
\begin{proof}
For each of the $H$ step,
\begin{itemize}
\item It takes $O(Kd^2A)$ to compute $\hat{V}_{h+1}(s_{h+1}^{\tau})$ for each state $s_{h+1}^{\tau}$.
\item It takes $O(Kd)$ to  sum up  $\phi(s^{\tau}_h,  a^{\tau}_h) \cdot ( r_h(s^{\tau}_h, a^{\tau}_h) + \hat{V}_{h+1} (s^{\tau}_{h+1}))$. 
\item It takes $O(d^2)$ to multiply $\Lambda$ with the sum of vectors. 
\item All other operations take $O(d)$.
\end{itemize}

Combining them together, we have $O(HKd^2A)$.
\end{proof}

\subsubsection{Sublinear LSVI-UCB}\label{sec:slsvi_ucb_time}
In this section, we show the runtime analysis of our Sublinear LSVI-UCB in Algorithm~\ref{alg:sublinearLSVI-UCB}.
\begin{lemma}
The running time of pre-computing $\Lambda^{-1}$ in Algorithm~\ref{alg:sublinearLSVI-UCB} takes
\begin{align*}
    O(K d^2 + d^{\omega})
\end{align*}
where $\omega \approx 2.373$ is the exponent of matrix multiplication \cite{w12,l14}.
\end{lemma}
\begin{proof}
It takes $O(Kd^2)$ to compute and sum up every $\phi(s_h^{\tau},a_h^{\tau})\phi(s_h^{\tau},a_h^{\tau})^\top$. Computing the inverse matrix of $\Lambda$ takes $O(d^\omega)$. All other operations take $O(d)$. Combining the complexity together, we obtain the pre-computing complexity $O(K d^2 + d^{\omega})$.
\end{proof}

\begin{lemma}
The running time of value iteration in Algorithm~\ref{alg:sublinearLSVI-UCB} takes
\begin{align*}
    O( H Kd^2A^{\rho}  )
\end{align*}

Further more,
\begin{itemize}
    \item If initialize the {\lsh} data-structure using Theorem~\ref{thm:ar15:formal}, $\rho=1-\frac{1}{4\sqrt{K}}$.
    \item If initialize the {\lsh} data-structure using Theorem~\ref{thm:ar17:formal}, $\rho=1-\frac{1}{8K}$.
\end{itemize}
\end{lemma}
\begin{proof}
For each of the $H$ step,
\begin{itemize}
\item It takes $O(Kd^2A^\rho )$ to compute $\hat{V}_{h+1}(s_{h+1}^{\tau})$ for each state $s_{h+1}^{\tau}$.  If we initialize the {\lsh} data-structure using Theorem~\ref{thm:ar15:formal}, we determine $\rho=1-\frac{1}{4\sqrt{K}}$ using Lemma~\ref{lem:ar15_rho_bound}. If we initialize the {\lsh} data-structure using Theorem~\ref{thm:ar17:formal}, we determine $\rho=1-\frac{1}{8K}$ using Lemma~\ref{lem:ar17_rho_bound}. 
\item It takes $O(Kd)$ to  sum up  $\phi(s^{\tau}_h,  a^{\tau}_h) \cdot ( r_h(s^{\tau}_h, a^{\tau}_h) + \hat{V}_{h+1} (s^{\tau}_{h+1}))$. 
\item It takes $O(d^2)$ to multiply $\Lambda$ with the sum of vectors. 
\item All other operations take $O(d)$.
\end{itemize}
\end{proof}

\subsection{Comparison}\label{sec:ucb_compare}

In this section, we show the comparison between our Sublinear LSVI-UCB with LSVI-UCB~\cite{jywj20}. We show the comparison results in Table~\ref{tab:LSVI-UCB_compare}.
\begin{table}[H]
    \centering
    \begin{tabular}{|l|l|l|l|} \hline
        {\bf Algorithm} & {\bf Preprocess} & {\bf \#Value Iteration} &  {\bf Regret}\\ \hline
        Ours & $O(Kd^2 A^{1+\rho_1})$ & $O( H  Kd^2A^{\rho_1}  )$ &  $O(C_{\beta}\sqrt{d^3H^4K\iota^2})$\\ \hline
        Ours & $O(Kd^2 A^{1+o(1)} )$ & $O( H  Kd^2A^{\rho_2}  )$ &  $O(C_{\beta}\sqrt{d^3H^4K\iota^2})$\\ \hline
         LSVI & 0 & $O( H  Kd^2A )$ & $O(C_{\beta}\sqrt{d^3H^4K\iota^2})$  \\ \hline
    \end{tabular}
    \caption{Comparison between Our Sublinear LSVI-UCB with LSVI-UCB \cite{jywj20}. Let $S$ denotes the quantity of available states and $A$ denotes the quantity of available actions. Let $d$ denotes the dimension of $\phi(s,a)$. Let $H$ denotes the number of steps per episode. Let $K$ denotes the total number of episodes. Let $\iota=\log(2Hd/p)$ and $p$ is the failure probability. Let $\rho_1= 1-\frac{1}{4\sqrt{K}}$ be the parameter determined by data structure in Theorem~\ref{thm:ar15:formal} and  $\rho_2=1-\frac{1}{8K}$  be the parameter determined by data structure Theorem~\ref{thm:ar17:formal}. Since $K>S$, we write the preprocessing time as $O(Kd^2A^{1+o(1)})$. This table is a detailed version of corresponding part of Table~\ref{tab:main_compare}.}

    \label{tab:LSVI-UCB_compare}
\end{table}

\section{Extension of Sublinear LSVI-UCB}\label{sec:extend}

This section extends the Sublinear LSVI-UCB with different settings.
\begin{itemize}
    \item In Section~\ref{sec:LGSC}, we introduce our Sublinear LSVI-UCB algorithm in the setting that the policy switch is limited.
    \item In Section~\ref{sec:MF}, we present the model-free version of Sublinear LSVI-UCB algorithm.
     \item In Section~\ref{sec:ext_compare}, we show that comparison of our algorithm with two LSVI-UCB extensions in terms of regret and value iteration complexity.
\end{itemize}

\subsection{LSVI-UCB Under Switch Limitation}\label{sec:LGSC}
In the limited switch setting, the number of modifications on the policy in reinforcment learning should not exceed a certain threshold. Therefore, we are required to bound the number of switches to achieve the optimal policy. \cite{gxdy21} proposes an approach to do it via LSVI-UCB. We denote this variantion as LGSC. The only different between the LGSC algorithm in \cite{gxdy21} and the LSVI-UCB algorithm in~\cite{jywj20} is that is rejects the updated policy if the change $\Lambda^k_h$ is below a threshold. Therefore, we could directly modify LGSC using the same way in Algorithm~\ref{alg:sublinearLSVI-UCB} and propose Sublinear LGSC. Moreover, we obtain the statement as follows:

 \begin{corollary}
[Convergence result of Sublinear LSVI-UCB-LGSC, an formal version of Corollary~\ref{coro:convergence_SMFLSVI-UCB:informal}] \label{coro:convergence_SLSVI-UCB_LSGSC:formal}
Let $\mdp({\cal S}, {\cal A}, H, \P, r)$ denotes a linear MDP. Given a fixed probability $p \in (0, 1)$, if we set LSVI-UCB parameter $\lambda = 1$, approximate $\maxip$ parameter $c= 1-\frac{1}{\sqrt{K}}$ and $ \iota = \log (2dT/p)$,   Sublinear LGSC cost
 has regret at most $O(\sqrt{d^3 H^3 T \iota^2})$ in total then with probability at least $1-p$.  Moreover, with $O(KA^{1+o(1)}+Kd^2A)$ preprocessing time and space, the value iteration complexity of Sublinear LGSC is $O(H Kd^2A^\rho)$, where $\rho = 1-1/K$. Moreover, the cost of global switching is at most $O(dH\log  K)$.
\end{corollary}

\begin{proof}

As the limitation on the policy switch does not affect the upper bound of regret, Sublinear LGSC have the same upper bound of regret as Sublinear LSVI-UCB.  We write the regret of LGSC following Eq.~\eqref{eq:bound_regret} in $O(\sqrt{d^3H^4K\iota^2})$. Meanwhile, the upper bound the cost of global switching is independent of the {\lsh} based $\maxip$ data structure. Therefore, Sublinear LGSC have the same upper bound in the cost of global switching, which is $O(dH\log  K)$. Moreover, as the value iteration of Sublinear LGSC is indentical to Sublinear LSVI-UCB, the query time, preprocessing time and space complexity could be determined.
\end{proof}
 
\subsection{Model-free LSVI-UCB}\label{sec:MF}

The major difference of model-free LSVI-UCB and model based LSVI-UCB is that the reward function $r$ remains to be estimated. We denote this method as MF. Therefore, MF in \cite{wdys20} contains two procedures. In the first procedure, MF performs the similar algorithm as Algorithm~\ref{alg:LSVI-UCB} except the reward function at each step of each episode is estimated by $\min\{\|\phi(s^{\tau}_{h+1},a)\|_{(\lambda_h^k)^{-1}},H\}$. Then, in the second procedure, the MF performs the same algorithm as LSVI-UCB based on the estimated reward. Accordingly, we could also propose a Sublinear MF. The Sublinear MF alternates the LSVI-UCB algorithm in the second procedure with the Sublinear LSVI-UCB. Threfore, we have the following statement.
\begin{corollary}
[Main result, convergence result of Model-free Sublinear LSVI-UCB (MF), an formal version of Corollary~\ref{coro:convergence_SMFLSVI-UCB:informal}] \label{coro:convergence_SMFLSVI-UCB:formal}
Let $\mdp({\cal S}, {\cal A}, H, \P, r)$ denotes a linear MDP. Given a fixed probability $p \in (0, 1)$, if we set LSVI-UCB parameter $\lambda = 1$, approximate $\maxip$ parameter $c= 1-\frac{1}{\sqrt{K}}$ and  $\beta =  \Theta ( d H \sqrt{\iota} ) $ with $ \iota = \log (2dT/p)$,  then using $O(KH^2\log(\sqrt{d^{-3}H^{-4}K})/\iota^2)$ episodes for exploration, MF has regret at most $O(\sqrt{d^3 H^4 K \iota^2})$ in total with probability at least $1-p$. Further more, with $O(KA^{1+o(1)}+Kd^2A)$ preprocessing time and space, the value iteration complexity of MF is $O(H Kd^2A^\rho)$, where $\rho = 1-1/K$.
\end{corollary} 
\begin{proof}
We start with several definitions. We denote $r^*:{\cal S}\times {\cal A} \rightarrow \R$ as the original reward function. We denote $r^1:{\cal S}\times {\cal A} \rightarrow \R$ as the reward function estimated by the in the exploration phase of MF.  Let $V^{*}_{1} (s_{1},r^*)$ denotes the optimal value function $\pi$ using reward $r^*$.  Let $V^{*}_{1} (s_{1},r^1)$ denotes the optimal value function$\pi$ using reward $r^1$. From \cite{wdys20}, we know that for any error $\epsilon>0$, with $O(d^3H^6\log(dHp^{-1}\epsilon^{-1})/\epsilon^2)$ episodes in exploration, $V^{*}_{1}(s^k_1, r^*) - V^{*}_{1}(s^k_1, r^1)\leq \epsilon$ for any episode $k$. Therefore if we pay $O(KH^2\log(\sqrt{d^{-3}H^{-4}K})/\iota^2)$ episodes, the would have $\epsilon\leq C_{\beta}\sqrt{d^3H^4\iota^2}/\sqrt{K}$, where $C_{\beta}\geq 100$.

Next, we upper bound the regret as:
\begin{align}
\text{Regret}(K) 
= & ~ \sum_{k=1}^{K} \left(V^{*}_{1}  (s^k_1) - V^{\pi_k}_{1} (s^{k}_{1})\right) \notag \\
= & ~ \sum_{k=1}^{K} \left(V^{*}_{1}(s^k_1, r^*) - V^{*}_{1}(s^k_1, r^1)+V^{*}_{1}(s^k_1, r^1) - V^{\pi_k}_{1} (s^{k}_{1})\right) \notag \\
= & ~ \sum_{k=1}^{K} \left(V^{*}_{1}(s^k_1, r^1) - V^{\pi_k}_{1} (s^{k}_{1})\right)+  \sum_{k=1}^{K} \left(V^{*}_{1}(s^k_1, r^*) - V^{*}_{1}(s^k_1, r^1)\right)\notag \\
= & ~  2C_{\beta}\sqrt{d^3H^4K\iota^2}+  \sum_{k=1}^{K} \left(V^{*}_{1}(s^k_1, r^*) - V^{*}_{1}(s^k_1, r^1)\right)\notag \\
= & ~ 2C_{\beta}\sqrt{d^3H^4K\iota^2}+  K\epsilon\notag \\
= & ~ 3C_{\beta}\sqrt{d^3H^4K\iota^2} \notag \\
\end{align}
where the second and third step are reorganizations, the forth step follows from Eq.~\eqref{eq:bound_regret}, the fifth step follows from $V^{*}_{1}(s^k_1, r^*) - V^{*}_{1}(s^k_1, r^1)\leq \epsilon$, the last step follows from $\epsilon\leq C_{\beta}\sqrt{d^3H^4\iota^2}/\sqrt{K}$. 

Therefore, we show that Sublinear MF achieves the same regret as LSVI-UCB and MF. Moreover, the preprocessing time, space and value iteration complexity of MF is as same as LSVI-UCB.

\subsection{Comparison}\label{sec:ext_compare}
In this section, we show the comparison between our Sublinear LSVI-UCB with LGSC~\cite{gxdy21} and MF~\cite{wdys20}. We show the comparison results in Table~\ref{tab:UCB_extend_compare}.
\begin{table}[H]
    \centering
    \begin{tabular}{|l|l|l|l|} \hline
        {\bf Algorithm} & {\bf Preprocess} & {\bf \#Value Iteration} &  {\bf Regret}\\ \hline\hline
        Ours & $O(Kd^2 A^{1+\rho_1})$ & $O( H Kd^2A^{\rho_1}  )$ &  $O(C_{\beta}\sqrt{d^3H^4K\iota^2})$\\ \hline
        Ours & $O(Kd^2 A^{1+o(1)} )$ & $O( H Kd^2A^{\rho_2} )$ &  $O(C_{\beta}\sqrt{d^3H^4K\iota^2})$\\ \hline
         LGSC & 0 & $O( HKd^2A )$ & $O(C_{\beta}\sqrt{d^3H^4K\iota^2})$  \\ \hline\hline
        Ours & $O(Kd^2 A^{1+\rho_1})$ & $O( HKd^2A^{\rho_1} )$ &  $O(C_{\beta}\sqrt{d^3H^4K\iota^2})$\\ \hline
        Ours & $O(Kd^2 A^{1+o(1)} )$ & $O( HKd^2A^{\rho_2} )$ &  $O(C_{\beta}\sqrt{d^3H^4K\iota^2})$\\ \hline
         MF & 0 & $O( HKd^2A )$ & $O(C_{\beta}\sqrt{d^3H^4K\iota^2})$  \\ \hline\hline
    \end{tabular}
    \caption{Comparison between Our Sublinear LSVI-UCB with LGSC~\cite{gxdy21} and MF~\cite{wdys20}. Let $S$ denotes the quantity of available states and $A$ denotes the quantity of available actions. Let $d$ denotes the dimension of $\phi(s,a)$. Let $H$ denotes the number of steps per episode. Let $K$ denotes the total number of episodes. Let $\iota=\log(2Hd/p)$ and $p$ is the failure probability. Let $\rho_1= 1-\frac{1}{4\sqrt{K}}$ be the parameter determined by data structure in Theorem~\ref{thm:ar15:formal} and  $\rho_2=1-\frac{1}{8K}$  be the parameter determined by data structure Theorem~\ref{thm:ar17:formal}. Since $K>S$, we write the preprocessing time as $O(Kd^2A^{1+o(1)})$. This table is a detailed version of corresponding part of Table~\ref{tab:main_compare}.}

    \label{tab:UCB_extend_compare}
\end{table}

\end{proof}

\section{More Data Structures: Adaptive $\maxip$ Queries}\label{sec:adaptive}
In this section, we show how to tackle the adaptive  $\maxip$ queries in RL.  In both Sublinear LSVI and Sublinear LSVI-UCB, the queries for $(c,\tau)$-$\maxip$ during the value iteration are adaptive but not arbitrary. Thus, we could not union bound the failure probability of {\lsh} for $(c,\tau)$-$\maxip$.  In this work, we present a quantization method to union bound the failure probability of adaptive  $\maxip$ queries.
This section is organized as:
\begin{itemize}
    \item In Section~\ref{sec:ada_maxip}, we introduce the {\lsh} data structure for adaptive $\maxip$ queries and theoretical guarantee of Sublinear LSVI with this data structure.
    \item In Section~\ref{sec:ada_maxmatnorm}, we present the {\lsh} data structure for adaptive $\maxmatnorm$ queries and theoretical guarantee of Sublinear LSVI-UCB with this data structure.
\end{itemize}

\subsection{Sublinear LSVI with Adaptive $\maxip$ Queries}\label{sec:ada_maxip}

In this section, we show how to tackle adaptive  $\maxip$ queries in Sublinear LSVI. We start with defining the quantized approximate $\maxip$.
\begin{definition}[Quantized approximate $\maxip$]\label{def:quantized_approximate_maxip}
Let $c \in (0,1)$ and $\tau \in (0,1)$. Let $\lambda \geq 0$.
Given an $n$-point dataset $Y \subset \mathbb{S}^{d-1}$, the goal of the $(c,\tau,\lambda)$-{$\maxip$} is to build a data structure that, given a query $x \in \mathbb{S}^{d-1}$ with the promise that there exists a datapoint $y \in Y$ with $\langle x , y \rangle \geq \tau$, it reports a datapoint $z \in Y$ with similarity $\langle x , z \rangle \geq c \cdot \maxip(x,Y) - \lambda$.
\end{definition}

Next, we show a standard way of performing approximate $\maxip$ via {\lsh}. We denote $Q$ as the convex hull of all queries for $(c,\tau)$-$\maxip$ and denote its maximum diameter in $\ell_2$ distance as $D_X$. Our quantization method\footnote{This is a standard trick in the field of sketching and streaming \cite{nsw19,bjwy20}.} contains two steps: (1) Preprocessing: we quantize $Q$ to a lattice $\hat{Q}$ with quantization error $\lambda/d$. In this way, each coordinate would be quantized into the multiples of $\lambda/d$.  (2) Query: given a query $x\in Q$, we first quantize it to the nearest $\hat{q}\in \hat{Q}$ and perform $(c,\tau)$-$\maxip$. As each $\hat{q}\in \hat{Q}$ is independent, we could union bound the failure probability of adaptive queries. On the other hand, this would generate an $\lambda$ additive error in the returned inner product.  

Next, we show our theorem for $(c,\tau,\lambda)$-{$\maxip$} over adaptive queries in Theorem~\ref{thm:maxip_lsh_adaptive}.
\begin{theorem}[A modified version of Theorem~\ref{thm:maxip_lsh}]\label{thm:maxip_lsh_adaptive} 
Let $c \in (0,1)$, $\tau \in(0,1)$ and $\lambda \in (0,1)$. Given a set of $n$-points $Y \subset \mathbb{S}^{d-1}$ on the sphere, one can construct a data structure with ${\cal T}_{\mathsf{init}}\cdot\kappa$ preprocessing time and ${\cal S}_{\mathsf{space}}\cdot\kappa$ space so that for every query $x\in \mathbb{S}^{d-1}$ in an adaptive sequence $X=\{x_1,x_2,\cdots,x_T\}$, we take query time complexity $O(dn^\rho\cdot\kappa)$:
\begin{itemize}
    \item if $ \maxip(x,Y)\geq\tau $, then we output a vector in $Y$ which is a $(c,\tau,\lambda)$-$\maxip$ with respect to $(x,Y)$ with probability at least $1-\delta$, where $\rho= f(c,\tau)+o(1)$.
    \item otherwise, we output {\fail}.
\end{itemize}
where $\kappa:= d\log (nd D_X / (\lambda \delta ) ) $ and $\rho\in (0,1)$.   We use $D_X$ to represent maximum diameter in $\ell_2$ distance of all queries in $X$.

Further more,
\begin{itemize}
    \item If ${\cal T}_{\mathsf{init}} = O( d n^{1+\rho})$ and ${\cal S}_{\mathsf{space}} = O(n^{1+\rho} + d n)$, then $f(c,\tau) = \frac{1-\tau}{1-2c\tau+\tau}$.
    \item If ${\cal T}_{\mathsf{init}} = O( d n^{1+o(1)})$ and ${\cal S}_{\mathsf{space}} = O( n^{1+o(1)} + d n)$, then $f(c,\tau) = \frac{2(1-\tau)^2}{(1-c\tau)^2}-\frac{(1-\tau)^4}{(1-c\tau)^4}$.
\end{itemize}
\end{theorem}

\begin{proof}
The failure probability for an adaptive sequence $X$ is equivalent to the probability that at least one query $\hat{q}\in \hat{Q}$ fail in solving all $\kappa$ number of $(c,\tau)$-{$\maxip$}. We bound this failure probability as 

\begin{align*}
    \Pr[\exists \hat{q}\in \hat{Q}~~~\textrm{s.t all } ~ (c,\tau)\textsc{-}{\maxip}~ \mathsf{fail} ]=n \cdot (\frac{dD_X}{\lambda})^d \cdot (1/10)^{\kappa}\leq \delta
\end{align*}
where the last step follows from $\kappa:= d\log (nd D_X / (\lambda \delta ) ) $.

For the success queries, it introduces a $\lambda$ error in the inner product. Thus, the results is $(c,\tau,\lambda)$-{$\maxip$}.

Then, following Theorem~\ref{thm:maxip_lsh}, we finish the proof.
\end{proof}

Next, we show a modified Version of Theorem~\ref{thm:convergence_SLSVI:formal} with $(c,\tau,\lambda)$-$\maxip$.

\begin{theorem}[Modified Version of Theorem~\ref{thm:convergence_SLSVI:formal}]\label{thm:convergence_LSVI_lsh_adpative}
Let $\mdp({\cal S}, {\cal A}, H, \P, r)$ denotes a linear MDP with core sets ${\cal S}_{\core}$, ${\cal A}_{\core}$ (see Definition~\ref{def:coresets:formal}) and span matrix $\Phi$ (see Definition~\ref{def:spansets:formal}). If we query each $\phi(s_j,a_j)$ in the $j$th row of $\Phi$ for $n=O(\epsilon^{-2} L^2 H^4 \iota)$ times, where $\iota=\log ( {Hd} / {p} )$, the output policy of Sublinear LSVI with  $(c,\tau,\lambda)$-$\maxip$ parameter $c= 1-  C_0 L\cdot \sqrt{ \iota / n} $ and $\lambda=C_0  LH\cdot \sqrt{ \iota / n} $ would be $\epsilon$-optimal with probability at least $1-p$. In other words, the regret of Sublinear LSVI is at most  $O(C_0LH^2\sqrt{\iota/n})$. Moreover, with ${\cal T}_{\mathsf{init}}\cdot\kappa$ preprocessing time and ${\cal S}_{\mathsf{space}}\cdot\kappa$ space, the value iteration complexity of Sublinear LSVI is $O( HS dA^\rho\cdot \kappa)$, where $\kappa:= d\log (nd D_X / (\lambda \delta ) ) $, $D_X$ is the maximum diameter of weight.

Further more,
\begin{itemize}
    \item If ${\cal T}_{\mathsf{init}} = O(SdA^{1+\rho})$ and ${\cal S}_{\mathsf{space}} = O(SA^{1+\rho}+SdA)$, then $\rho=1-\frac{C_0L\sqrt{\iota/n}}{4}$.
    \item If ${\cal T}_{\mathsf{init}} = O(SdA^{1+o(1)})$ and ${\cal S}_{\mathsf{space}} = O(SA^{1+o(1)}+SdA)$, then $\rho=1-\frac{C_0^2L^2\iota}{8n}$.
\end{itemize}
\end{theorem}

\begin{proof}

We start with showing the modified version of value difference. Because the quantization transforms $(c,\tau)$-{$\maxip$} into a $(c,\tau,\lambda)$-{$\maxip$} with a $\lambda$ additive error, we rewrite the value difference as:
\begin{align}\label{eq:value_diff_lsh_adaptive}
    V_1^{*}(s)-\hat{V}_1(s)
    \leq&~ \E_{\pi^*}\Big[\sum_{h=1}^{H}[(\P_h-\hat{\P}_h)\hat{V}_{h+1}](s_h,a_h) ~\Big|~s_1=s\Big]+(1-c) \sum_{h=1}^{H}(H+1-h) +\lambda\cdot H\notag\\
    =&~ \E_{\pi^*}\Big[\sum_{h=1}^{H}[(\P_h-\hat{\P}_h)\hat{V}_{h+1}](s_h,a_h) ~\Big|~s_1=s\Big]+ \frac{1-c}{2}\cdot H(H+1)+\lambda\cdot H
\end{align}
where the first step adds $\lambda$ error over each step based on Lemma~\ref{lem:value_diff_lsh}, and the second step is a reorganization.

Next, we bound the $V_1^{*}(s)-\hat{V}_1(s)$ as:
\begin{align*}
    V_1^{*}(s)-\hat{V}_1(s)
    \leq &~ 
    \E_{\pi^*}\Big[\sum_{h=1}^{H}[(\P_h-\hat{\P}_h)\hat{V}_{h+1}](s_h,a_h)|s_1=s\Big]+\frac{1-c}{2}\cdot H(H+1)+\lambda\cdot H\\
    \leq&~H \cdot L\cdot C_0\cdot H \cdot \sqrt{ \iota / n}+\frac{1-c}{2}\cdot H(H+1)+\lambda\cdot H \\
    =&~ L\cdot C_0\cdot H^2 \cdot \sqrt{ \iota / n} +\frac{1-c}{2}\cdot H(H+1)+\lambda\cdot H\\
     \leq &~ L\cdot C_0\cdot H^2 \cdot \sqrt{ \iota / n} +(1-c)H^2+\lambda\cdot H\\
    \leq &~ 2C_0LH^2\sqrt{\iota/n}+\lambda\cdot H\\
    \leq &~ 3C_0LH^2\sqrt{\iota/n}\\
    \leq & ~ \epsilon
\end{align*}
where the first step follows from Eq.~\eqref{eq:value_diff_lsh_adaptive}, the second step  follows the upper bound of $[(\P_h-\hat{\P}_h)\hat{V}_{h+1}](s,a)$ in Eq.~\eqref{eq:thm1_pp_h_lsh}, the third step is an reorganization, the forth step follows from $H\geq 1$ so that $H^2\geq H$, the fifth step follows from $1-c= C_0 L\sqrt{\iota/n}$, the sixth step follows from $\lambda=C_0 LH\cdot \sqrt{\iota/n}$, the seventh step follows from $n= O( C_0^2 \cdot \epsilon^{-2} L^2 H^4 \iota)$.

Using Theorem~\ref{thm:maxip_lsh_adaptive}, we derive the preprocessing time, space and query time for value iteration in Sublinear LSVI. Because the value iteration complexity dominates Sublinear LSVI, the final runtime complexity is  $O(HS dA^\rho\cdot \kappa)$ with $\rho$ strictly smaller than $1$. 
\end{proof}

\subsection{Sublinear LSVI-UCB with Adaptive $\maxmatnorm$ Queries}\label{sec:ada_maxmatnorm}
In this section, we show how to tackle adaptive  $\maxmatnorm$ queries in sublinear LSVI-UCB.

 We start with defining the quantized approximate $\maxmatnorm$.
\begin{definition}[Quantized Approximate $\maxmatnorm$]\label{def:quantized_approximate_maxmatnorm}
Let $c \in (0,1)$ and $\tau \in (0,1)$. Let $\lambda \geq 0$.
Given an $n$-point dataset $Y \subset \mathbb{S}^{d-1}$, the goal of the $(c,\tau,\lambda)$-{$\maxmatnorm$} is to build a data structure that, given a query $x \in \mathbb{S}^{d-1}$ with the promise that there exists a datapoint $y \in Y$ with $\langle x , y \rangle \geq \tau$, it reports a datapoint $z \in Y$ with similarity $\langle x , z \rangle \geq c \cdot \maxmatnorm(x,Y) - \lambda$.
\end{definition}

Next, we present how to extend quantized approximate $\maxip$ to approximate $\maxmatnorm$.
\begin{theorem}[A modified version of Theorem~\ref{thm:maxmatnorm_lsh}]\label{thm:maxmatnorm_lsh_adaptive}
Let $c \in (0,1)$, $\tau \in(0,1)$ and $\lambda \in (0,1)$. Let $vec$ denotes the vectorization of ${d\times d}$ matrix into a $d^2$ vector. Given a set of $n$-points $Y$ and $yy^\top\in \mathbb{S}^{d^2-1}$ for all $y\in Y$, one can construct a data structure with with ${\cal T}_{\mathsf{init}}\cdot\kappa$ preprocessing time and ${\cal S}_{\mathsf{space}}\cdot\kappa$ space so that for every query $x \in \R^{d\times d}$ with $\vect(x)\in \mathbb{S}^{d^2-1}$ in an adaptive sequence $X=\{x_1,x_2,\cdots,x_T\}$, we take query time $O( d^2 n^{\rho}\cdot\kappa)$:
\begin{itemize}
    \item if $ \maxmatnorm(x,Y)\geq\tau $, then we output a vector in $Y$ which is a $(c,\tau,\lambda)$-$\maxmatnorm$ with respect to $(x,Y)$ with probability at least $\delta$, where $\rho:= f(c,\tau) + o(1)$.
    \item otherwise, we output {\fail}.
\end{itemize}
where $\kappa:= d\log (nd D_X / (\lambda \delta ) ) $ and $\rho\in (0,1)$.   We use $D_X$ to represent maximum diameter in $\ell_2$ distance of all queries in $X$ after vectorization.

Further more,
\begin{itemize}
    \item If ${\cal T}_{\mathsf{init}} = O(d^2n^{1+\rho} )$ and ${\cal S}_{\mathsf{space}} = O(n^{1+\rho} + d^2 n)$, then $f(c,\tau) = \frac{1-\tau^2}{1-c^2\tau^2+\tau^2}$.
    \item If ${\cal T}_{\mathsf{init}} = O( d^2n^{1+o(1)})$ and ${\cal S}_{\mathsf{space}} = O( n^{1+o(1)} + d^2 n)$, then $f(c,\tau) = \frac{2(1-\tau^2)^2}{(1-c^2\tau^2)^2}-\frac{(1-\tau^2)^4}{(1-c^2\tau^2)^4}$.
\end{itemize}
\end{theorem}

\begin{proof}

We start with applying $(c^2,\tau^2,\lambda)$-$\maxip$ data structure over $\vect(x)$ and $\vect(YY^\top)$. Then, we would obtain a $z\in Y$ that

\begin{align}\label{eq:matmaxnorm_c_adaptive}
    \langle \vect(x),\vect(z z^\top) \rangle \geq  c^2 \max_{y \in Y} \langle \vect(x),\vect(y y^\top)-\lambda \rangle
\end{align}
we could use it and derive the following propriety for $z$:

\begin{align*}
    \|z\|_{x}
    =&~\sqrt{\langle \vect(x),\vect(z z^\top) \rangle}\\
    \geq &~\sqrt{c^2 \max_{y \in Y} \langle \vect(x),\vect(y y^\top) \rangle-\lambda} \\
    \geq &~\sqrt{c^2 \max_{y \in Y} \langle \vect(x),\vect(y y^\top) \rangle}-\sqrt{\lambda}\\
    \geq &~c \max_{y \in Y}\sqrt{ \langle \vect(x),\vect(y y^\top) \rangle}-\lambda\\
    =&~ c\max_{y \in Y}  \|y\|_{x}-\lambda
\end{align*}
where the second step follows from Eq.~\eqref{eq:matmaxnorm_c_adaptive}, the third step follows from Cauchy-Schwartz inequality, the forth follows from $\lambda\in (0,1)$, the last step is a reorganization.

Thus, $z$ is the solution for $(c,\tau,\lambda)$-$\maxmatnorm(x,Y)$. Next, applying Theorem~\ref{thm:maxip_lsh_adaptive} , we finish the proof.

\end{proof}

\begin{theorem}[Modified Version of Theorem~\ref{thm:convergence_SLSVI-UCB:formal}] \label{thm:convergence_SLSVI-UCB_adaptive}
Let $\mdp({\cal S}, {\cal A}, H, \P, r)$ denotes a linear MDP. For any probability $p \in (0, 1)$ that is fixed, if we set approximate $\maxmatnorm$ parameter $c= 1-\frac{\iota}{\sqrt{K}}$,  quantization error $\lambda\leq \sqrt{H^2K} $ and Sublinear LSVI-UCB parameter $\beta =  \Theta ( d H \sqrt{\iota} ) $ with $ \iota = \log (2dT/p)$, then the Sublinear LSVI-UCB (Algorithm \ref{alg:sublinearLSVI-UCB}) has regret at most $O(C_\beta\cdot \sqrt{d^3 H^4 K \iota^2})$ with probability $1-p$.  Moreover, with ${\cal T}_{\mathsf{init}}\cdot\kappa$ preprocessing time and ${\cal S}_{\mathsf{space}}\cdot\kappa$ space, the value iteration complexity of Sublinear LSVI-UCB is $O(H Kd^2A^\rho\cdot \kappa)$, where $\kappa:= d\log (nd D_X / (\lambda \delta ) ) $, $D_X$ is the maximum diameter of weight.

Further more
\begin{itemize}
    \item If ${\cal T}_{\mathsf{init}} = O(Kd^2A^{1+\rho})$ and ${\cal S}_{\mathsf{space}} = O(KA^{1+\rho}+Kd^2A)$, then $\rho=1-\frac{1}{4\sqrt{K}}$.
    \item If ${\cal T}_{\mathsf{init}} = O(Kd^2A^{1+o(1)})$ and ${\cal S}_{\mathsf{space}} = O(KA^{1+o(1)}+Kd^2A)$, then $\rho=1-\frac{1}{8K}$.
\end{itemize}
\end{theorem}

\begin{proof}
We start with showing the modified version of Q-function difference $Q_{1}^*(s,a)-Q^k_{1}(s,a)$. Because the quantization transforms $(c,\tau)$-{$\maxip$} into a $(c,\tau,\lambda)$-{$\maxip$} with a $\lambda$ additive error, we rewrite the $Q_{1}^*(s,a)-Q^k_{1}(s,a)$ as:
\begin{align*}
Q_{1}^*(s,a)-Q^k_{1}(s,a) \leq (H-c\frac{1-c^{H}}{1-c})+H\lambda
\end{align*}

Next, we could upper bound the regret with probability $1-p$ as:
  \begin{align*}
  \text{Regret}(K)  
  \leq&~ 2K\gamma H^2 +2 H \sqrt{T \iota} +  \beta H\sqrt{ 2 dK \iota}+H\lambda\\
  =&~ 2K\gamma H^2 +2 H \sqrt{T \iota} +  C_{\beta} \cdot\sqrt{ 2 d^3H^4 K \iota^2} +H\lambda\\
   =&~ 2\sqrt{H^4K\iota^2} +2  \sqrt{H^3K \iota} +  C_{\beta} \cdot\sqrt{ 2 d^3H^4 K \iota^2}+H\lambda\\
   =&~ 3\sqrt{H^4K} +2  \sqrt{H^3K \iota} +  C_{\beta} \cdot\sqrt{ 2 d^3H^4 K \iota^2}\\
  \leq &~ 2C_{\beta}\sqrt{d^3H^4K\iota^2}
  \end{align*}
  where the first step follows from   Eq.~\eqref{eq:bound_regret}, the second step follows from $\beta= C_{\beta} \cdot d H \sqrt{\iota}$, the third step follows from $\gamma= \frac{1}{\sqrt{K}}$, the forth step is a reorganization follows from $\lambda\leq \sqrt{H^2K}$, the last step follows from $C_\beta\geq 100$.

 Using Theorem~\ref{thm:maxmatnorm_lsh_adaptive}, we derive the preprocessing time, space and query time for value iteration in  Sublinear LSVI-UCB. Because the value iteration complexity dominates LSVI-UCB, the final runtime complexity is  $O(H Kd^2A^\rho\cdot\kappa)$ with $\rho$ strictly smaller than $1$.  We alternate the $S$ in preprocessing and space by $K$ since $K>S$. Note that to let $\rho$ strict less than $1$. We set $c^2\in [0.5,1)$ and $\tau^2\in [0.5,1)$.
\end{proof}

\section*{Acknowledgements}

The authors would like to thank Lijie Chen for very useful discussions about the literature of hardness results of $\maxip$ problems. The authors would like to thank Yihe Dong, Ilya Razenshteyn, Tal Wagner, and Peilin Zhong for helpful discussion on locality sensitive hashing ({\lsh}).  The authors would like to thank Wen Sun for helpful discussions on shifting reward function. The authors would like to thank Simon Du for useful discussions about reinforcement learning literature. The authors would like to thank Rajesh Jayaram for discussing adaptive queries.

\ifdefined\isarxiv
\newpage
\addcontentsline{toc}{section}{References}
\bibliographystyle{alpha}
\bibliography{ref}
\else
\fi

\end{document}